# TEM analyses of in situ presolar grains from unequilibrated ordinary chondrite LL3.0 Semarkona

S. A. Singerling[1*], L. R. Nittler[2], J. Barosch[2], E. Dobrică[3], A. J. Brearley[4], and R. M. Stroud[1]

[1]U.S. Naval Research Laboratory, Code 6366, Washington, DC 20375, USA
[2]Carnegie Institution of Washington, Washington, DC 20015, USA
[3]University of Hawai'i at Mānoa, Honolulu, HI, 96822, USA
[4]University of New Mexico, Albuquerque, NM, 87131, USA

*NRC Postdoc; corresponding author current institution: Virginia Polytechnic Institute and State University, Blacksburg, VA 24061, USA; email: ssheryl@vt.edu

## Abstract

We investigated six presolar grains from very primitive regions of the matrix in the unequilibrated ordinary chondrite Semarkona with transmission electron microscopy (TEM). These grains include one SiC, one oxide (Mg-Al spinel), and four silicates. This is the first TEM investigation of presolar grains within an ordinary chondrite host (in situ) and the first TEM study to report on any presolar silicates (in or ex situ) from an ordinary chondrite. Structural and elemental compositional studies of presolar grains located within their meteorite hosts have the potential to provide information on conditions and processes throughout the grains' histories.

Our analyses show that the SiC and spinel grains are stoichiometric and well crystallized. In contrast, the majority of the silicate grains are non-stoichiometric and poorly crystallized. These findings are consistent with previous TEM studies of presolar grains from interplanetary dust particles and chondritic meteorites. The individual silicates have Mg#'s ranging from 15 to 98. Internal compositional heterogeneities were observed in several grains, including Al in the SiC, Mg and Al in the spinel, and Mg, Si, Al, and/or Cr in two silicates. We interpret the poorly crystalline nature, non-stoichiometry, more Fe- rather than Mg-rich compositions, and/or compositional heterogeneities as features of the formation by condensation under non-equilibrium conditions.

Evidence for parent body alteration includes rims with mobile elements (S or Fe) on the SiC grain and one silicate grain. Other features characteristic of secondary processing in the interstellar medium, the solar nebula, and/or on parent bodies, were not observed or are better explained by processes operating in circumstellar envelopes. In general, there was very little overprinting of primary features of the presolar grains by secondary processes (e.g., ion irradiation, grain-grain collisions, thermal metamorphism, aqueous alteration). This finding underlines the need for additional TEM studies of presolar grains located in the primitive matrix regions of Semarkona, to address gaps in our knowledge of presolar grain populations accreted to ordinary chondrites.

# 1. INTRODUCTION

Presolar grains are among the original building blocks of our solar system and predate its formation by up to several billion years (Heck et al., 2020). This extrasolar dust is identified by its anomalous isotopic compositions, the ratios of which are sometimes orders of magnitude different from all other material in the solar system. Presolar grains are nm to µm in size and include phases of varying mineralogy, such as carbonaceous phases (i.e., graphite, diamond, silicon carbide), oxides, and silicates (Lodders and Amari, 2005; Zinner, 2014; Nittler and Ciesla, 2016). In this study, we primarily focus on presolar silicon carbides (SiC), oxides, and silicates. Presolar SiC grains can form in a number of different polytypes, which differ in the stacking order of their Si-C bond pairs, but are primarily 3C (cubic) and 2H (hexagonal) (Daulton et al., 2003). Presolar oxides include corundum ($Al_2O_3$), Mg-Al spinel ($MgAl_2O_4$), and less commonly titanium dioxide ($TiO_2$), chromite ($FeCr_2O_4$), magnetite ($Fe^{2+}Fe^{3+}_2O_4$), and hibonite ($CaAl_{12}O_{19}$). Presolar silicates include olivine ($(Mg,Fe)_2SiO_4$), pyroxene ($Ca_{0\text{-}1}(Mg,Fe)_{1\text{-}2}Si_2O_6$), non-stoichiometric materials, and less commonly silica ($SiO_2$) and silicate perovskite ($MgSiO_3$) (Lodders and Amari, 2005; Zinner, 2014; Floss and Haenecour, 2016; Nittler and Ciesla, 2016). Composite grains have also been identified. Some of these contain refractory phases and are reminiscent of some of the earliest formed solids in the solar system, such as calcium-aluminum-rich inclusions (CAIs) and amoeboid olivine aggregates (AOAs) (Vollmer et al., 2009b; Nguyen et al., 2010; Floss and Stadermann, 2012; Vollmer et al., 2013; Nguyen et al., 2014; Leitner et al., 2018; Nittler et al., 2018).

Presolar grains are divided into groups based on their isotopic compositions, as these provide insights into their stellar origins. For SiC grains, those groups are based on C, N, and Si isotopes and include mainstream, X, Y, Z, AB, C, and PNG grains (Lodders and Amari, 2005; Zinner, 2014; Nittler and Ciesla, 2016). For oxides and silicates, the groups are based on O isotopes and include groups 1, 2, 3, and 4 (Nittler et al., 1997), though grains outside these groupings exist (e.g., $^{16}O$-rich grains, extreme Group 1 grains) (Nittler et al., 2008; Vollmer et al., 2008). The anomalous isotopic compositions of presolar grains are inconsistent with formation by any known mechanisms in our solar system, but rather, are the product of formation in circumstellar environments ranging from the atmospheres of evolved low mass stars to the ejecta produced during the explosive deaths of massive stars. For presolar SiC, the mainstream, Y, and Z grains are purported to form around evolved low mass stars during their asymptotic giant branch phase (AGB stars), the X and C grains in the ejecta of core-collapse supernovae, the AB grains around more exotic C-rich stars, and the PNG grains in the ejecta of novae. For presolar oxides and silicates, most Groups 1, 2, and 3 are theorized to have formed around AGB stars, with the different groups reflecting extra mixing (cool bottom processing and/or hot bottom burning) and/or different masses or metallicities of the stars. The origins of Group 4 grains are less well-constrained, but many likely formed in the ejecta of core-collapse supernovae (e.g., Zinner, 2014).

Although presolar grains are rare, they are preserved to varying degrees in primitive meteorites (ppb to 100s of ppm, matrix-normalized), interplanetary dust particles (IDPs) (up to 1 wt.%, normalized), and samples from primitive bodies returned by spacecraft, such as JAXA's Hayabusa2 (Lodders and Amari, 2005; Zinner, 2014; Nittler and Ciesla, 2016; Nittler et al., 2022). Unequilibrated ordinary chondrites are a class of primitive meteorites that have the potential to retain pristine features, including presolar grains. This class of meteorites is composed primarily of chondrules (up to ~80 vol%) and fine-grained matrix materials (often ≤15 vol%) (Scott and Krot, 2005). Although unequilibrated ordinary chondrites are primitive, they have experienced secondary processes to varying degrees, including thermal metamorphism from heating (maximum temperatures of up to 600°C) (Huss et al., 2006) and aqueous alteration from interaction with fluids, mostly in their asteroidal parent bodies. The LL3.0 unequilibrated ordinary chondrite Semarkona escaped significant heating but underwent aqueous alteration (e.g., Hutchison et al., 1987; Alexander et al., 1989; Krot et al., 1997; Grossman and Brearley, 2005). While large parts of the matrix are highly altered, a recent TEM study identified amorphous silicate-rich regions of relatively pristine matrix material which escaped such alteration in Semarkona (Dobrică and Brearley, 2020). A NanoSIMS (secondary ion mass spectroscopy) search of these regions identified abundant presolar grains, consistent with minimal secondary processing (Barosch et al., 2021).

The study of presolar grains still contained within their meteorite host (hereafter, in situ) using focused ion beam-transmission electron microscopy (FIB-TEM) is a powerful approach to elucidate the grains' microstructural characteristics, elemental compositional properties, and their association with solar system materials. Early TEM analyses of presolar grains were ex situ, wherein the presolar grains were separated from their meteorite host by acid dissolution and density separation techniques (Amari et al., 1994). The propensity for ex situ work was mostly due to instrumental challenges presented by the sub-micron size and low relative abundance of the grains compared to similar matrix materials of solar system origin. The introduction of the NanoSIMS enabled isotopic measurements with 100-nm resolution, which made in situ studies of smaller (~200 nm) presolar grains possible. Additionally, presolar silicates are destroyed during acid dissolution, so in situ NanoSIMS studies allowed for the first measurements of presolar silicates (Messenger et al., 2003).

In situ studies allow us to better trace entire histories of individual presolar grains, from their formation in stellar environments, through their residence in the interstellar medium and protosolar disk, and their incorporation into meteorite parent bodies. A number of in situ studies have analyzed presolar oxides and silicates from carbonaceous chondrites (CCs) and IDPs (Messenger et al., 2003; Messenger et al., 2005; Nguyen et al., 2007; Vollmer et al., 2007; Busemann et al., 2009; Stroud et al., 2009; Vollmer et al., 2009b; Nguyen et al., 2010; Bose et al., 2012; Leitner et al., 2012; Vollmer et al., 2013; Zega and Floss, 2013; Nguyen et al., 2014; Zega et al., 2015; Nittler et al., 2018; Leitner et al., 2020; Nittler et al., 2020). Our current study is the first to employ FIB-TEM to investigate the structural and elemental compositional characteristics of in situ presolar grains from an unequilibrated ordinary chondrite, in an effort to: 1) determine what the properties of the grains can tell us about their histories from condensation conditions in the progenitor AGB stars to the interstellar medium, solar nebula, and asteroidal parent body, and 2) compare the characteristics of presolar grains from ordinary chondrites (OCs) to those from other planetary materials (i.e., CCs and IDPs).

# 2. METHODS

## 2.1 NanoSIMS Isotopic Measurements

We report data for six presolar grains: one SiC (F2-30a), one oxide (F2-37), and four silicates (F2-9, F2-30b, F2-23, F1-1). An additional, larger composite presolar grain (F2-8) with silicate and oxide components was also extracted and studied, but will be treated in a separate paper given its unique characteristics. The grains are a subset of presolar grains identified by their highly anomalous C and/or O isotopic compositions during an automated NanoSIMS isotopic imaging search (Barosch et al., 2021) of the Semarkona thin section (UNM 102) first studied by Dobrică and Brearley (2020). The NanoSIMS mapping methods were essentially identical to that used by Nittler et al. (2021) with a <150-nm $Cs^+$ beam being scanned over 10×10 $\mu m^2$ areas with simultaneous detection of negative secondary ions of the C and O isotopes as well as $^{28}Si$ and $^{27}Al^{16}O$ to aid in mineralogical characterization. Following their identification by automated imaging, the grains were re-located and analyzed for their Mg and Si isotopic ratios. A 2.4 pA $O^-$ beam of ~120 nm diameter was rastered over the grains (5×5 $\mu m^2$ or smaller) to produce positive secondary ions of the Mg and Si isotopes as well as $^{27}Al$ in multi-collection imaging mode. Measurement conditions were similar to those reported by Hoppe et al. (2018), and we used the surrounding matrix as a normalization standard for Mg and Si isotopic ratios. All images were processed with the L'image software (L. Nittler, Carnegie Institution).

In addition to the initial isotopic measurements, we performed further NanoSIMS C- and O-isotopic mapping on the FIB sections following the TEM analyses. The FIB sections were transferred from the original Cu TEM half grids, Pt-welded onto a Si substrate, and sputter-coated with a thin coating of Au. NanoSIMS measurement conditions were essentially identical to the first round of mapping, except that the raster size was set independently for each FIB section and the measurements were manually monitored. This second round of NanoSIMS measurements allowed us to determine if we indeed successfully sampled the presolar grains within the FIB sections. This is significant because it is, in general, unknown how much material from a presolar grain remains below the surface after NanoSIMS measurements. Additionally, alignment of NanoSIMS and FIB-SEM images can be difficult. To our knowledge, no previous TEM studies of in situ grains have included multiple rounds of NanoSIMS analyses both before and after TEM analyses with the exception of Floss et al. (2014), who performed coordinated NanoSIMS-FIB-TEM analyses on some organic grains which were isotopically anomalous. Ideally, this second round of isotopic measurements should be performed in all in situ TEM studies of presolar grains, but the time-consuming nature of such techniques makes this approach unrealistic. Rather than trying to apply this technique for all studies of presolar grains, it can instead be employed when the presence of a presolar grain is uncertain (for example when no clear grain boundary of an expected size is seen in the TEM images). NanoSIMS mapping of the FIB sections can provide evidence for the absence or presence of isotopic anomalies which would, in turn, confirm or refute identification of presolar grains in the TEM data.

### 2.2 Focused Ion Beam Sample Preparation

We used the focused ion beam (FIB) sample preparation technique to prepare the presolar grains and surrounding matrix material for transmission electron microscopy (TEM) analyses. We used a FEI Helios G3 SEM-FIB to perform in situ liftout of five FIB sections containing the six presolar grains; one section contained two grains that were in close proximity—SiC F2-30a and silicate F2-30b. In order to identify the locations of presolar grains for liftout, we aligned SIMS isotope images and SEM-FIB secondary electron images as image overlays, and deposited C and Pt fiducial markers on the grain centers and directions of thinning. FIB section preparation involved: 1) depositing fiducial markers on the presolar grains followed by a protective C or Pt strap using a FEI Multichem Gas Injection System; 2) cutting a lamella of the grain from the surrounding matrix with a $Ga^+$ ion beam; 3) extracting the lamella using a FEI EasyLift NanoManipulator System; 4) welding the lamella onto a TEM Cu half grid with C and/or Pt deposition; and 5) thinning the lamella to electron transparency (i.e., 80–100 nm) using progressively smaller ion beam currents and accelerating voltages. The working distance was 4 mm. Imaging with the electron beam was done at 5 kV and 1.6 nA. Conditions for the ion beam varied—30 kV and 25 or 40 pA for imaging, 30 kV and 0.79 or 2.5 nA for milling, 30 kV and 40 or 80 pA for deposition, 30 kV and 0.23 nA for initial thinning, and 16 kV and 50 down to 23 pA for final thinning. After our TEM analyses were complete, we used the FIB to cut the grain lamellae from the Cu grids and welded them to a Si substrate for the second round of NanoSIMS measurements.

### 2.3 Transmission Electron Microscopy

All TEM data was collected between the two sets of NanoSIMS analyses—initial isotopic measurements on the grains in the polished thin section and re-analyses of isotopic data on the remounted FIB sections. Given the complicated nature of remounting and damage from the ion beam, we did not perform TEM analyses after the second set of NanoSIMS measurements.

We analyzed the presolar grains for structural and elemental compositional information on two TEMs—a JEOL 2200FS TEM and a Nion UltraSTEM-200X, both operating at 200 kV with double-tilt holders. The JEOL was primarily used for microstructural work, whereas the Nion was used for compositional work, including EDS and EELS. On the JEOL, a Gatan OneView® camera was used to collect bright-field (BF), high-resolution (HR), and dark-field (DF) images, and selected area electron diffraction (SAED) patterns. To collect the DF images, we placed a high contrast objective aperture over a diffraction spot in diffraction mode, then switched to imaging mode and adjusted the brightness to get a representative DF image. Image magnification calibrations were performed with a Multi-Cal Si-Ge lattice standard, whereas the diffraction camera lengths were calibrated with a polycrystalline Al diffraction standard. In most cases, the small sizes of the presolar grains precluded the use of SAED patterns for meaningful microstructural information. We instead performed Fast Fourier Transformations (FFTs) on representative regions within HRTEM images in order to identify phases as well as get more detailed information on the crystallinity of the grains. For phase identification, we measured the d-spacings of diffraction spots and compared those to the literature values of phases with appropriate compositions and that have been observed in presolar materials (Table S1). Given the nanocrystalline nature of most of the grains, more detailed electron diffraction indexing (i.e., tilting the sample to get at multiple zone axes) was not practical. Gatan DigitalMicrograph® software was used for image processing, including diffraction patterns.

On the Nion, a high-angle annular detector with an inner angle of 66 mrad was used to collect scanning transmission electron microscope (STEM) high-angle annular dark-field (HAADF) images. A windowless 0.7 sr Bruker Xflash® SDD detector was used to collect energy dispersive X-ray spectroscopy (EDS) maps at a nominal current of 65 pA. EDS maps were collected for 15–20 mins with a resolution of 4 nm/pixel and over 492–851 frames. For processing of extracted EDS spectra, we used the Bruker ESPRIT software to calibrate spectra, correct the background, and check the deconvolution. Quantification was done using the Cliff Lorimer method with detector-specific computed k factors. We excluded any thickness correction for the Cliff Lorimer quantification given the unknown densities of many of the phases, with the exception of the SiC analyses. In the spectra of SiC analyses, we noted that the strobe peak (0 keV) was lower than the Si Kα peak, indicating the sample was thick causing multiple scattering. For the quantification of SiC, we performed thickness corrections to get 1:1 ratio for Si:C. For all EDS data quantification, the following elements were included only for deconvolution purposes, as they represent contamination from sample preparation or components within the TEM or holder: Cu, Pt, Ga, Cs, and Zr. We noted excess O in all EDS analyses, which was likely due to a combination of overlapping phases in the thickness of the section, incipient hydration of the presolar grains, and oxidation of $Ga^{+}$ irradiated FIB sections surfaces. Owing to the difficulty in determining the relative contributions of those effects, we opted to use atomic ratios for determining stoichiometry and comparing compositional information. All EDS raw data are summarized in Table S2. Electron energy loss spectroscopy (EELS) analyses of vesicles was performed at 60 kV with a Gatan Enfinium ER EELS spectrometer and a 3 mm aperture. The EELS data were collected as spectrum images (SIs), with separate SIs for the zero-loss peak (ZLP) and the O K core loss edge (532 eV). The SIs were collected with a 0.02 eV dispersion and as 50 by 50 pixel maps with pixel dwell times of 0.1 s/pixel, exposures of 0.0005 s (ZLP) and 0.1 s (core loss), and for total times of 728–732 s. The ZLP SI was used to correct for energy drift during acquisition of the core loss SI. All EELS analyses were performed in Gatan DigitalMicrograph®.

# 3. RESULTS

### 3.1 NanoSIMS Results

The O- and Mg-isotopic compositions of the O-bearing presolar grains from Semarkona are summarized in Table 1; Si isotopic ratios were close to solar with relatively large errors. The grains fall into the $^{17}$O-rich and normal-to-depleted $^{18}$O Groups 1 and 2 (Section 1 above; Nittler et al., 2008), whose isotopic compositions point to origins in low-mass AGB stars (e.g., Nittler et al., 2008). However, recent studies of Mg isotopes in presolar silicates suggest other possible sources for some Group 1 grains (Leitner and Hoppe, 2019; Hoppe et al., 2021). None of the grains studied here show the unusually high $^{25}$Mg excesses suggested by Leitner and Hoppe (2019) to be a supernova signature. Based on the combined O and Mg isotope data, four of the grains almost certainly formed in AGB stars. The fifth, Group 1 grain F1-1, falls in the "$^{26}$Mg-rich Group 1" category of Hoppe et al. (2021), who argued that the isotopic characteristics of such grains are most consistent with an origin in massive stars, either in red supergiant (RSG) winds or during a supernova explosion. We favor a RSG origin for this grain as it does not require ad hoc mixing of supernova ejecta with a pre-existing dense shell of circumstellar matter as proposed by Hoppe et al. (2021). We note that the silicate dust condensation conditions in RSG winds may be similar to those experienced in AGB star outflows (Cherchneff, 2013).

The NanoSIMS re-analysis of the FIB sections confirmed the presence of four of the six presolar grains—SiC grain F2-30a, oxide grain F2-37, and silicate grains F2-9 and F1-1—whereas data on the other two grains, silicates F2-23 and F2-30b, were inconclusive. We show isotopic maps and associated STEM HAADF images of the grains in Figures 1 and A1; SiC grain F2-30a is excluded owing to overlapping FIB-deposited C. Aligning STEM HAADF images to NanoSIMS isotopic maps is not straightforward given the differences in resolution between the two techniques (100 nm for NanoSIMS and 1 nm for TEM) and the possibility of distortions due to tilting in either instrument. Instead, the C fiducial markers can be used to help determine the relative location of the presolar grain to the marker. In the figures with isotopic maps, these markers are represented by white outlines to decrease clutter. The confirmed presolar grains are identified by their O isotopic anomalies (F2-37 in Fig. 1a–c, F2-9 in Fig. S1a–b, and F1-1 in Fig. S1f–h). On the other hand, grains F2-30b (Fig. S1c–e) and F2-23 showed inconclusive results. For F2-23, an O-isotopic anomaly was present in one map (Fig. 1e) but then absent in a second map (Fig. 1f). It is likely that only a small amount of the original presolar material survived in the FIB section and was destroyed during the first NanoSIMS re-analysis map. Similarly, in the case of F2-30b, too little material from the original grain itself may remain in the FIB section for the spatial resolution of the NanoSIMS to detect it as anomalous. If the grains were sputtered away by NanoSIMS re-analysis, it could indicate that they were smaller than the thickness of the FIB section (~100 nm).

If this was the case, the elemental compositional data obtained from EDS analyses would include adjacent meteorite matrix materials. We will discuss the implications of this within the TEM data results sections for F2-30b and F2-23.

Note also that the NanoSIMS map of the F2-37 FIB section revealed an additional $^{13}C$-rich presolar grain (F2-37* in Table 1), with $^{13}C/^{12}C$ at least eight times the terrestrial value. This $^{13}C$-rich grain was very close to the targeted presolar spinel that appears as a hotspot in $AlO^-$ ions (Fig. 1b) and the $^{17}O/^{16}O$ ratio (Fig. 1c). The $^{13}C$-rich grain was below the original surface of the Semarkona section and thus was not visible during the initial NanoSIMS mapping. It is likely to be SiC, based on this being the most abundant $^{13}C$-rich presolar phase in meteorites. However, it was clearly smaller than the 100 nm beam size of the NanoSIMS and was not visible in TEM EDS data of the same area.

**Table 1.** Isotopic compositions of Semarkona presolar grains studied here. Errors are 1-$\sigma$, based on counting statistics.

| **O-bearing Grains** | | | | | | |
|---|---|---|---|---|---|---|
| **Grain** | **Progenitor** | **Isotopic Group** | **$^{17}O/^{16}O$ ($\times 10^{-4}$)** | **$^{18}O/^{16}O$ ($\times 10^{-3}$)** | **$\delta^{25}Mg/^{24}Mg$** | **$\delta^{26}Mg/^{24}Mg$** |
| F2-37 | AGB star | Group 1 | 7.41 ± 0.29 | 2.02 ± 0.05 | 136 ± 20 | 200 ± 20 |
| F2-9 | AGB star | Group 2 | 9.34 ± 0.33 | 0.80 ± 0.05 | 16 ± 5 | 90 ± 5 |
| F2-30b | AGB star | Group 1 | 10.14 ± 0.36 | 2.02 ± 0.05 | 3 ± 9 | 7 ± 8 |
| F2-23 | AGB star | Group 1 | 6.40 ± 0.24 | 1.80 ± 0.04 | -36 ± 7 | -38 ± 6 |
| F1-1 | RSG star | $^{26}Mg$-rich Group 1 | 5.90 ± 0.13 | 1.57 ± 0.02 | -68 ± 7 | 136 ± 8 |
| **SiC Grains** | | | | | | |
| **Grain** | **Progenitor** | **Isotopic Group** | **$^{12}C/^{13}C$** | **$\delta^{29}Si/^{28}Si$ (‰)** | **$\delta^{30}Si/^{28}Si$ (‰)** | |
| F2-30a | AGB star | Mainstream | 53.5 ± -1.3 | 100 ± -9 | 115 ± -6 | |
| F2-37* | AGB star | AB | 11 ± -1 | N/A | N/A | |

AGB – asymptotic giant branch, RSG – red supergiant, N/A – system not measured.

In addition to imparting information on the presolar grains, the follow-up NanoSIMS analyses identified another interesting feature of the FIB sections, namely the presence of a thin $^{18}O$-depleted layer at top of several of the sections. This appears to correspond to elliptical, vesicle-like features noted in all FIB sections studied. These vesicles occur near the interface between the meteoritic material and the C and/or Pt deposited during sample preparation. They range in size from 100 by 60 nm to about 10 by 10 nm, although the majority tend to be 70 to 30 nm. Figure 2a–b illustrates examples of these vesicles (indicated by the arrows) near presolar grains F2-37 and F2-9 (approximate location indicated by the yellow-shaded regions); additional images and data are shown in Figure S2. In some cases, the vesicles are located to the sides of the presolar grains (Fig. 2b), but in others, they overlap the grains themselves (Fig. 2a). The vesicles are similar to those observed with TEM from experimental ion irradiation done on meteoritic materials (e.g., Laczniak et al., 2021), but these features have not been previously described in studies of in situ presolar grains. As such, it is unlikely that FIB sectioning is the cause of the vesicle formation, especially since our sample preparation protocol follows those of similar studies (e.g., Stroud, 2003). The vesicles show greater (O+C)/cation ratios than adjacent materials (control) (Fig. 2c). The ratio allows us to better correct for the dependence of the Cliff Lorimer quantification method on sample thickness and density of the phases present. To get around the effects of C contamination from FIB-deposited protective straps/fiducial markers, the abundances of O and C are summed. Large vesicles may have been transected during FIB sectioning, exposing the vesicles to vacuum and causing the entrapped gases to escape. Smaller vesicles, however, remained intact. When measuring the compositions of the vesicles using EDS, we were likely sampling both the gases entrapped in the vesicles as well as any adjacent material within the thickness of the FIB section.

In addition to EDS analyses, we collected EELS spectral images (SIs) of two vesicles, but did not find any significant differences between the vesicles and adjacent material, either near the O K edge or the loss low region (Fig. S3). This could be because the SIs' energy resolution was not high enough to allow for the detection of the $O_2$ or $H_2O$ features, the vesicles analyzed were transected and the gases escaped, and/or because the O contamination from oxidation during or following FIB sample preparation, which we described in the Methods section, was obscuring O features. The most likely source of these vesicles is from the NanoSIMS $O^-$ beam as discussed below in Section 4.1.

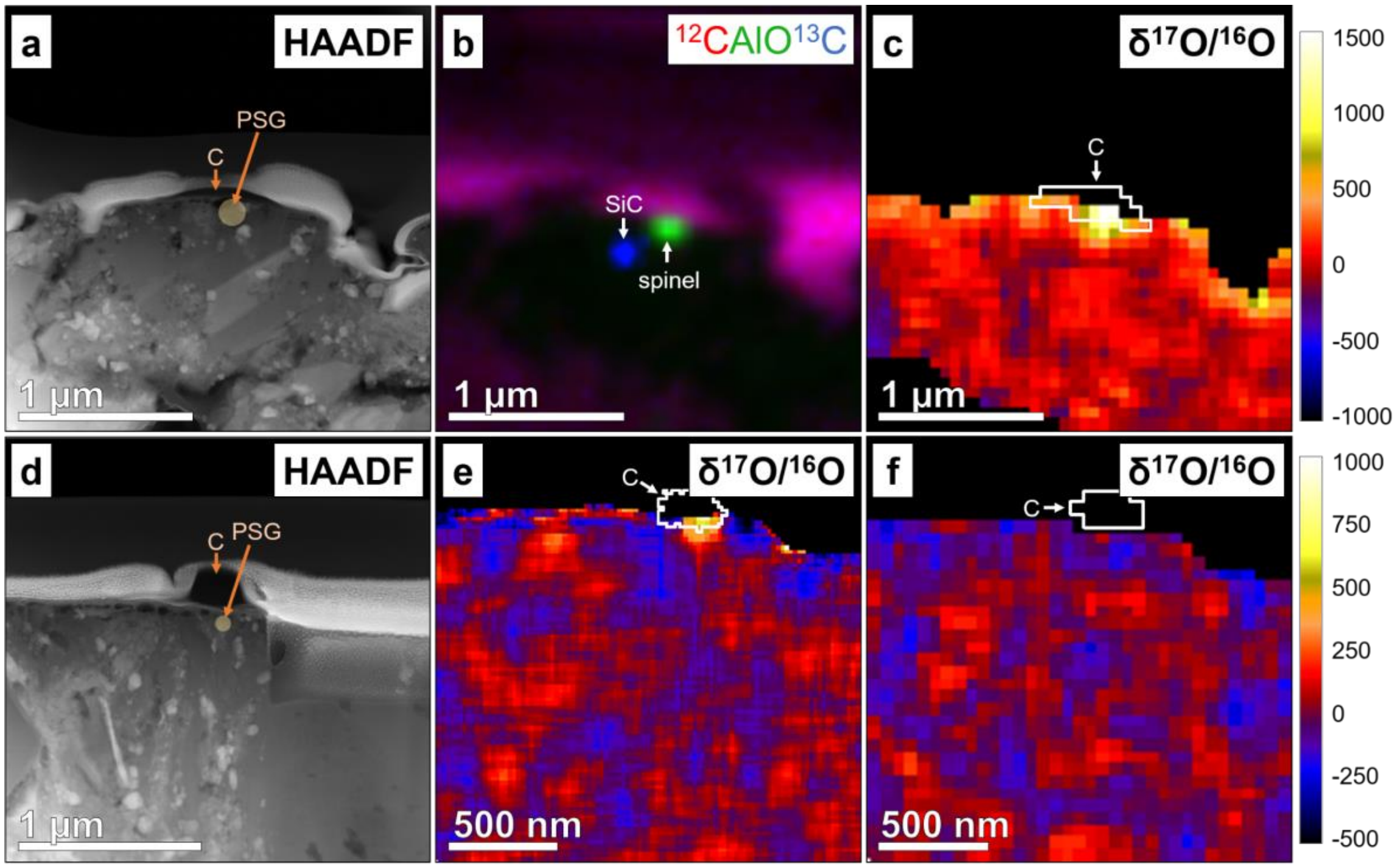

**Figure 1.** STEM HAADF images and NanoSIMS isotopic maps of two FIB sections of oxide F2-37 (a–c) and silicate F2-23 (d–f). Panel b is a composite map with red = $^{12}C$, green = $^{27}Al^{16}O$, and blue = $^{13}C$, whereas the other NanoSIMS panels show $^{17}O/^{16}O$ ratios images expressed as δ-values. The suspected locations of the presolar grains are indicated by the yellow ellipses and labeled PSG in the STEM HAADF images. The C fiducial markers are represented by white outlines and labeled in the isotopic ratio maps. The post-TEM NanoSIMS analysis of F2-37 found a $^{17}O$-enrichment associated with Al, confirming the presence of presolar spinel and also identified a highly $^{13}C$-rich grain a few hundred nm away, most likely a <100 nm presolar SiC grain. The first $\delta^{17}O/^{16}O$ map collected on grain F2-23 (e) shows an anomalous region just below the C fiducial marker; however a longer raster map shows no such anomaly (f), so there remains ambiguity as to whether the FIB section contained material from the originally identified presolar grain.

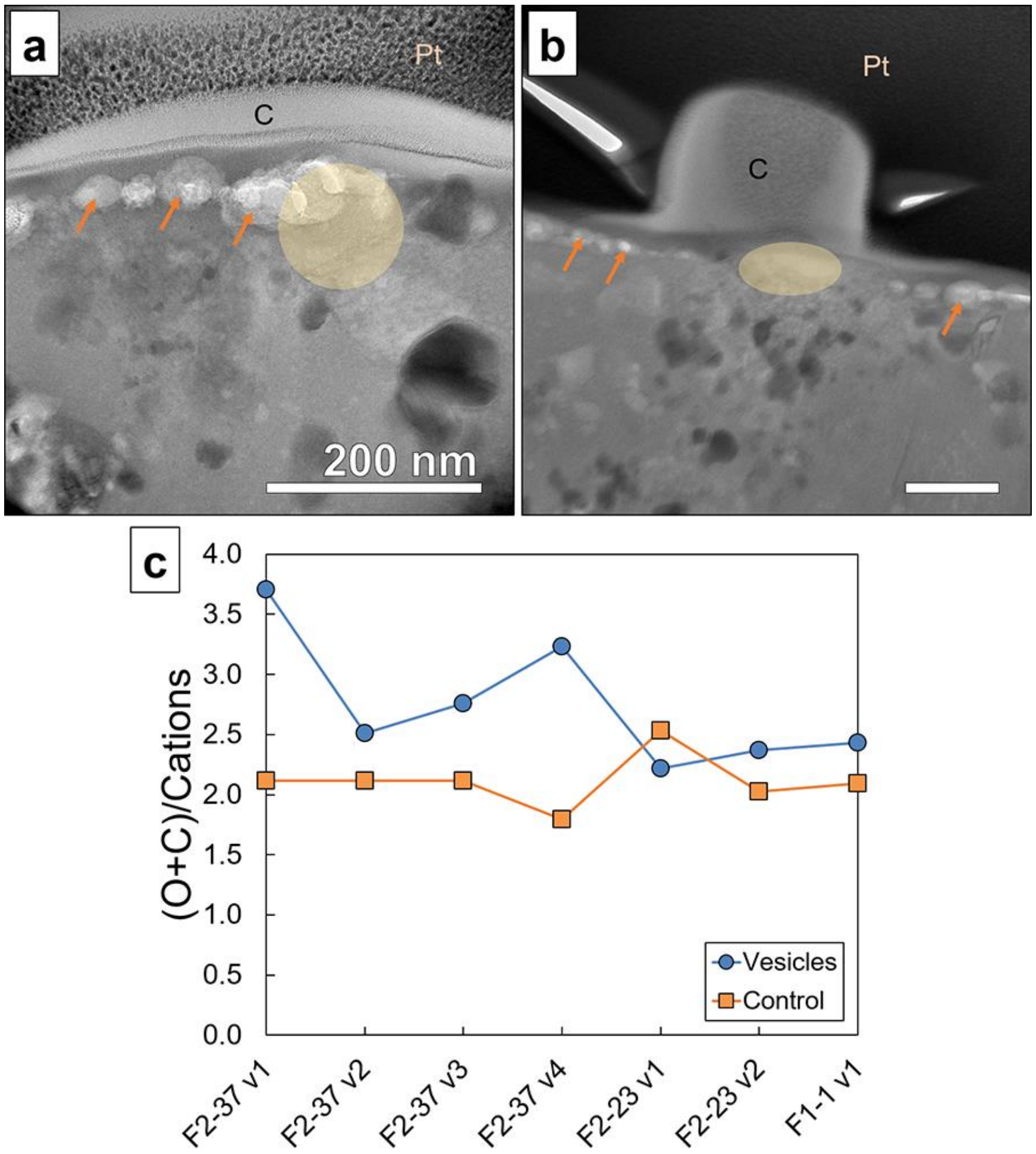

**Figure 2.** Bright-field TEM images (a–b) and a plot (c) of vesicles (denoted v1–v4 in the x-axis) observed in FIB sections with presolar grains. Scale bars are 200 nm. Example BF images of (a) F2-37 and (b) F2-9 show the presence of vesicles (orange arrows) (a) overlapping the presolar grain and (b) distant from it, but in all cases occurring at the interface between the meteoritic material and the deposited material from FIB sample preparation (i.e., C and Pt). The locations of the presolar grains are shown by the yellow ellipses. Compositional information (c) shows that in all but one case, the vesicles contain higher abundances of O+C ratioed to total cations as compared to adjacent material (control).

## 3.2 Transmission Electron Microscopy Findings

### *3.2.1 Matrix Materials*

In addition to elucidating information about the presolar grains themselves, the FIB sections allow us to also make observations of the adjacent non-presolar materials. Figure 3 shows the meteorite matrix material adjacent to the presolar grains in the five FIB sections. The C fiducial markers indicate the location of the presolar grains, with the arrows pointing to smaller C fiducial markers that lie just on top of the grains, in some cases (Fig. 3a, b, e). We observed three main textural types for the matrix materials: 1) fibrous phyllosilicates, 2) organic materials, and 3) amorphous silicates. Type 1 (blue outline) is fibrous in texture and is likely composed of phyllosilicates. Type 2 (purple outline) is C-rich, implying that it is composed of organic materials, and was only observed in one FIB section (Fig. 3e). Type 3 can be subdivided into amorphous silicate with and without nanoscale high Z phases. In the case of abundant high Z phases, the amorphous silicate material is embedded with nm-scale Fe(Ni) sulfides (pyrrhotite and pentlandite) and Fe carbides, the latter having magnetite rims. Type 3 makes up the majority of the matrix material in the FIB sections and, as such, is the only type not outlined in Figure 3a–e.

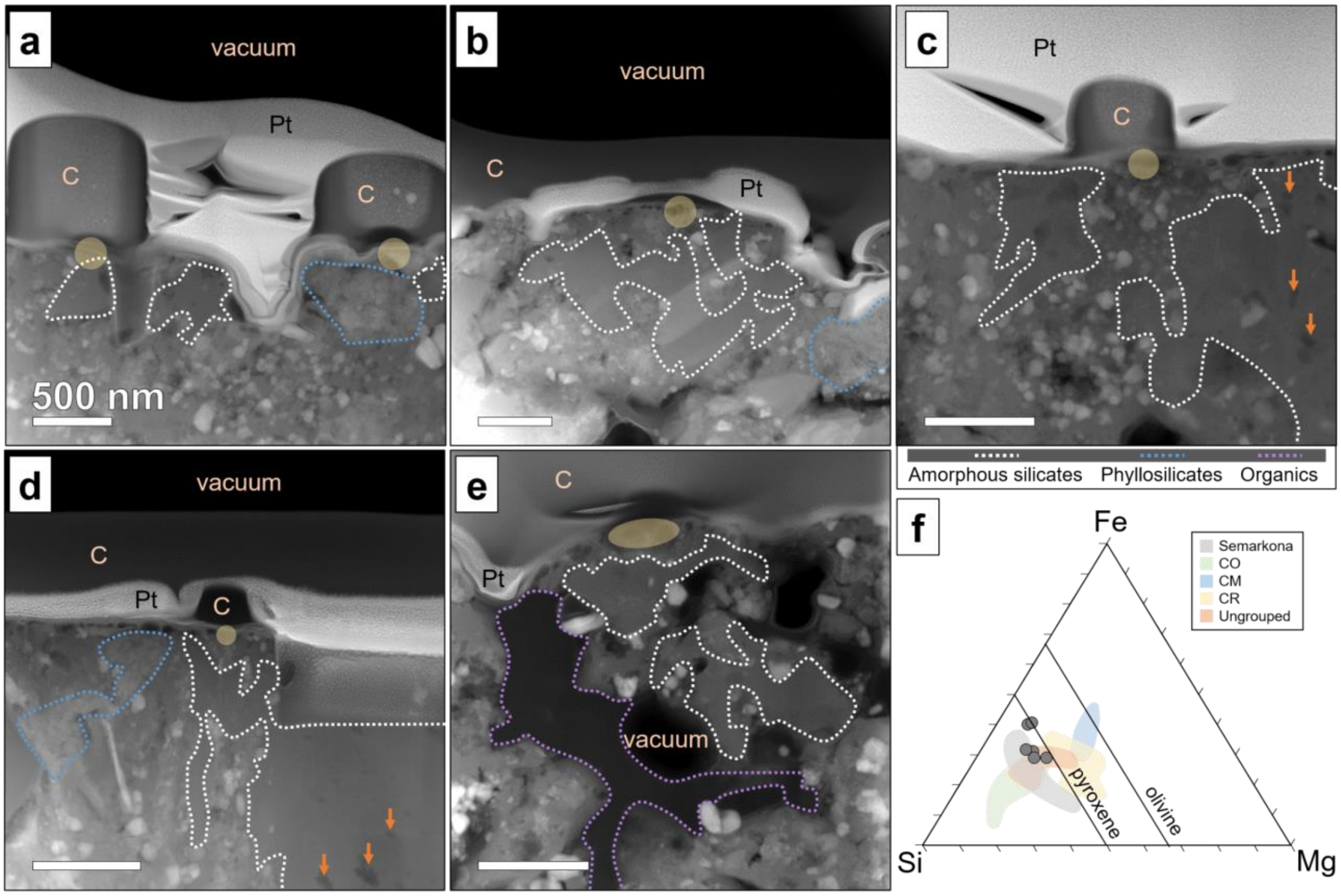

**Figure 3.** STEM HAADF images (a–e) and Si-Fe-Mg ternary diagram (in atomic %) (f) of the matrix material in FIB sections with presolar grains: (a) F2-30a and F2-30b, (b) F2-37, (c) F2-9, (d) F2-23, and (e) F1-1. Scale bars are 500 nm. C and Pt fiducial markers are labeled, and the rough locations of the presolar grains are indicated by the yellow ellipses. The dashed lines in (a–e) show regions of different meteorite matrix materials, including phyllosilicates (blue, type 1), organics (purple, type 2), and amorphous silicates (type 3) both with (not outlined) and without (white) embedded nanophase sulfides and/or carbides. The majority of the matrix is dominated by amorphous silicate with embedded sulfides and/or carbides; however, the presolar grains are often found associated with amorphous silicate material free of sulfides and/or carbides. Compositional data on these amorphous silicates (f) show that they are pyroxene-like in terms of Si, Mg, and Fe and tend to be more Fe than Mg-rich. Orange arrows indicate examples of pores. The envelopes show data for amorphous silicates in various meteorite groups; see Table S3 for the data and sources (Brearley, 1993; Greshake, 1997; Chizmadia, 2004; Chizmadia and Brearley, 2008; Le Guillou and Brearley, 2014; Dobrică and Brearley, 2020).

**Table 2.** TEM EDS compositions (atomic %, ratio, and Mg#) of the amorphous silicate matrix materials adjacent to the presolar grains from Semarkona

| Sample | O | Na | Mg | Al | Si | S | K | Ca | Fe | (Ca+Mg+Fe)/Si | Mg#[1] |
|---|---|---|---|---|---|---|---|---|---|---|---|
| F2-30a | 56 ± 1 | 1.08 ± 0.08 | 3.4 ± 0.2 | 0.77 ± 0.08 | 20.7 ± 0.3 | 1.0 ± 0.1 | 0.28 ± 0.07 | 0.35 ± 0.07 | 16 ± 1 | 0.96 | 18 |
| F2-37 | 69 ± 2 | 1.06 ± 0.09 | 4.0 ± 0.2 | 0.68 ± 0.08 | 15.1 ± 0.3 | 0.56 ± 0.08 | 0.36 ± 0.08 | 0.22 ± 0.07 | 8.6 ± 0.8 | 0.85 | 32 |
| F2-9 | 63 ± 1 | 1.30 ± 0.09 | 5.7 ± 0.2 | 3.0 ± 0.2 | 15.6 ± 0.2 | 1.3 ± 0.1 | 0.45 ± 0.07 | 0.24 ± 0.06 | 8.7 ± 0.7 | 0.94 | 40 |
| F2-30b | 61 ± 1 | 1.05 ± 0.09 | 3.4 ± 0.2 | 0.74 ± 0.09 | 18.0 ± 0.3 | 0.56 ± 0.09 | 0.19 ± 0.06 | 0.36 ± 0.08 | 15 ± 1 | 1.02 | 19 |
| F2-23 | 66 ± 1 | 2.5 ± 0.1 | 4.7 ± 0.2 | 0.84 ± 0.08 | 16.4 ± 0.2 | 0.53 ± 0.07 | 0.21 ± 0.06 | 0.18 ± 0.06 | 8.6 ± 0.7 | 0.82 | 35 |
| F1-1 | 67 ± 1 | 0.88 ± 0.08 | 3.7 ± 0.2 | 0.61 ± 0.07 | 16.9 ± 0.3 | 0.51 ± 0.07 | 0.7 ± 0.1 | 0.26 ± 0.06 | 9.6 ± 0.8 | 0.80 | 28 |

Errors are 1σ. Elements with abundances <0.2 at.% for all grains are excluded.

[1]Mg# = (Mg/(Mg+Fe)) × 100.

**Table 3.** Textural and compositional information on the presolar grains from Semarkona

| Sample | Isotopic Group | Type | Phases | Size (nm) | Morphology | Crystallinity | Heterogeneities |
|---|---|---|---|---|---|---|---|
| F2-30a | Mainstream | SiC | SiC | 210 × 90 | Euhedral | Single xl (amorphous rim) | Al |
| F2-37 | Group 1 | Oxide | Mg-Al sp | 230 × 150 | Anhedral | Single xl | Mg, Al |
| F2-9 | Group 2 | Silicate | NS silicates | 370 × 120 | Subhedral | Weakly nano-xl | Mg, Si |
| F2-30b | Group 1 | Silicate | NS silicates | 200 × 90 | Subhedral | Weakly nano-xl | None |
| F2-23 | Group 1 | Silicate | NS silicates | 160 × 80 | Anhedral | Amorphous | None |
| F1-1 | $^{26}$Mg-rich Group 1 | Silicate | NS silicates | 550 × 90 | Anhedral | Weakly nano-xl | Mg, Al, Cr, Si |

Sp – spinel, NS – non-stoichiometric, xl – crystal/crystalline.

**Table 4.** TEM EDS compositional ratios of the presolar grains from Semarkona

| **Sample** | **Description** | **(Ca+Mg+Fe)/Si** | **Mg#[1]** | **Mg/Al** |
|---|---|---|---|---|
| F2-37 | Mg-Al spinel | N/A | N/A | 0.54 |
| | Al-rich spinel | N/A | N/A | 0.33 |
| | Mg-rich spinel | N/A | N/A | 1.28 |
| F2-9 | Si-rich silicate | 0.38 | 94 | N/A |
| | Mg-rich silicate | 1.69 | 98 | N/A |
| | Mg-rich rim | 2.38 | 63 | N/A |
| F2-30b | Silicate | 1.38 | 15 | N/A |
| F2-23 | Silicate | 1.44 | 27 | N/A |
| F1-1 | Al-rich silicate | 0.43 | 84 | N/A |
| | Cr-rich silicate | 1.25 | 57 | N/A |
| | Mg-rich silicate | 1.39 | 75 | N/A |
| | Si-rich silicate | 0.23 | 89 | N/A |

Errors are 1σ. Elements with abundances <0.2 at.% for all grains are excluded. nd – not detected, N/A – not applicable.
[1]Mg# = (Mg/(Mg+Fe)) × 100.

The presolar grains tend to occur in association with amorphous silicates with a low abundance of nanoscale sulfides and carbides (white outlined textural type) . This textural type occurs in irregularly shaped patches that range from 100s of nm to microns in size and can contain pores (e.g., orange arrows in Fig. 3c, d) as well as the occasional Fe carbide and/or Fe sulfide. Of the three types of amorphous silicates identified by Dobrică and Brearley (2020) in Semarkona, our type 3 is texturally most similar to their "continuous groundmass with no clear boundaries." The elemental compositions of the amorphous silicates in association with the presolar grains are presented in Table 2. The (Ca+Mg+Fe)/Si ratio acts as a proxy for the stoichiometry of silicates, where the ratio is 2 for olivine and 1 for pyroxene. The Mg number (Mg#) illustrates whether a material is more Fe-rich or Mg-rich and is calculated as (Mg/(Mg+Fe))×100. For the matrix material adjacent to the presolar grains, the (Ca+Mg+Fe)/Si ratio ranges from 0.8 to 1.02 and the Mg numbers from 18 to 40. In general, the matrix material is pyroxene-like in composition (Fig. 3f), although it tends to be enriched in Si, is more Fe than Mg-rich, and has minor amounts of the volatile elements Na, K, and S. This is consistent with the analyses of "continuous groundmass with no clear boundaries" from Dobrică and Brearley (2020).

*3.2.2 Presolar Grains*

The six presolar grains studied here were determined to be a SiC (1), an oxide (1), and silicates (4). Textural and qualitative compositional information on the grains, including the phases present, grain size and morphology, crystallinity, and presence of heterogeneities, are summarized in Table 3. Quantitative compositional information from EDS analyses are summarized in Table 4 and present data on different regions of the grains that show heterogeneities.

*SiC F2-30a*

Grain F2-30a was the only presolar SiC grain studied in this work; the probable additional $^{13}$C-rich SiC grain identified by the post-TEM analysis of F2-37 (Fig. 1b) was too small to confirm from the TEM EDS data. F2-30a measures 210 by 90 nm in size, as determined from BF-TEM images, and is triangular and euhedral in morphology, as viewed in cross section. Well-developed faces are visible in Figure 4c, and the grain is distinct from the surrounding matrix material. The SiC grain is a single crystal surrounded by an amorphous rim with the crystalline portion appearing as the higher contrast region in the BF image (Fig. 4a). The crystalline SiC has a diffraction pattern with well-defined spots, whereas the amorphous rim has a diffraction pattern with only faint spots and a diffuse disk (Fig. 4b). The d-spacings of the diffraction spots match both the 3C and 2H polytypes (Table S2), the two most commonly observed polytypes for presolar SiC. The amorphous rim ranges in thickness from 20 nm at the top of the grain to 10 nm along the bottom edges; however, NanoSIMS analyses can amorphize the top 10–30 nm of material exposed to the ion beam, so the 20 nm thick rim near the top of the SiC grain is likely an analytical artifact. The amorphous rim near the bottom, though, is likely intrinsic to the sample.

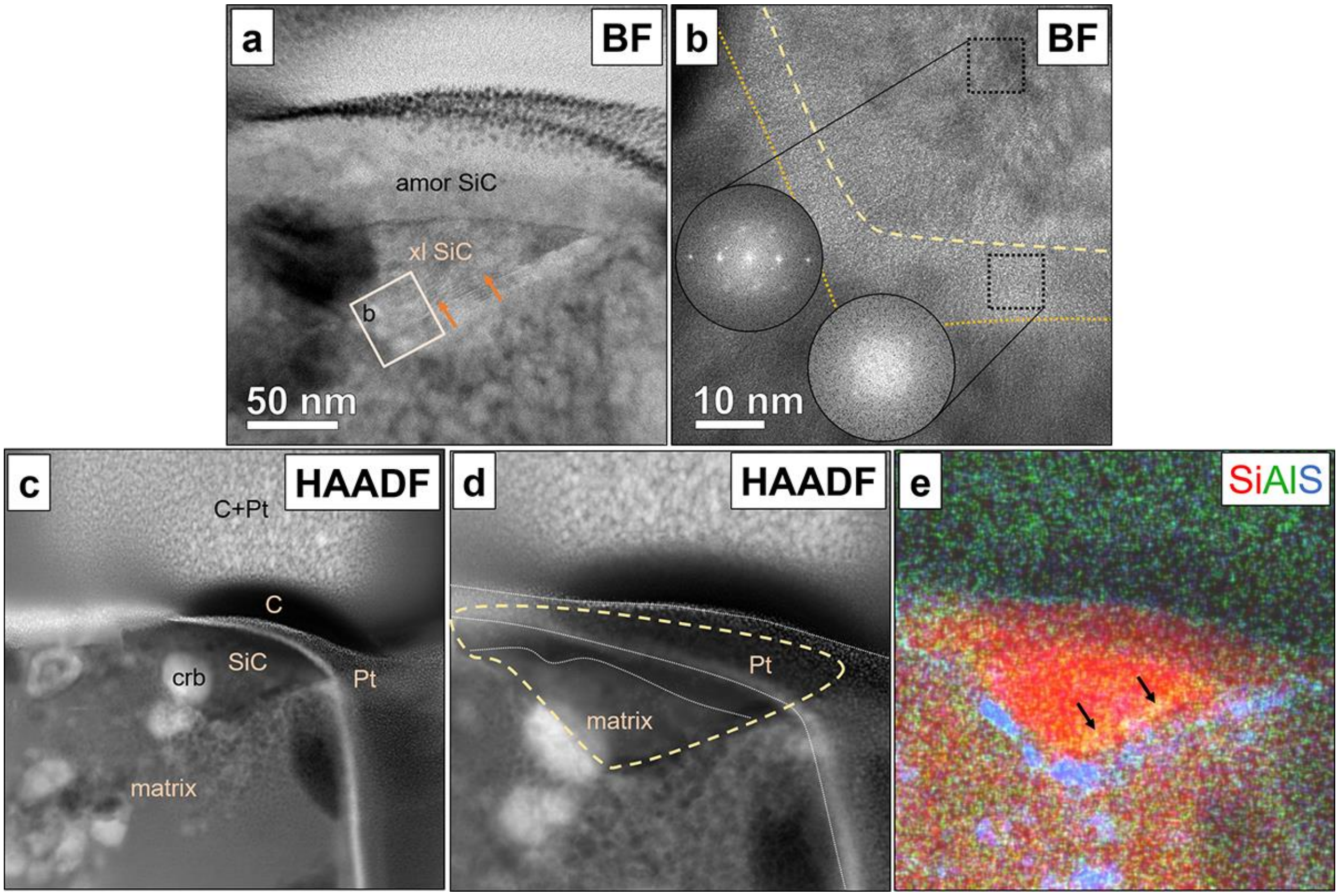

**Figure 4.** Bright-field (BF) TEM (a), high-resolution TEM (b), and STEM HAADF (c–d) images as well as a Si-Al-S EDS composite X-ray map (e) of presolar SiC grain F2-30a. The BF images (a–b) show the presence of both amorphous and crystalline SiC. The former is present as a rim around the grain and has low contrast in (a) and diffuse diffractogram spots in (b). The SiC grain, including the amorphous regions, appears to have well-defined crystal faces in (c). A higher magnification HAADF image (d) shows overlapping matrix material as well as Pt from FIB sample preparation. The EDS composite X-ray map (e) shows the presence of Al-rich regions (black arrows pointing to yellowish bands), which are associated with high-density stacking faults (orange arrows in a), as well as a S-rich rim (blue). Amor – amorphous, xl – crystalline, crb – Fe carbide.

In terms of other heterogeneities in the SiC, there is an Al enrichment in one region of the grain (indicated by the arrows in Fig. 4e). This is also a region with numerous crystal defects, specifically high-density stacking faults (indicated by the arrows in Fig. 4a and shown in more detail in Fig. S4). The high defect SiC contains 1.5 at.% Al, as compared to the low defect SiC with 0.31 at.% Al (Table S2). The EDS analyses of the SiC as a whole also show minor amounts of Mg, S, and Fe, but these are likely contributions from overlapping adjacent meteorite matrix materials (Fig. 4d). The Si-Al-S EDS composite map (Fig. 4e) shows the presence of a S-rich rim on the bottom edges of the grain, but this appears to be on the outside of the SiC grain. No subgrains or voids were identified, but the overlapping matrix material makes observations of such features in the EDS maps and HAADF STEM images of the SiC grain difficult.

*Oxide F2-37*

Grain F2-37 was the only presolar oxide grain studied in this work. It measures 230 by 150 nm in size and is irregular and anhedral in morphology. The grain boundary is not well defined in either the BF or STEM HAADF images (Fig. 5a, c), but it is a single crystal, as shown by the dark field (DF) image (Fig. 5b). The crystallographic orientation of the upper portions of the grain seems to have been affected by the formation of vesicles and so do not appear bright in the DF image (Fig. 5a, b). Based on electron diffraction patterns (Table S1) and EDS analyses (Table S2), the grain is Mg-Al spinel ($MgAl_2O_4$). Although the overall grain is stoichiometric (Mg/Al = 0.54), the Al-Fe-Mg EDS composite map (Fig. 5d) shows some heterogeneities in the Mg and Al contents, with Mg-rich regions having a Mg/Al ratio of 1.28 and Al-rich regions a Mg/Al ratio of 0.33. The overall grain contains minor amounts of Na, Si, S, Ca, and Fe (Table S2). No voids, with the exception of the vesicles, were identified within F2-37.

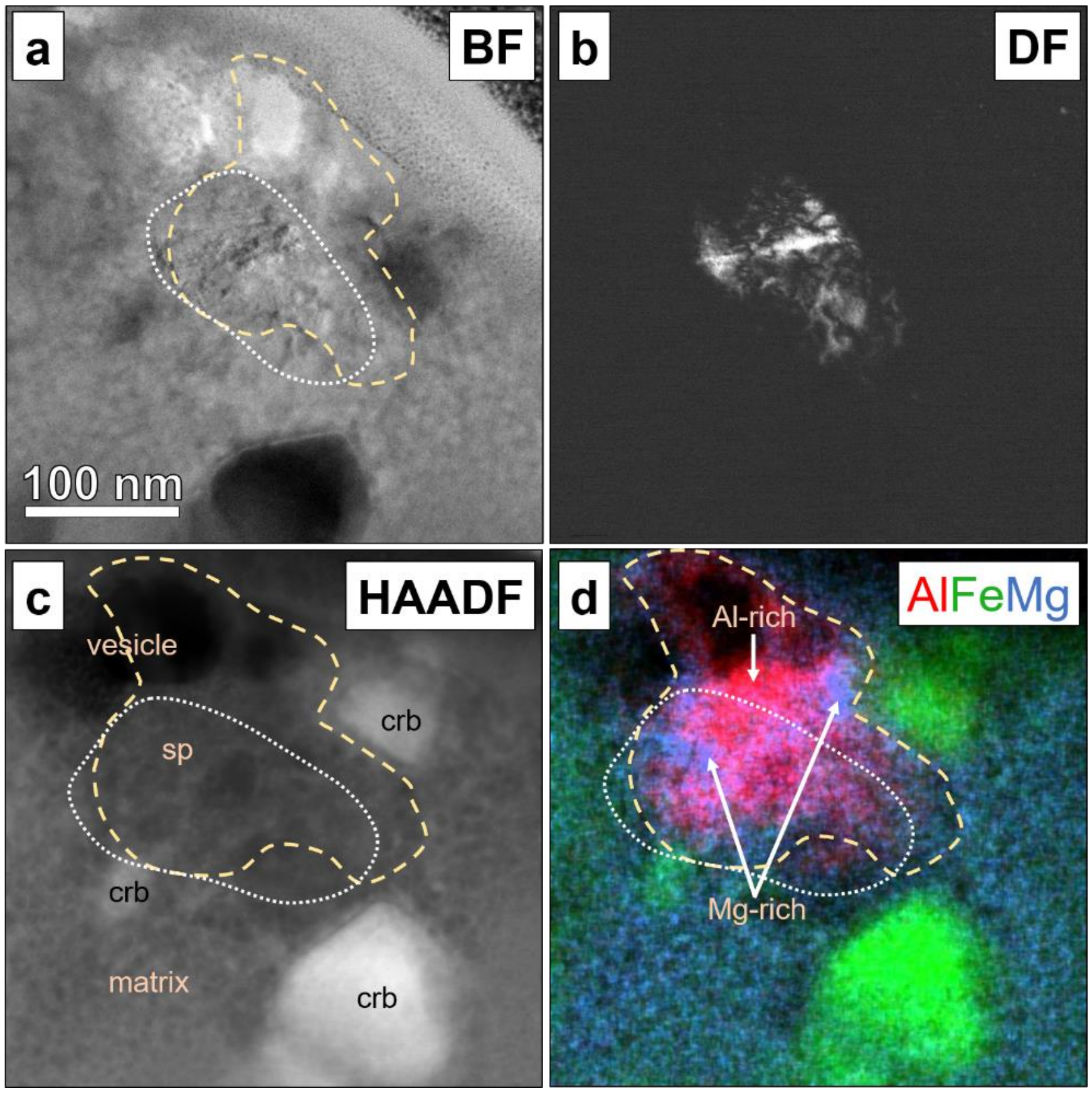

**Figure 5.** Bright-field (BF) TEM (a), dark-field (DF) TEM (b), and STEM HAADF (c) images as well as an Al-Fe-Mg EDS composite X-ray map (d) of presolar spinel grain F2-37. The BF image shows a higher contrast region of the grain (white dashed outline) that appears as a single crystal domain in (b). The crystallinity of the grain near the top of the section was likely damaged owing to vesicle formation. The HAADF image (c) does not show any obvious grain boundary between the spinel and the matrix. The EDS composite X-ray map (d) shows heterogeneities in Al (red) and Mg (blue). Sp – spinel, crb – Fe carbide.

*Silicate F2-9*

Grain F2-9 is a silicate that measures 310 by 90 nm in size with an ellipsoidal and subhedral morphology. Although the grain boundary is not obvious in the STEM HAADF image (Fig. 6c), a higher contrast band in the BF image (Fig. 6a) is visible and corresponds to a 30 to 40 nm thick rim that is enriched in Mg (Fig. 6d). In terms of crystallinity, F2-9 contains regions of crystallinity (with well-defined diffraction spots) adjacent to amorphous patches (with diffuse to no diffraction spots) on a highly localized scale (what we refer to as weakly nanocrystalline) (Fig. 6b). Electron diffraction patterns of the crystalline regions do not definitively match any particular common presolar silicate (e.g., olivine, pyroxene). Additionally, as the Si-Fe-Mg EDS composite map illustrates (Fig. 6d), the grain shows Mg and Si heterogeneities, with Mg-rich regions (24 at.% Mg, 14.4 at.% Si) and Si-rich regions (9.7 at.% Mg, 27.3 at.% Si). As mentioned previously, F2-9 has a Mg-rich rim that is also enriched in Fe compared to the grain itself (16.9 at.% Mg, 9.8 at.% Fe). The Mg-rich and Si-rich regions are non-stoichiometric, with (Ca+Mg+Fe)/Si ratios of 1.69 (Mg-rich) and 0.38 (Si-rich); more Mg than Fe-rich, with Mg#'s of 98 (Mg-rich) and 94 (Si-rich); and have minor amounts Na, Al, S, and Fe (Table 4). The Mg-rich rim is also non-stoichiometric ((Ca+Mg+Fe)/Si ratio of 2.38); more Mg than Fe-rich (Mg# of 63) although with a higher Fe content than the interior; and has minor amounts of Na, Al, S as well as K, Ca, Cr, and Ni (Table S2). No voids or subgrains were observed within F2-9.

*Silicate F2-30b*

Grain F2-30b is a silicate that measures 200 by 90 nm in size with a triangular and subhedral morphology. The grain boundary is not obvious in the STEM HAADF image (Fig. 7c), largely due to overlapping meteorite matrix material, namely Fe carbide best viewed in the STEM HAADF image and the Si-Fe-S EDS composite map (Fig. 7d). However, the grain is visible as a more Mg-rich region in the Mg EDS map (Fig. 7e). This silicate grain is also weakly nanocrystalline, as illustrated by the HR BF image (Fig. 7b), and electron diffraction patterns of the crystalline regions are not consistent with any particular silicate. Heterogeneities in the silicate were not identified, but the overlapping Fe carbide may be obscuring any compositional variations. F2-30b is non-stoichiometric, with an (Ca+Mg+Fe)/Si ratio of 1.38; more Fe than Mg-rich, with an Mg# of 15; and contains minor amounts of Na and S (Table 4). If F2-30b does not extend through the entire thickness of the FIB section, as mentioned in section 3.1, adjacent meteoritic materials may contribute to elevated concentrations of Na, S, and Fe. No voids or subgrains were identified in F2-30b, but the overlapping matrix material makes observations of such features in the EDS maps and HAADF STEM images difficult.

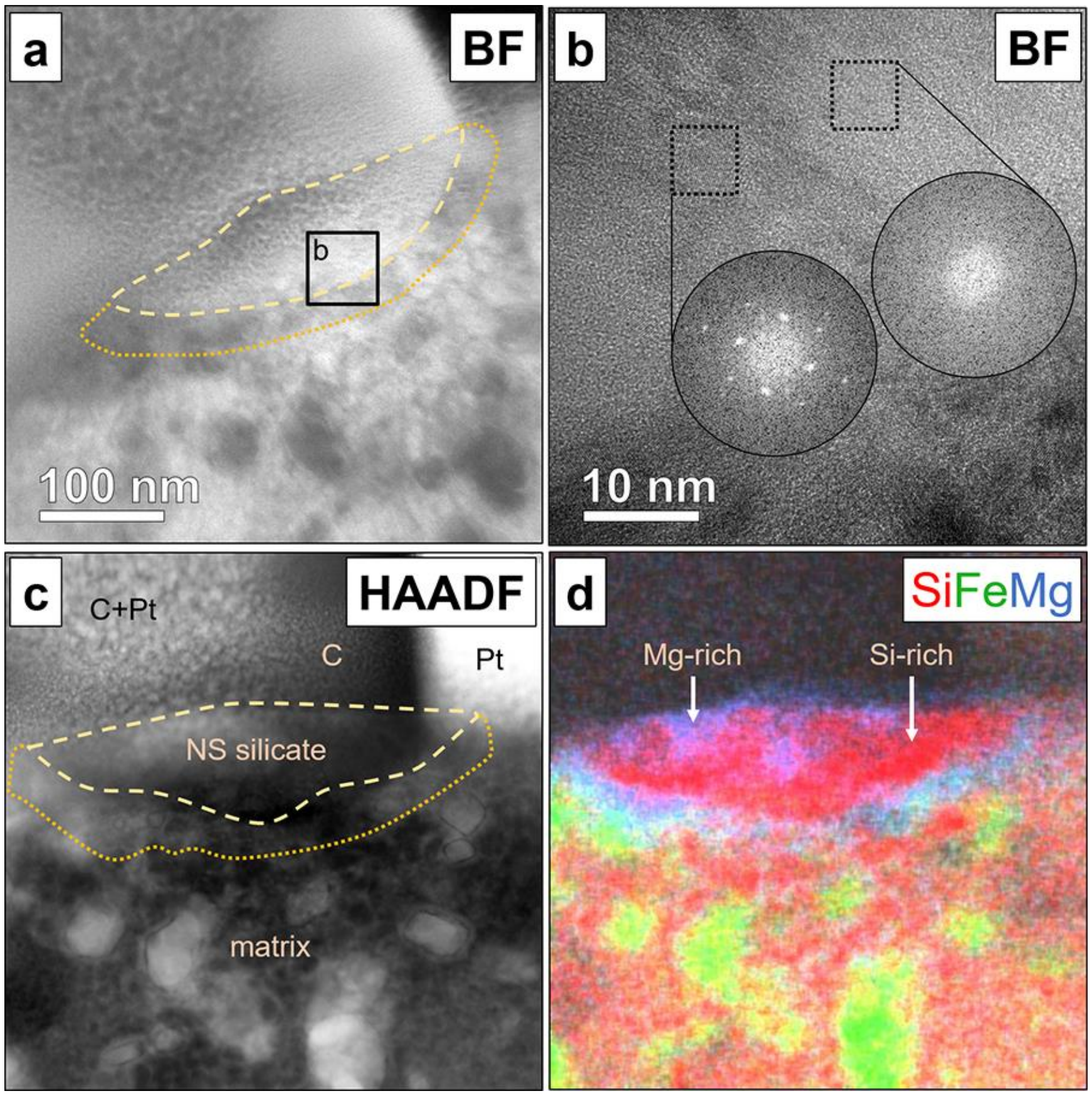

**Figure 6.** Bright-field (BF) TEM (a), high-resolution TEM (b), and STEM HAADF (c) images as well as a Si-Fe-Mg EDS composite X-ray map (d) of presolar silicate grain F2-9. The BF image shows a higher contrast rim on the grain (orange dashed outline). The high-resolution image (b) shows that the grain contains regions of crystallinity (with well-defined diffraction spots) adjacent to amorphous patches (with diffuse to no diffraction spots) on a highly localized scale (what we refer to as weakly nanocrystalline). The HAADF image (c) does not show any obvious grain boundary between the silicate and the matrix, but the silicate is clearly visible in the EDS composite X-ray map (d) as the region rich in Si (red) and Mg (blue) and poor in Fe (green). The map also shows heterogeneity in the Mg content of the silicate as well as the presence of a Mg-rich rim. NS – non-stoichiometric.

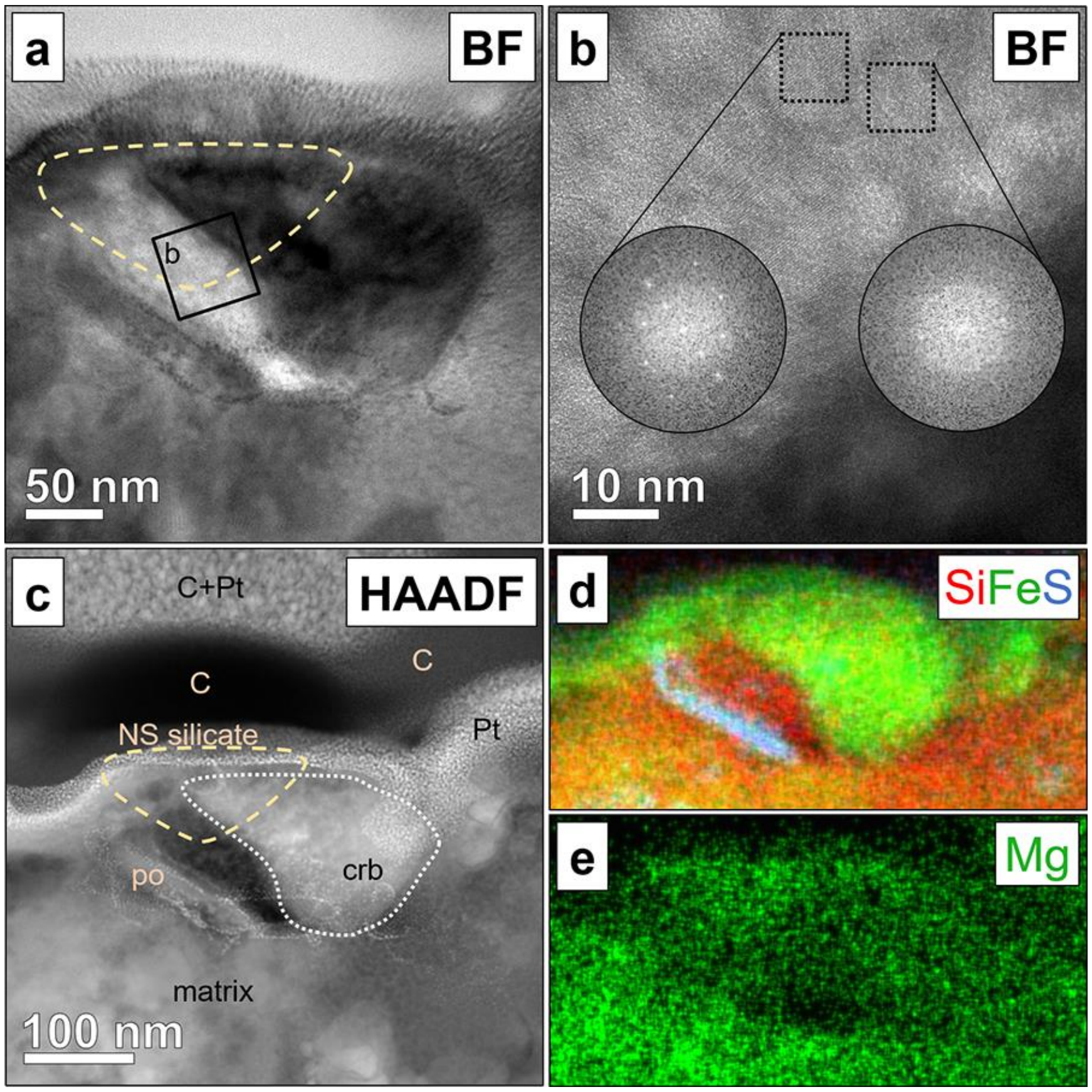

**Figure 7.** Bright-field (BF) TEM (a), high-resolution TEM (b), and STEM HAADF (c) images as well as EDS X-ray maps (d–e) of presolar silicate grain F2-30b. The BF image shows the presolar grain (outlined in yellow) which appears to be overlapping with meteorite matrix materials. The high-resolution image (b) shows that the grain contains regions of crystallinity (with well-defined diffraction spots) adjacent to amorphous patches (with diffuse to no diffraction spots) on a highly localized scale (what we refer to as weakly nanocrystalline). The HAADF image (c) does not show any obvious grain boundary between the silicate and the matrix. Instead, the overlapping Fe carbide material dominates the Z-contrast. The Si-Fe-S EDS composite X-ray map (d) shows the extent of the carbide (green), whereas the Mg map (e) better represents the presolar grain location. NS – non-stoichiometric, po – pyrrhotite, crb – Fe carbide.

*Silicate F2-23*

Grain F2-23 is a silicate, although it is not readily visible in BF or STEM HAADF images (Fig. 8a, c). In lieu of obvious grain boundaries in TEM images, the $\delta^{17}O/^{16}O$ anomaly from the follow-up NanoSIMS analyses measures 160 by 80 nm in size and has an ellipsoidal and anhedral morphology. This topmost portion of the FIB section is dominated by the products of the first round of NanoSIMS analyses, namely vesicles and amorphized phases. However, even further below the top of the FIB section, beyond the penetration range of the NanoSIMS ion beam, the silicates are amorphous (Fig. 8b). The silicate is non-stoichiometric, with a (Ca+Mg+Fe)/Si ratio of 1.44; more Fe than Mg-rich, with an Mg# of 27; and has minor amounts of Na, Al, S, K, Ca, Cr, and Ni (Table S2). If F2-23 does not extend through the entire thickness of the FIB section, as mentioned in section 3.1, adjacent meteoritic materials may contribute to elevated concentrations of Na, S, Fe, K, Ca, Cr, and Ni.

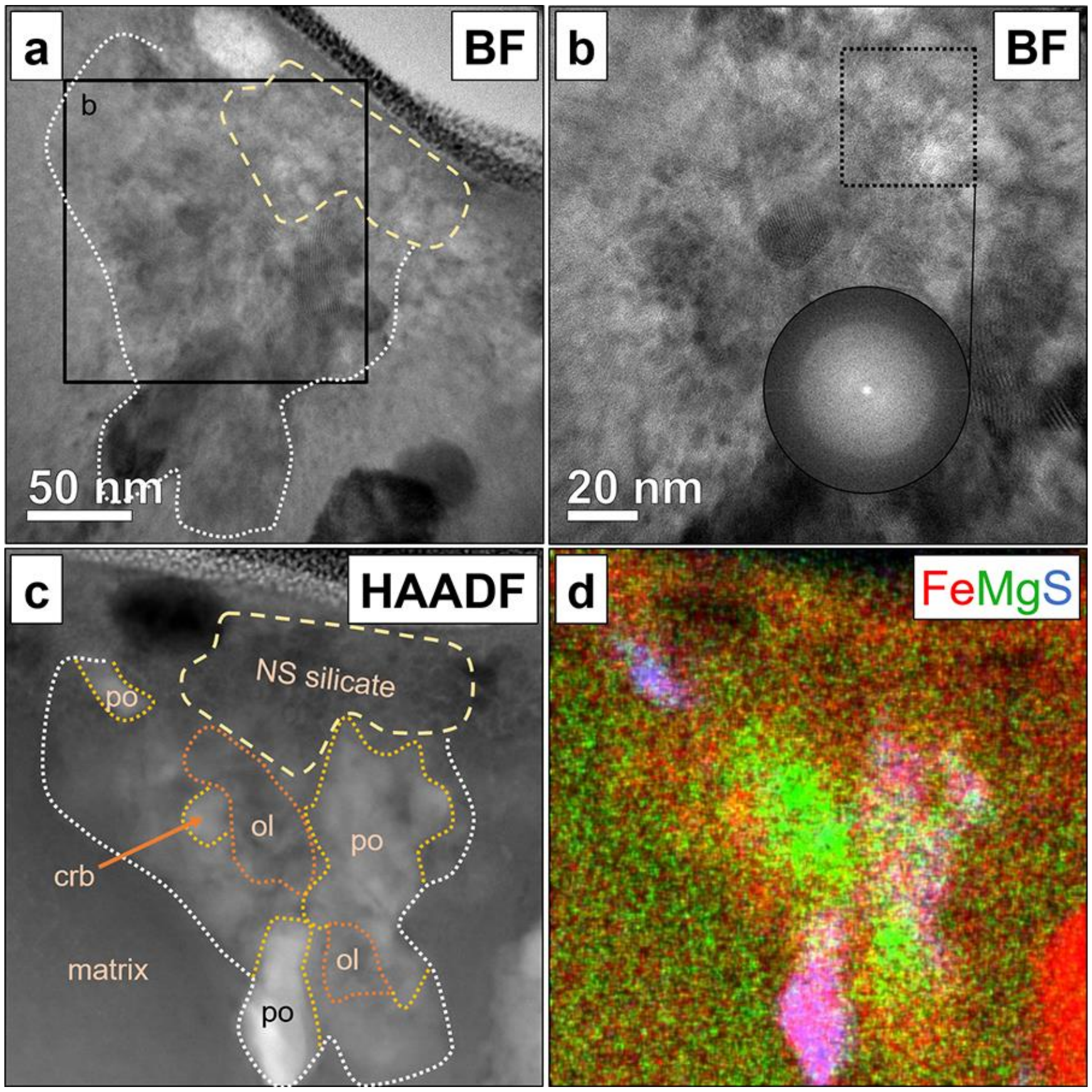

**Figure 8.** Bright-field (BF) TEM (a), high-resolution TEM (b), and STEM HAADF (c) images as well as an Fe-Mg-S EDS composite X-ray map (d) of presolar silicate grain F2-23. The BF image shows a higher contrast region (white dashed outline) surrounded by amorphous silicate material. The location of the presolar grain (outlined in yellow) is based on NanoSIMS isotopic mapping but avoids the pyrrhotite regions. The high-resolution image (b) shows that the presolar silicate region is amorphous. The HAADF image (c) better illustrates the multiple phases present nearby, but there does not appear to be any obvious grain boundary between the presolar grain and adjacent materials. The EDS composite X-ray map (d) shows the presence of the olivine as the two Mg-rich (green) regions as well as sulfides (purple). The map also shows that the presolar region and the interstitial silicates are more enriched in Fe (red) compared to the adjacent amorphous silicate material. NS – non-stoichiometric, ol – olivine, po – pyrrhotite, crb – Fe carbide.

The adjacent assemblage (outlined in white in Fig. 8a, c) consists of Fe sulfides, an Fe carbide, and silicates. Electron diffraction patterns of Mg-rich regions (green in Fig. 8d) indicate polycrystallinity and are consistent with olivine as are their stoichiometries, with an average (Ca+Mg+Fe)/Si ratio of 1.95 and Mg# 54. In addition to the two olivine regions, there are also interstitial silicates which are non-stoichiometric, with a (Ca+Mg+Fe)/Si ratio of 1.32; more Fe than Mg-rich, (Mg# of 29); and have minor amounts of Na, Al, and S. These interstitial silicates are compositionally distinct from surrounding amorphous silicates, being more Fe rather than Mg rich (Fig. 8d). The aggregate itself is likely not presolar, owing to the presence of Fe sulfides and Fe carbides which are common parent body alteration products in the matrix of Semarkona, as well as the fact that the NanoSIMS maps did not show anomalous material extend to such a depth in the FIB section.

*Silicate F1-1*

Grain F1-1 is composed of silicates that measure 550 by 90 nm in size with an elongate and anhedral morphology. It is surrounded by amorphous silicate material ranging from 760 to 260 nm in thickness. The grain contains a silicate core, with variable Mg and Si contents (Fig. 9f), which is flanked on both sides by more Al-rich silicates and on one side by a Cr-rich silicate (Fig. 9c–d). All regions are Ca-bearing, with Ca content ranging from 2.5 to 5.3 at.% (Table 4), and are weakly nanocrystalline (Fig. 9b). The electron diffraction patterns of the crystalline regions are not consistent with any particular silicate (Table S1). The sizes of each region are 170 by 70 nm for the left Al-rich region, 130 by 40 nm for the right Al-rich region, 100 by 50 nm for the Cr-rich region, and 240 by 80 nm for the Mg,Si-rich core. All regions are non-stoichiometric, with (Ca+Mg+Fe)/Si ratios ranging from 0.23 (Si-rich) to 0.43 (Al-rich) to 1.25 (Cr-rich) to 1.39 (Mg-rich). The Cr-rich region has an intermediate Mg# (57), but the remaining regions are more Mg than Fe-rich, with Mg#'s of 75 (Mg-rich), 84 (Al-rich), and 89 (Si-rich). Major elements (defined as >1 at.%) include Al, Mg, Ca, and Fe for the Al-rich regions; Mg, Al, Ca, Cr, and Fe for the Cr-rich region; Mg, Ca, and Fe for the Mg-rich regions; and Mg, Al, and Ca for Si-rich regions. The Cr-rich region also has minor amounts of Na and S, the Mg-rich region Al and S, and the Si-rich region Fe (Table S2). No voids or subgrains were identified in the grain.

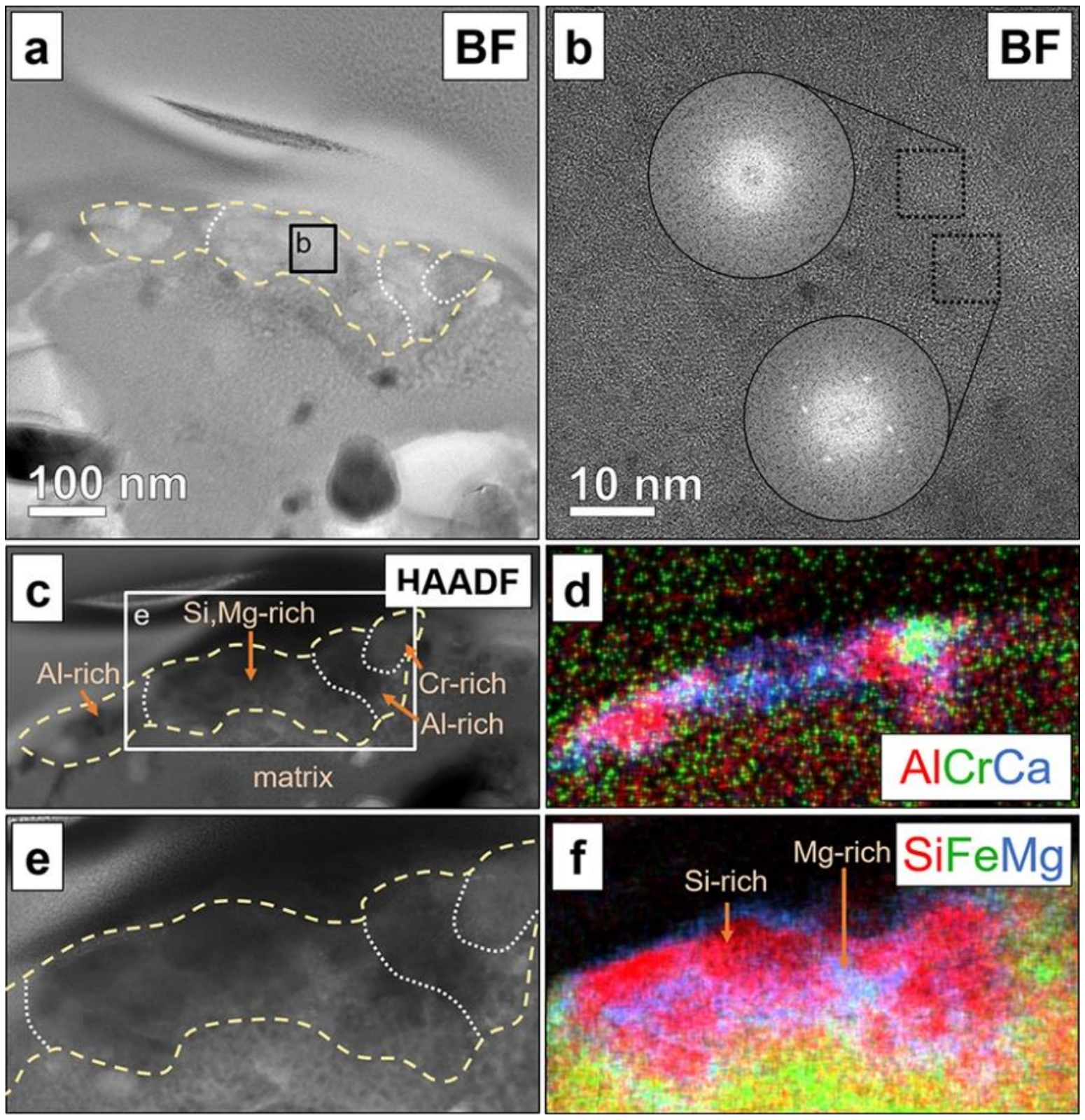

**Figure 9.** Bright-field (BF) TEM (a), high-resolution TEM (b), and STEM HAADF (c, e) images as well as EDS composite X-ray maps (d, f) of presolar silicate grain F1-1. The BF image shows the overall grain surrounded by amorphous silicate materials. The high-resolution image (b) shows one example of the weakly nanocrystalline nature observed in all regions of the grain. The HAADF images (c, e) show only slight differences in Z-contrast between the grain and the matrix. The Al-Cr-Ca EDS composite X-ray map (d) shows the presence of compositional heterogeneities, including two Al-rich regions and one Cr-rich region. The Si-Fe-Mg EDS composite X-ray map (f) illustrates the heterogeneity of Mg and Si in the core of the grain.

Although the post-TEM NanoSIMS maps of F1-1 show less isotopically anomalous O in the Al regions (Fig. S1h), we argue that these regions are still likely to be presolar; including the Al-rich regions of the grain gives a similar grain size between the two rounds of NanoSIMS analyses. Additionally, the Al-rich and Mg,Si-rich regions of the grain have similar Ca contents. It is difficult to imagine a scenario where only the central, Mg,Si-rich portion is presolar despite the similarities in Ca contents and the fact that the whole assemblage shows such a distinct EDS signature (Fig. 9d) as compared to adjacent meteorite matrix material.

# 4. DISCUSSION

In the following section, we will discuss possible laboratory artifacts in the samples analyzed, then look at the grains' histories from their formation in winds from AGB (or RSG, in the case of grain F1-1) stars, to their residence in the interstellar medium (ISM), to their time in the solar nebula, and later their asteroidal parent body. Lastly, we will compare the structural and elemental compositional characteristics of presolar grains from different planetary materials using data from previous research in addition to our own work and discuss similarities and differences.

## 4.1 Laboratory Artifacts

Rather than being features inherent to the sample, we argue that the vesicles located near the top of the FIB sections formed as a result of ion beam imaging from the NanoSIMS isotopic mapping, namely from the $O^-$ beam used to measure Mg and Si isotopes. To our knowledge, no previous FIB-TEM data have been reported for in-situ presolar grains that had undergone $O^-$ ion beam NanoSIMS analyses. Zega et al. (2011) and Zega et al. (2014) performed $O^-$ ion beam analyses with subsequent FIB-TEM ex situ work on presolar oxides, and Zega et al. (2011) found evidence for $^{16}O$ implantation in the form of a darker region in a HAADF image of one grain. However, vesicles were not observed in either study. Most likely, these vesicles are a feature of $O^-$ ion beam implantation into fine-grained silicate matrix that was already beam-damaged by the $Cs^+$ NanoSIMS measurements. Indeed, the greater (O+C)/cation ratios of the vesicles compared to adjacent, vesicle-free materials is consistent with this hypothesis. Researchers planning future in situ studies of presolar grains using FIB-TEM should be aware of the potential for vesicle formation from NanoSIMS $O^-$ ion beam analyses, especially since, in our study, the vesicles partly formed at the expense of oxide and silicate material, including presolar grains in some instances. This is an example of the tradeoffs inherent to coordinated in situ isotopic and structural studies of presolar grains.

## 4.2 Grain Condensation in Evolved Stellar Winds

As discussed in Section 3, based on their isotopic compositions, all but one of the presolar grains studied here are inferred to have most likely formed in asymptotic giant branch (AGB) stars, with the exception likely forming in a red supergiant (RSG) wind. AGB stars represent a late stage in the evolution of low- and intermediate-mass ($M < 8\ M_\odot$) stars. They consist of an inert C- and O-rich (O and Ne for the most massive AGB stars) core, surrounded by thin shells of He and H which alternately undergo nuclear burning, surrounded by a very large convective envelope. They are observed to be prodigious producers of dust grains, which drive strong mass-loss, ultimately resulting in a planetary nebula and a cooling white dwarf remnant. RSGs are evolved massive stars that similarly have strong dusty winds from large H envelopes. However, their cores are not inert, but rather experience enhanced stages of nuclear burning of heavier and heavier fuels until core-collapse drives a supernova explosion. The conditions in RSG envelopes are not identical to those in AGB envelopes. Nevertheless, the dust condensation process in the two types of stars likely share similarities (Verhoelst et al., 2009; Cherchneff, 2013).

Astronomical observations provide a great deal of information about dust formation in AGB envelopes, though much uncertainty remains as to the exact species present and the details of their formation. In this regard, TEM analysis of bona fide presolar AGB stardust grains is one of the most powerful methods for obtaining a better understanding of circumstellar conditions, given its unique ability to provide both structural and elemental compositional information on presolar grains, both of which record the physical and chemical conditions of formation.

Various approaches have been taken historically towards understanding grain condensation in AGB stars, but these can essentially be categorized as thermodynamic equilibrium condensation (e.g., Lodders and Fegley, 1999) and non-equilibrium kinetic models of nucleation and growth. Thermodynamic equilibrium calculations determine which phases are stable in a cooling gas at a specific temperature and pressure and of a given composition, whereas kinetic models consider the actual atom-by-atom growth of dust grains, taking into account, for instance, whether there is sufficient time to reach equilibrium during growth. Spectral features of both amorphous and crystalline silicates are commonly seen in O-rich AGB stars (Molster et al., 2002). It is important to note that the resolution of these observational techniques may prevent researchers from being able to distinguish between amorphous and weakly nanocrystalline materials. As such, we treat weakly nanocrystalline materials similarly to amorphous materials in the ensuing discussion. Thermodynamic equilibrium does not predict the formation of amorphous phases, so the presence of amorphous silicates, seen both in the astronomical data and the presolar silicate TEM data, indicates that AGB grain formation does not always occur under equilibrium conditions (e.g., Gail and Sedlmayr, 1999; Ferrarotti and Gail, 2001; Gail et al., 2009; Nagahara and Ozawa, 2009; Bose et al., 2012). Nonetheless, crystalline grains may well have formed by equilibrium condensation, and equilibrium phase stability certainly influences what phases condense even if full thermodynamic equilibrium is not achieved.

Condensation under equilibrium conditions is expected to yield stoichiometric, crystalline phases (e.g., Floss and Haenecour, 2016). As an O-rich AGB envelope cools, the first phases to condense in equilibrium are refractory oxides like corundum (>1600 K at $10^{-5}$ bars), followed by Al-rich silicates (e.g., melilite), and then, between 1300 and 1400 K, the Mg-rich silicates forsterite and enstatite as well as Fe metal (e.g., Sharp and Wasserburg, 1995; Lodders and Fegley, 1999). Iron is not expected to condense into silicates until much lower temperatures. Indeed, most astronomical studies of crystalline silicates in AGB stars have concluded that they are dominated by Mg-rich grains (Molster et al., 2002; Jones et al., 2012), though some low-mass-loss-rate AGB stars appear to have Fe-rich ones as well (e.g., Guha Niyogi et al., 2011). Non-equilibrium condensation, on the other hand, can result in non-stoichiometric phases with short to intermediate scale nanocrystalline order or amorphous grains (Floss and Haenecour, 2016). Silicates which form under such conditions may also be more Fe-rich in composition (Mg# <90), as the Fe content is not restricted to the solubility limit of specific crystal structures (Nguyen et al., 2016). In this scenario, a wider range of initial compositions are possible for the presolar grains, being more dependent on the gas kinetics and local composition rather than crystal chemical controls (Nagahara et al., 2009).

In addition to forming phases not predicted in condensation calculations, non-equilibrium condensation can also promote heterogeneities within presolar grains (Floss and Haenecour, 2016).

Although the gas composition only changes slowly in AGB star envelopes, the observation of complex Al and N zonation textures in a presolar SiC grain (G619) from an AGB star (Singerling et al., 2021) illustrates that conditions clearly do change on the timescales that SiC grains, at least, form. To summarize, grains that are stoichiometric, crystalline, Mg-rich (silicates only), and homogeneous are consistent with formation under equilibrium conditions, whereas those that are non-stoichiometric, amorphous/weakly nanocrystalline, Fe-rich (silicates only), and heterogeneous likely formed under non-equilibrium conditions. We will now discuss the features of each grain which impart information on circumstellar conditions.

*4.2.1 SiC*

Grain F2-30a is SiC with diffraction data consistent with either of the 3C or 2H polytypes or a 3C+2H mixture, which are the most common polytypes observed for presolar SiC (e.g., Daulton et al., 2003). The euhedral morphology of the grain is consistent with gas-phase condensation, and the lack of abundant defects and voids implies that the grain did not cool rapidly. The higher Al abundance in the region of high stacking fault density could indicate a change in gas composition during condensation, but might also have resulted from gradual diffusion of the Al to energetically favorable defect sites (e.g., the stacking fault boundaries) over the lifetime of the grain. Indeed, Al-rich SiC in association with stacking fault-rich regions was observed in a grain from an AGB star (G312) by Singerling et al. (2021).

*4.2.2 Spinel*

Grain F2-37 is stoichiometric and crystalline and, as such, this suggests that it formed under close to equilibrium conditions. If so, the fact that it is a Mg-Al spinel, rather than an Fe-Cr spinel, requires it to be a higher temperature condensate (Mg-Al spinel ≈1500 K and Fe-Cr spinel ≈1200 K for $P = 10^{-3}$ bars for solar system composition gas) (Yoneda and Grossman, 1995; Ebel, 2006). However, the observed Mg and Al heterogeneities within the grain argue for non-equilibrium condensation or post-condensation alteration.

*4.2.3 Silicates*

Grains F2-9, F2-30b, F2-23, and F1-1 all contain silicate components. As mentioned previously, amorphous or weakly nanocrystalline silicates that are non-stoichiometric and more Fe- rather than Mg-rich favor an origin under non-equilibrium conditions. Stoichiometric amorphous silicates likely formed as crystalline grains from equilibrium condensation and subsequently underwent processing (i.e., ion irradiation) in the ISM which caused them to lose their crystallinity (Nguyen et al., 2016). Processing in the ISM will be discussed in section 4.3. Figure 10 illustrates the stoichiometries, represented by the (Ca+Mg+Fe)/Si ratio, and the Fe-Mg content, represented by the Mg#, of the silicates. The ideal ratios for olivine and pyroxene are included as dashed horizontal lines, and higher Mg#'s indicate more Mg-rich compositions. The grains with observed heterogeneities have multiple data points, and the amorphous silicate matrix material is also included for comparison. Overall, the presolar silicate grains tend to be non-stoichiometric, with (Ca+Mg+Fe)/Si ratios ranging from 0.23 to 2.38. Although a few regions of the silicate grains are Mg-rich, the majority are more Fe-rich than equilibrium condensation calculations predict, which is essentially pure forsterite ($Fo_{100}$) (e.g., Sharp and Wasserburg, 1995; Lodders and Fegley, 1999).

Grain F2-9 is weakly nanocrystalline and non-stoichiometric and is likely a product of non-equilibrium condensation. The heterogeneities (Mg-rich, Si-rich, and Mg-rich rim) are visible in Figure 10 as different data points and show the variation in stoichiometry from intermediate between olivine and pyroxene to more Si-rich than pyroxene for the grain itself. The Mg-rich rim is more Si-poor than olivine and also notably more Fe-rich. If the rim is a primary feature, having formed in the circumstellar environment, it reflects a shift to more Si-poor and Fe-rich gas compositions. An alternative formation mechanism for the rim involves parent body processing which will be discussed in section 4.4. In either case, the interior regions of the grain are consistent with variable compositions and a lack of crystallinity, all consistent with condensation under non-equilibrium conditions.

Grain F2-30b is weakly nanocrystalline, non-stoichiometric, and Fe-rich. It is the most Fe-rich silicate observed in this study and likely a product of non-equilibrium condensation. Although theoretical studies do not predict the condensation of Fe-rich silicates at the high temperatures at which AGB dust forms, there is evidence for the presence of Fe-rich crystalline silicates in some O-rich AGB stars (Guha Niyogi et al., 2011). Again, the non-stoichiometric nature of this grain is consistent with non-equilibrium condensation. Grain F2-23 is similar to F2-30b, in that it is weakly nanocrystalline, non-stoichiometric, and Fe-rich, the second-most Fe-rich silicate observed in this study. It is also likely a product of non-equilibrium condensation. Although F2-30b and F2-23 may contain elevated concentrations of Fe from adjacent meteoritic materials, they are both even more Fe-rich than the amorphous matrix materials, implying that the Fe-rich compositions cannot simply be attributed to contamination.

Grain F1-1 is composed of several different regions, all of which are inferred to be silicates based on their Si and O contents. These include Mg-rich silicates with Si-rich pockets, Al-rich silicates, and a Cr-rich silicate. All four types are weakly nanocrystalline and non-stoichiometric, implying formation by non-equilibrium condensation. The co-existence of Mg-rich and Si-rich silicates is not consistent with equilibrium condensation, given that the formation of Mg-bearing silicates, such as forsterite or enstatite, is more thermodynamically favorable. Instead, more Si-rich silicates can form under non-equilibrium conditions around stars with low Mg/Si ratios and low mass-loss rates, both of which result in excess Si (e.g., Gail and Sedlmayr, 1999; Ferrarotti and Gail, 2001). The Al-rich and Cr-rich regions are not likely to be oxides, given that the stoichiometries are not appropriate. The Ca- and Al-rich silicates are instead reminiscent of calcium-aluminum-rich inclusions (CAIs) in primitive meteorites, although CAIs are the product of equilibrium or fractional condensation. Presolar grains containing Ca-Al-rich phases with ferromagnesian silicates have been reported by TEM (Nguyen et al., 2010; Vollmer et al., 2013; Nguyen et al., 2014; Nittler et al., 2018) and Auger electron spectroscopy (Vollmer et al., 2009a; Floss and Stadermann, 2012; Leitner et al., 2018) previously.

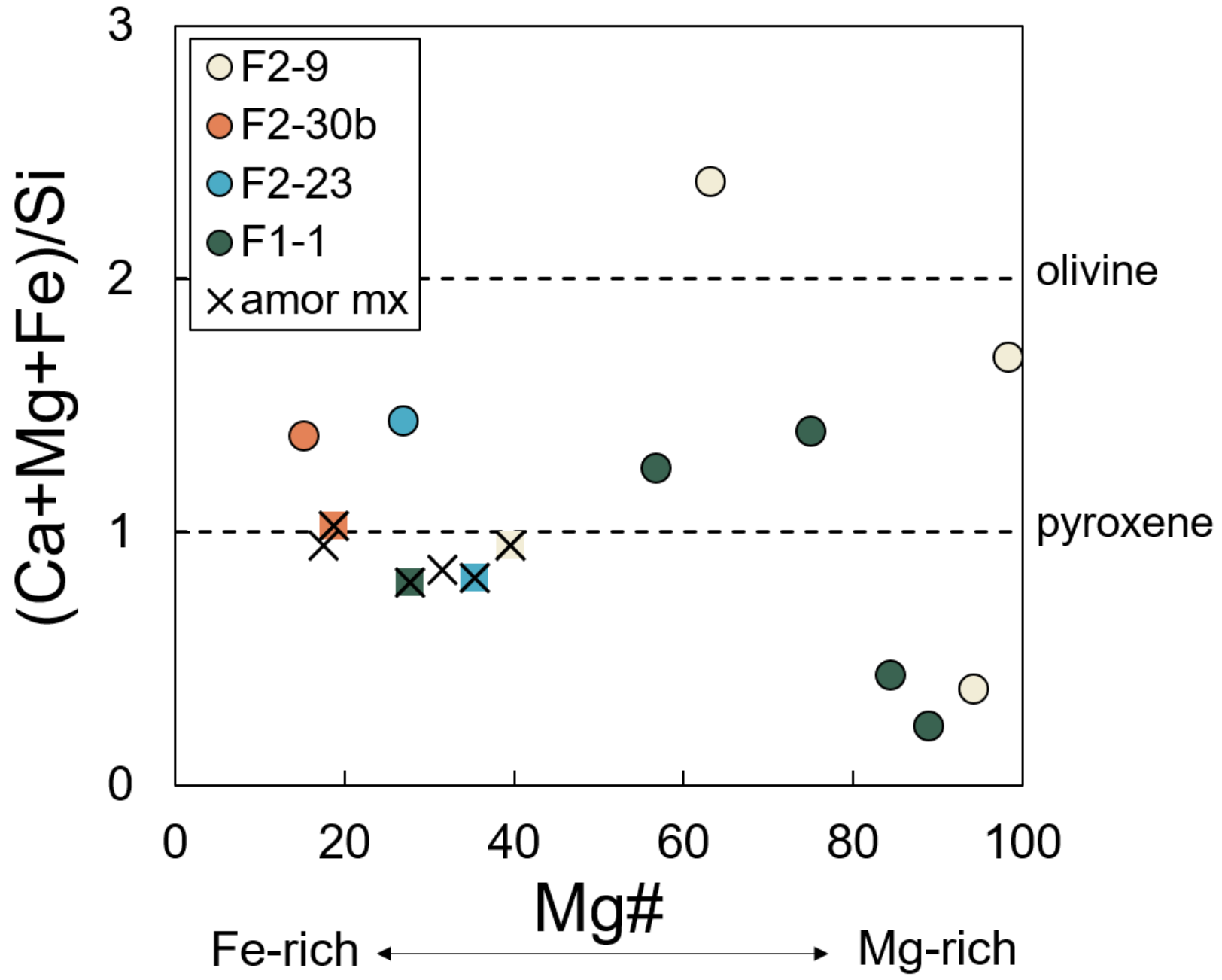

**Figure 10.** Compositional data on the presolar silicates and the matrix amorphous silicate material. Presolar silicates with compositional heterogeneities are plotted separately (e.g., F2-9 = Mg-rich, Si-rich, and rim regions; see Table 4 for data). The matrix amorphous silicates (amor mx) are also included from all samples studied; those from the presolar silicates are color-coded by sample. The ratio (Ca+Mg+Fe)/Si is used as a proxy for the stoichimetry of the silicates, with pyroxene having a ratio of 1.0 and olivine 2.0. The presolar silicates have values that vary from 0.23 to 2.38, with no presolar silicates having a ratio consistent with olivine or pyroxene. The matrix amorphous silicates have ratios similar to pyroxene. The Mg# illustrates whether the material is more Mg-rich or Fe-rich. As the plot shows, most of the presolar silicates are much more Fe-rich than predicted from equilibrium condensation calculations. The matrix amorphous silicates are also Fe-rich.

### 4.3 Interstellar Medium (ISM) Processing

Following their formation in circumstellar envelopes, radiation pressure on presolar grains drives a strong stellar wind, propelling the grains away from their progenitor stars to become part of the diffuse ISM and eventually the molecular cloud from which the Sun formed. Dust in the ISM can be affected by a number of energetic processes, such as acceleration by supernova shock leading to sputtering by gas atoms and/or destruction in grain-grain collisions.

Ion irradiation and grain-grain collisions can be simulated in the laboratory with ion implantation and pulsed laser methods, respectively. In general, these processes tend to amorphize silicates (e.g., Wang et al., 2000; Demyk et al., 2001; Leroux, 2001; Carrez et al., 2002; Jäger et al., 2003; Demyk et al., 2004; Weber et al., 2020; Laczniak et al., 2021). Astronomical observations also show a decrease in the proportion of crystalline to amorphous silicates from circumstellar envelopes, where 10 to 20% of silicates are crystalline, to the ISM, where less than 2% of silicates are crystalline (Kemper et al., 2004, 2005). Clearly, processes within the ISM, such as sputtering from acceleration by supernova shocks and ion irradiation, bombardment by low energy cosmic rays, and grain-grain collisions, are responsible for this amorphization of crystalline silicates, which can happen on timescales of about 10 Ma (Jones et al., 1994; Tielens et al., 1994; van Dishoeck, 2004; Bringa et al., 2007; Jones and Nuth, 2011).

Observations of presolar silicate grains also show a high proportion of amorphous material, where about 2/3 of the grains from chondrites and IDPs have been found to be amorphous (Nittler et al., 2018). The higher proportion of crystalline materials in presolar grains (33%) as compared to the observations of the dust in the ISM (<2%) may partly be explained by the fact that not all grains are affected by processes in the ISM equally or at the same rate. Alternatively, a portion of the amorphous dust observed in the ISM may have formed there at low temperature. When discussing presolar grains from solar system materials, there is the added complicating factor of the effects of protoplanetary disk, solar nebula, and parent body processes. Thermal processes, such as annealing, melting, and vaporization, tend to generate more crystalline materials, and interactions with fluids tend to create phyllosilicates or other hydrous phases. Still, while these processes are unlikely to generate amorphized silicates, they could very well preferentially destroy preexisting amorphous silicates.

In principle, one could use presolar grains to investigate ISM amorphization processes, but how can one distinguish primary formation as an amorphous grain from formation as a crystalline grain that was subsequently amorphized? One main difference between primary amorphous (circumstellar) and secondary amorphous (ISM) silicates is related to their compositions, specifically their stoichiometries. Since equilibrium condensation results in stoichiometric, crystalline phases, whereas non-equilibrium condensation processes can clearly lead to a wide range of non-stoichiometric compositions, grain compositions may provide an answer. Namely, grains with close to pure stoichiometry of high-T phases, like forsterite, would have likely formed as crystalline materials, and if found to be amorphous, could be inferred to have been amorphized following their formation (Nguyen et al., 2016). An analogy is the presolar $MgSiO_3$ silicate grain with perovskite structure reported by Vollmer et al. (2007); this grain likely formed as enstatite in its parent AGB star and later transformed in the high-pressure perovskite structure by grain-to-grain collisions caused by supernova shocks. None of the amorphous and/or weakly nanocrystalline presolar silicates reported here have stoichiometric mineral compositions. It is important to point out that ion irradiation can result in compositional changes along with amorphization. Carrez et al. (2002) performed experiments in which olivine was irradiated by $He^+$ at low energy (4–50 keV) and observed decreases in O/Si and Mg/Si ratios. Iron concentrations, on the other hand, remained unchanged, with the Fe forming metallic nanoparticles. However, these effects were most significant for thin materials (<100 nm). Given that the silicate grains in this study were ≥80 nm, such compositional changes would be relatively minor. Additionally, Carrez et al. (2002) noted the presence of nm-scale bubbles adjacent to amorphized materials. Given the larger sizes of the presolar grains of this study, ion irradiation at these lower energies ought to result in partly amorphous material adjacent to bubbles-rich regions. The fact that we do not observe such textural features and that no metallic nanophase Fe was observed imply that the non-stoichiometries of the grains are not the result of compositional changes from ion irradiation.

Another related feature tied to processing in the ISM from previous studies involves compositional heterogeneities. However, there is conflicting information in the literature on how ISM processing might affect compositional distributions within presolar grains. Sun et al. (2004) argue that annealing and/or radiation processing in the ISM would cause heterogeneities by promoting internal diffusion and solid-state precipitation of nanoscale subgrains. On the other hand, Keller et al. (2005) argue that amorphization in the ISM would likely lead to chemical homogenization. Further experimental work has the potential to resolve this discrepancy, especially as the response to ISM processes would depend on the original compositions of the grains as well as their radiation resistance. For the grains of this study with heterogeneities (F2-37, F2-9, and F1-1), we argue that these are likely a product of either non-equilibrium condensation or parent body processes, given the minimal evidence for ISM process overprinting related to amorphization as discussed above. Note that some ISM processes (e.g., ion implantation, accretion of grain mantles) could in principle change the chemical composition of a circumstellar grain, but in so doing would likely substantially dilute or erase the isotopic signatures that are used to identify it as presolar.

### 4.4 Parent Body Alteration

After their transit through the ISM, presolar grains were eventually incorporated into the protoplanetary disk that formed our solar system. During the early stages of the disk, nebular processes including ion irradiation from the young Sun, heating, and oxidation could have affected the presolar grains; however, we did not find any evidence for such nebular processes. During the early stages of planet formation, presolar grains that survived nebular processes were incorporated, along with a much larger amount of material that formed in the disk itself, into asteroids and comets, including the Semarkona asteroidal parent body. Numerous processes occur on asteroids with the potential for affecting presolar grains and overprinting their isotopic, chemical, and microstructural properties, namely shock metamorphism from impacts, thermal metamorphism from heating by decay of short-lived radionuclides (i.e., $^{26}$Al, $^{60}$Fe), and aqueous alteration from the interaction of fluids with rocky materials. As mentioned previously, Semarkona experienced minimal thermal processing compared to other chondritic meteorites, but it shows evidence for aqueous alteration as the dominant secondary process. However, the areas we analyzed were targeted because they show considerably less aqueous alteration than much of the meteorite (Dobrică and Brearley, 2020).

Our TEM images of Semarkona presolar grains demonstrate a lack of obvious boundaries between the presolar grains and adjacent meteorite materials. This is especially true for oxide F2-37, silicates F2-9, F2-30b, F2-23, and F1-1. This is at odds with previous studies of presolar grains from CCs with minimal aqueous alteration, where the boundaries are often more distinct (e.g., Vollmer et al., 2009a). One potential explanation for the lack of distinct boundaries for the Semarkona grains is that this reflects parent body processes, specifically annealing or interaction with fluids. In general, CCs experienced less heating than OCs, although there are exceptions. More pertinent to this discussion is the relative heating of Semarkona as compared to the CCs investigated in previous TEM presolar grains studies, especially the highly unaltered ungrouped C2 Acfer 094 and CO3.0 Dominion Range (DOM) 08006 (Vollmer et al., 2009a; Nittler et al., 2018), which along with CM Asuka 12169 (Nittler et al., 2021), contain the highest abundances of presolar silicates of any chondrites. Grossman and Brearley (2005) found that both Semarkona and Acfer 094 are highly primitive (3.00), having escaped significant heating, as determined from the Cr contents of FeO-rich olivines in Type IIA chondrules. Davidson et al. (2019) found that DOM 08006 was similarly primitive (3.00), based on the Cr content of the olivine as well. Given the similarities in the degree of heating between Semarkona and the three CCs, it is unlikely that the lack of grain boundaries was a product of annealing from thermal metamorphism. Instead, differences in the degree of aqueous alteration of the meteorites might explain our observations concerning grain boundaries. However, given the brecciated nature of chondrites, the role of thermal metamorphism in erasing grain boundaries between presolar and meteoritic components cannot be entirely discounted, although aqueous alteration is a more likely explanation.

Another potential indicator for parent body processing is Fe enrichment of silicates. As discussed above, equilibrium condensation calculations indicate that circumstellar mafic silicates should be very Mg-rich (i.e., Mg# near 100). In our study, all grains or regions within grains are more Fe rich than expected from equilibrium condensation which is likely a primary feature reflecting condensation under non-equilibrium conditions as concluded above from their non-stoichiometric compositions. However, parent-body effects should also be considered. Indeed, some previous studies have invoked parent body alteration to explain the presence of more Fe-rich presolar silicates (e.g., Nguyen and Zinner, 2004; Nguyen et al., 2007; Vollmer et al., 2009b; Floss and Stadermann, 2012). If parent body alteration were indeed responsible for the Fe enrichment, we might expect the grains' isotopic anomalies to be reduced by infiltration of parent-body fluids. Still, this is difficult to assess on an individual grain basis, as the initial anomaly of a given grain could have been greater than that observed; reduced anomalies only become apparent in comparing grain populations. Additionally, if the Fe-rich presolar silicates were the product of aqueous alteration on the parent body, solar silicates ought to also show Fe enrichments, especially on their grain boundaries for incipient aqueous alteration. However, Dobrică and Brearley (2020) did not note such enrichments on the pyroxenes in Semarkona. This is an example of how we can use compositional and textural features in conjunction with Fe content to better assess whether parent body processes altered presolar silicates.

One last feature of the presolar grains that could reflect parent body alteration is the presence of compositional heterogeneities within the grains. Heterogeneities in Mg and Al, which are not readily mobile during parent body alteration (Brearley, 2006), are likely primary features and the product of non-equilibrium condensation. However, elements like S and Fe are more mobile during parent body processes (Brearley, 2006). The S-rich rim on SiC F2-30a (Fig. 4e) is likely a secondary parent body alteration feature, with the S-rich material having come from the matrix and been mobilized during alteration before concentrating near the surface of the presolar grain. In contrast, the Mg-rich rim on silicate F2-9 may have initially formed in the circumstellar environment, consistent with the heterogeneity, non-stoichiometry, and weakly nanocrystalline nature in the interior of the grain. However, the rim is also enriched in Fe (Fig. 6d), with an Fe content of 9.8 at.% compared to Fe contents ranging from 0.4 to 0.6 at.% from the interior (Mg- and Si-rich) regions of the grain. In this instance, the Fe enrichment seems to be a product of leaching or infiltration of Fe into the presolar grain from the adjacent matrix material.

In short, although there is evidence for parent body processing in Semarkona as a whole, the presolar grains show minimal overprinting. This holds true even for the presolar silicates, which are exceptionally sensitive to alteration. The fact that the presolar grains have survived and that they show minimal overprinting demonstrates the effectiveness and utility across chondrite groups of the criteria for identifying pristine matrix, as outlined in numerous studies (Greshake, 1997; Chizmadia and Brearley, 2008; Abreu and Brearley, 2010; Davidson et al., 2014; Le Guillou and Brearley, 2014; Leroux et al., 2015; Dobrică and Brearley, 2020).

### 4.5 Comparisons of Presolar Oxide and Silicate Grains in IDPs, CCs, and OCs

Presolar grains have been observed in both asteroidal and cometary materials, the former hosted in primitive meteorites (i.e., chondrites) and the latter hosted in chondritic porous IDPs (hereafter, IDPs), ultracarbonaceous Antarctic micrometeorites (UCAMMs), and samples returned from the Stardust mission (Nittler and Ciesla, 2016). Moreover, there is mounting evidence that asteroidal materials can be divided into two isotopic reservoirs, with the carbonaceous (CC) reservoir originating farther from the Sun than the non-carbonaceous one (NC) (Warren, 2011;

Kruijer et al., 2020). Studies of the presolar components present in samples from each of these reservoirs could potentially answer lingering questions related to processes in the early solar system, such as the presence or absence of large-scale mixing and inheritances of heterogeneities from the solar system's parent molecular cloud. Given the time-intensive nature of TEM studies of presolar grains, there is currently a limited amount of data available for statistically significant comparisons between inner and outer solar system materials.

Previous TEM studies have characterized presolar oxide and silicate grains in IDPs (in situ silicates), CCs (in and ex situ oxides and silicates), and OCs (ex situ oxides) (Messenger et al., 2003; Stroud et al., 2004; Messenger et al., 2005; Nguyen et al., 2007; Stroud et al., 2007; Vollmer et al., 2007; Busemann et al., 2009; Stroud et al., 2009; Vollmer et al., 2009b; Nguyen et al., 2010; Zega et al., 2011; Bose et al., 2012; Leitner et al., 2012; Vollmer et al., 2013; Zega and Floss, 2013; Nguyen et al., 2014; Takigawa et al., 2014; Zega et al., 2014; Zega et al., 2015; Nguyen et al., 2016; Nittler et al., 2018; Takigawa et al., 2018; Leitner et al., 2020; Nittler et al., 2020). The current study adds new data for presolar oxides and silicates from an OC. The data from previous studies are summarized in Table S4.

Figure 11 graphically presents the crystallinity (single crystal, polycrystalline, nanocrystalline/amorphous) and stoichiometry of the grains. The oxides and silicates are shown separately and grouped by host material. Each data point corresponds to a grain, with the color indicating whether the grain is stochiometric (blue) or not (red). In general, the oxides are crystalline (single crystal or polycrystalline) and stoichiometric for CCs and OCs. The silicates are predominantly poorly crystallized (amorphous, nanocrystalline, and/or weakly nanocrystalline) and non-stoichiometric for CCs. The minimal data for silicates from IDP and OC makes it difficult to observe any trends regarding crystallinity and stoichiometry, but the data from the current work implies this propensity for poor crystallization and non-stoichiometry also holds true for silicates in OCs. A detailed comparison of isotopic characteristics of these grain populations will be presented elsewhere.

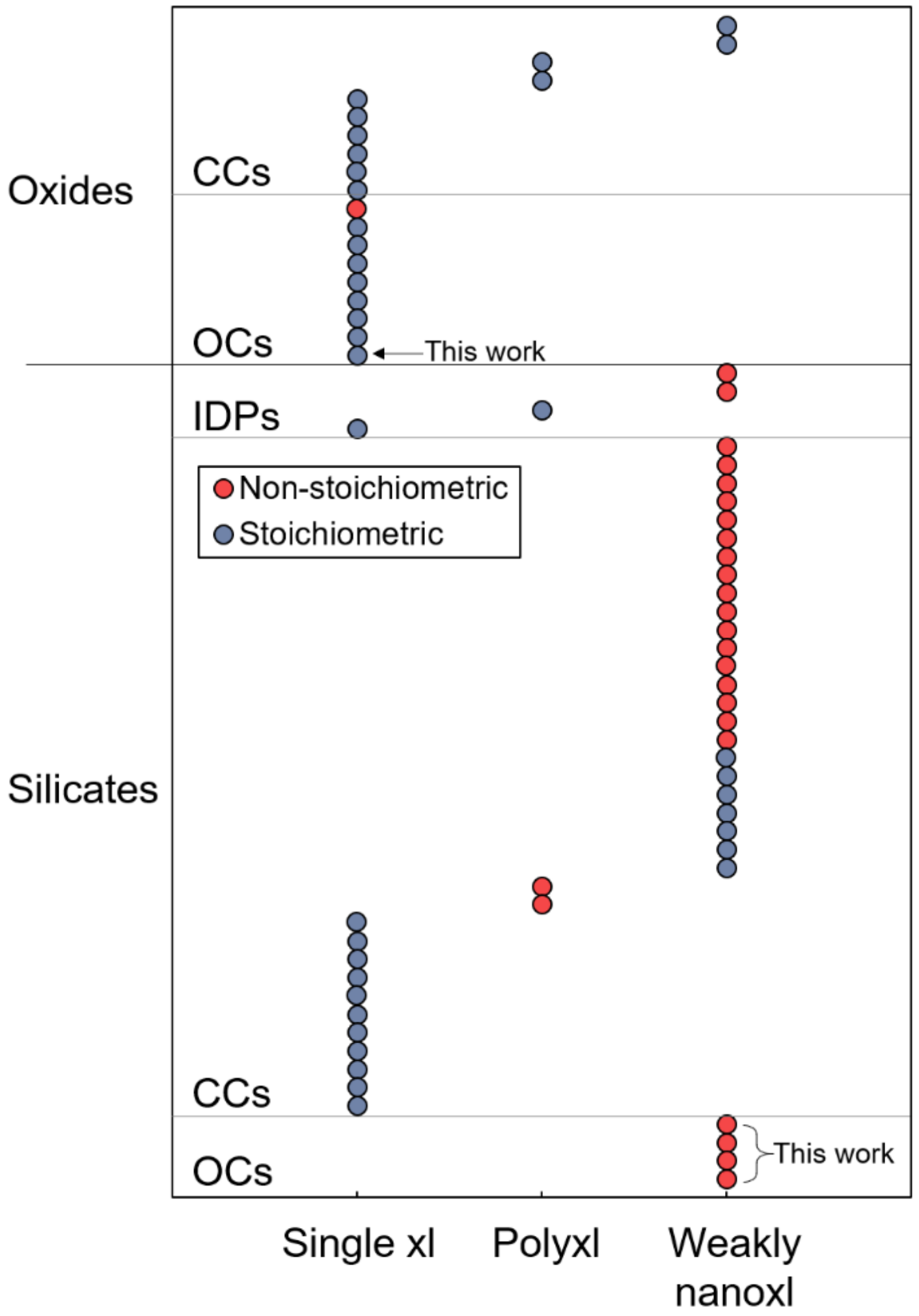

**Figure 11.** Comparisons of crystallinity and stoichiometry between presolar oxide and silicate grains from different hosts, including IDPs, CCs, and OCs,; see Table S4 for the data and sources (Messenger et al., 2003; Stroud et al., 2004; Messenger et al., 2005; Nguyen et al., 2007; Stroud et al., 2007; Vollmer et al., 2007; Busemann et al., 2009; Stroud et al., 2009; Vollmer et al., 2009a; Nguyen et al., 2010; Zega et al., 2011; Bose et al., 2012; Leitner et al., 2012; Vollmer et al., 2013; Zega and Floss, 2013; Nguyen et al., 2014; Takigawa et al., 2014; Zega et al., 2014; Zega et al., 2015; Nguyen et al., 2016; Nittler et al., 2018; Takigawa et al., 2018; Leitner et al., 2020; Nittler et al., 2020; Sanghani et al., 2022). The presolar grains featured in this study are labeled "This work". The crystallinity of each grain is shown along with the stoichiometry of the phases. Oxides tend to be crystalline—single crystal (xl) or polycrystalline (polyxl)—and stoichiometric, whereas silicates are predominantly weakly nanocrystalline (weakly nanoxl) and non-stoichiometric. With the limited data currently available, these trends are true for the different host materials (e.g., IDPs, CCs, and OCs).

## 5. CONCLUSIONS

We have presented an in situ TEM study of six presolar grains from the unequilibrated ordinary chondrite Semarkona. We characterized the microstructures and elemental compositions of one SiC, one oxide (Mg-Al spinel), and four silicate grains. The SiC and spinel grains were single crystals and stoichiometric, whereas the silicates tended to be weakly nanocrystalline and non-stoichiometric. Our findings of stochiometric, well crystallized oxides and non-stoichiometric, poorly crystalized silicates are consistent with previous TEM studies of presolar grains from other planetary materials (IDPs, CCs, and OCs), although this study is the first to analyze grains in situ from an OC.

The in situ TEM studies reveal features in each grain that impart information on circumstellar conditions, processing in the ISM, and parent body alteration. In general, the presolar silicates are consistent with gas-phase condensation under non-equilibrium conditions, based on their weakly nanocrystalline nature, non-stoichiometries, and high Fe (Mg# <90) contents. In the grains studied, we found no evidence for processing in the ISM or the solar nebula. Similarly, very little overprinting from parent body alteration was observed with the exception of a S-rich rim on SiC grain F2-30a and Fe infiltration into the rim of silicate grain F2-9. As such, future studies of presolar grains in pristine materials can provide further insights into circumstellar conditions and interstellar medium processing. Additional in situ TEM studies of presolar grains of all types from various meteorite groups and planetary materials are required to better compare their grain populations. There is a particular lack of data for oxides both in IDPs and CCs, as well as silicates in IDPs and OCs.

## Acknowledgements

We thank Associate Editor Martin Lee and three anonymous reviewers for excellent feedback and suggestions for improving the manuscript. This research was supported by NASA Emerging Worlds grants 80HQTR19T0038 (RMS) and 80NSSC20K0340 (LRN) and NASA Cosmochemistry grant NNX15AD28G (AJB).

## References

Abreu N. M. and Brearley A. J. (2010) Early solar system processes recorded in the matrices of two highly pristine CR3 carbonaceous chondrites, MET 00426 and QUE 99177. *Geochim. Cosmochim. Acta* **74**, 1146–1171.

Alexander C. M. O'D., Barber D. J., and Hutchison R. (1989) The microstructure of Semarkona and Bishunpur. *Geochim. Cosmochim. Acta* **53**, 3045–3057.

Amari S., Lewis R. S., and Anders E. (1994) Interstellar grains in meteorites: I. Isolation of SiC, graphite, and diamond; size distributions of SiC and graphite. *Geochim. Cosmochim. Acta* **58**, 459–470.

Barosch J., Nittler L. R., Dobrică E., Brearley A. J., Hezel D. C., and Alexander C. M. O'D. (2021) Presolar O- and C-anomalous grains in pristine ordinary chondrite matrices, 84th Annual Meeting of the Meteoritical Society. *Meteor. Planet. Sci.*, p. 6137.

Bernatowicz T. J., Messenger S., Pravdivtseva O., Swan P., and Walker R. M. (2003) Pristine presolar silicon carbide. *Geochim. Cosmochim. Acta* **67**, 4679–4691.

Bose M., Floss C., Stadermann F. J., Stroud R. M., and Speck A. K. (2012) Circumstellar and interstellar material in the CO3 chondrite ALHA77307: An isotopic and elemental investigation. *Geochim. Cosmochim. Acta* **93**, 77–101.

Brearley A. J. (1993) Matrix and fine-grained rims in the unequilibrated CO3 chondrite, ALHA77307: Origins and evidence for diverse, primitive nebular dust components. *Geochim. Cosmochim. Acta* **57**, 1521–1550.

Brearley A. J. (2006) The action of water. *Meteorites and the early solar system II* **1**, 587–624.

Bringa E. M., Kucheyev S. O., Loeffler M. J., Baragiola R. A., Tielens A. G. G. M., Dai Z. R., Graham G., Bajt S., Bradley J. P., Dukes C. A., Felter T. E., Torres D. F., and van Breugel W. (2007) Energetic processing of interstellar silicate grains by cosmic rays. *Astrophys. Jour.* **662**, 372–378.

Busemann H., Nguyen A. N., Cody G. D., Hoppe P., Kilcoyne A. L. D., Stroud R. M., Zega T. J., and Nittler L. R. (2009) Ultra-primitive interplanetary dust particles from the comet 26P/Grigg-Skjellerup dust stream collection. *Earth Planet. Sci. Lett.* **288**, 44–57.

Carrez P., Demyk K., Cordier P., Gengembre L., Grimblot J., D'Hendecourt L., Jones A. P., and Leroux H. (2002) Low-energy helium ion irradiation-induced amorphization and chemical changes in olivine: Insights for silicate dust evolution in the interstellar medium. *Meteor. Planet. Sci.* **37**, 1599–1614.

Cherchneff, I. (2013). The chemistry of dust formation in red supergiants. *Eur. Astro. Soc. Pub. Ser.* **60**, 175–184.

Chizmadia L. J. (2004) Aqueous alteration of CM carbonaceous chondrites: Constraints on early solar system processes. Ph. D. thesis, Univ. of New Mexico.

Chizmadia L. J. and Brearley A. J. (2008) Mineralogy, aqueous alteration, and primitive textural characteristics of fine-grained rims in the Y-791198 CM2 carbonaceous chondrite: TEM observations and comparison to ALHA81002. *Geochim. Cosmochim. Acta* **72**, 602–625.

Daulton T. L., Bernatowicz T. J., Lewis R. S., Messenger S., Stadermann F. J., and Amari S. (2003) Polytype distribution of circumstellar silicon carbide: Microstructural characterization by transmission electron microscopy. *Geochim. Cosmochim. Acta* **67**, 4743–4767.

Davidson J., Busemann H., Nittler L. R., Alexander C. M. O'D., Orthous-Daunay F.-R., Franchi I. A., and Hoppe P. (2014) Abundances of presolar silicon carbide grains in primitive meteorites determined by NanoSIMS. *Geochim. Cosmochim. Acta* **139**, 248–266.

Davidson J., Alexander C. M. O'D., Stroud R. M., Busemann H., and Nittler L. R. (2019) Mineralogy and petrology of Dominion Range 08006: A very primitive CO3 carbonaceous chondrite. *Geochim. Cosmochim. Acta* **265**, 259–278.

Demyk K., Carrez P., Leroux H., Cordier P., Jones A. P., Borg J., Quirico E., Raynal P. I., and d'Hendecourt L. (2001) Structural and chemical alteration of crystalline olivine under low energy He+ irradiation. *Astron. Astrophys.* **368**, L38–L41.

Demyk K., d'Hendecourt L., Leroux H., Jones A. P., and Borg J. (2004) IR spectroscopic study of olivine, enstatite and diopside irradiated with low energy H+ and He+ ions. *Astron. Astrophys.* **420**, 233–243.

Dobrică E. and Brearley A. J. (2020) Amorphous silicates in the matrix of Semarkona: The first evidence for the localized preservation of pristine matrix materials in the most unequilibrated ordinary chondrites. *Meteor. Planet. Sci.* **55**, 649–668.

Ebel D. S. (2006) Condensation of rocky material in astrophysical environments. *Meteorites and the early solar system II* **1**, 253–277.

Ferrarotti A. S. and Gail H. P. (2001) Mineral formation in stellar winds - II. Effects of Mg/Si abundance variations on dust composition in AGB stars. *Astron. Astrophys.* **371**, 133–151.

Floss C. and Stadermann F. J. (2012) Presolar silicate and oxide abundances and compositions in the ungrouped carbonaceous chondrite Adelaide and the K chondrite Kakangari: The effects of secondary processing. *Meteor. Planet. Sci.* **47**, 992–1009.

Floss C., Le Guillou C., and Brearley A. J. (2014) Coordinated NanoSIMS and FIB-TEM analyses of organic matter and associated matrix materials in CR3 chondrites. *Geochim. Cosmochim. Acta* **139**, 1–25.

Floss C. and Haenecour P. (2016) Presolar silicate grains: Abundances, isotopic and elemental compositions, and the effects of secondary processing. *Geochem. J.* **50**, 3–25.

Gail H. P. and Sedlmayr E. (1999) Mineral formation in stellar winds - I. Condensation sequence of silicate and iron grains in stationary oxygen rich outflows. *Astron. Astrophys.* **347**, 594–616.

Gail H. P., Zhukovska S. V., Hoppe P., and Trieloff M. (2009) Stardust from asymptotic giant branch stars. *Astrophys. Jour.* **698**, 1136–1154.

Gosling J. T. (2007) The solar wind. In *Encyclopedia of the Solar System* (eds. L. A. McFadden, P. R. Weissman, and T. V. Johnson). Academic Press-Elsevier, Amsterdam. pp. 99–116.

Greshake A. (1997) The primitive matrix components of the unique carbonaceous chondrite Acfer 094: A TEM study. *Geochim. Cosmochim. Acta* **61**, 437–452.

Grossman J. N. and Brearley A. J. (2005) The onset of metamorphism in ordinary and carbonaceous chondrites. *Meteor. Planet. Sci.* **40**, 87–122.

Guha Niyogi S., Speck A. K., and Onaka T. (2011) A temporal study of the oxygen-rich pulsating variable asymptotic giant branch star, T Cep: Investigation on dust formation and dust properties. *Astrophys. Jour.* **733**.

Haenecour P. and Floss C. (2013) Presolar silica grains in meteorites: Identification of a supernova silica grain in the CO3.0 chondrite Lapaz 031117. *Lunar Planet. Sci. XLIV.* Lunar Planet. Inst., Houston, #1024(abstr.).

Heck P. R., Greer J., Kööp L., Trappitsch R., Gyngard F., Busemann H., Maden C., Ávila J. N., Davis A. M., and Wieler R. (2020) Lifetimes of interstellar dust from cosmic ray exposure ages of presolar silicon carbide. *Proc. Nat Acad. Sci.* **117**, 1884–1889.

Honda M., Kataza H., Okamoto Y. K., Miyata T., Yamashita T., Sako S., Takubo S., and Onaka T. (2003) Detection of crystalline silicates around the T Tauri star Hen 3-600A. *Astrophys. Jour.* **585**, L59.

Hoppe P., Leitner J., and Kodolányi J. (2018) New insights into the Galactic Chemical Evolution of magnesium and silicon isotopes from studies of silicate stardust. *Astrophys. Jour.* **869**, 47.

Hoppe P., Leitner J., Kodolányi J., and Vollmer C. (2021) Isotope systematics of presolar silicate grains: New insights from magnesium and silicon. *Astrophys. Jour.* **913**, 10.

Huss G. R., Rubin A. E., and Grossman J. N. (2006) Thermal metamorphism in chondrites. *Meteorites and the early solar system II* **943**, 567–586.

Hutchison R., Alexander C. M. O'D., and Barber D. J. (1987) The Semarkona meteorite - First recorded occurrence of smectite in an ordinary chondrite, and its implications. *Geochim. Cosmochim. Acta* **51**, 1875–1882.

Jäger C., Fabian D., Schrempel F., Dorschner J., Henning T., and Wesch W. (2003) Structural processing of enstatite by ion bombardment. *Astron. Astrophys.* **401**, 57–65.

Jones A. P., Tielens A. G. G. M., Hollenbach D. J., and Mckee C. F. (1994) Grain destruction in shocks in the interstellar-medium. *Astrophys. Jour.* **433**, 797–810.

Jones A. P. and Nuth J. A. (2011) Dust destruction in the ISM: A re-evaluation of dust lifetimes. *Astron. Astrophys.* **530**, A44.

Jones O. C., Kemper F., Sargent B. A., McDonald I., Gielen C., Woods P. M., Sloan G. C., Boyer M. L., Zijlstra A. A., and Clayton G.C. (2012) On the metallicity dependence of crystalline silicates in oxygen-rich asymptotic giant branch stars and red supergiants. *Mon. Not. R. Astron. Soc.* **427**, 3209–3229.

Keller L. P., Messenger S., and Christoffersen R. (2005) GEMS revealed: Spectrum imaging of aggregate grains in interplanetary dust. *Lunar Planet. Sci. XXXVI.* Lunar Planet. Inst., Houston, #2088(abstr.).

Kemper F., Vriend W. J., and Tielens A. G. G. M. (2004) The absence of crystalline silicates in the diffuse interstellar medium. *Astrophys. Jour.* **609**, 826–837.

Kemper F., Vriend W. J., and Tielens A. G. G. M. (2005) The absence of crystalline silicates in the diffuse interstellar medium (vol 609, pg 826, 2004). *Astrophys. Jour.* **633**, 534.

Krot A. N., Zolensky M. E., Wasson J. T., Scott E. R. D., Keil K., and Ohsumi K. (1997) Carbide-magnetite assemblages in type-3 ordinary chondrites. *Geochim. Cosmochim. Acta* **61**, 219–237.

Kruijer T. S., Kleine T., and Borg L. E. (2020) The great isotopic dichotomy of the early Solar System. *Nat. Astron.* **4**, 32–40.

Laczniak D. L., Thompson M. S., Christoffersen R., Dukes C. A., Clement S. J., Morris R. V., and Keller L. P. (2021) Characterizing the spectral, microstructural, and chemical effects of solar wind irradiation on the Murchison carbonaceous chondrite through coordinated analyses. *Icar.* **364**.

Le Guillou C. and Brearley A. J. (2014) Relationships between organics, water and early stages of aqueous alteration in the pristine CR3. 0 chondrite MET 00426. *Geochim. Cosmochim. Acta* **131**, 344–367.

Leitner J., Vollmer C., Hoppe P., and Zipfel J. (2012) Characterization of presolar material in the CR chondrite Northwest Africa 852. *Astrophys. Jour.* **745**.

Leitner J., Hoppe P., Floss C., Hillion F., and Henkel T. (2018) Correlated nanoscale characterization of a unique complex oxygen-rich stardust grain: Implications for circumstellar dust formation. *Geochim. Cosmochim. Acta* **221**, 255–274.
Leitner J. and Hoppe P. (2019) A new population of dust from stellar explosions among meteoritic stardust. *Nat. Astro.* **3**, 725–729.
Leitner J., Metzler K., Vollmer C., Floss C., Haenecour P., Kodolányi J., Harries D., and Hoppe P. (2020) The presolar grain inventory of fine-grained chondrule rims in the Mighei-type (CM) chondrites. *Meteor. Planet. Sci.* **55**, 1176–1206.
Leroux H. (2001) Microstructural shock signatures of major minerals in meteorites. *Eur. J. Mine.* **13**, 253–272.
Leroux H., Cuvillier P., Zanda B., and Hewins R. H. (2015) GEMS-like material in the matrix of the Paris meteorite and the early stages of alteration of CM chondrites. *Geochim. Cosmochim. Acta* **170**, 247–265.
Lodders K. and Fegley B. (1999) Condensation chemistry of circumstellar grains. In *I.A.U. Symposium* (eds. T. Le Berte and A. C. W. Lebre). Cambridge University Press, pp. 279–290.
Lodders K. and Amari S. (2005) Presolar grains from meteorites: Remnants from the early times of the solar system. *Chem. Erde-Geoche.* **65**, 93–166.
Mendybaev R. A., Beckett J. R., Grossman L., Stolper E., Cooper R. F., and Bradley J. P. (2002) Volatilization kinetics of silicon carbide in reducing gases: An experimental study with applications to the survival of presolar grains in the solar nebula. *Geochim. Cosmochim. Acta* **66**, 661–682.
Messenger S., Keller L. P., Stadermann F. J., Walker R. M., and Zinner E. (2003) Samples of stars beyond the solar system: Silicate grains in interplanetary dust. *Science* **300**, 105–108.
Messenger S., Keller L. P., and Lauretta D. S. (2005) Supernova olivine from cometary dust. *Science* **309**, 737–741.
Molster F. J., Waters L. B. F. M., and Tielens A. G. G. M. (2002) Crystalline silicate dust around evolved stars-II. The crystalline silicate complexes. *Astron. Astrophys.* **382**, 222–240.
Nagahara H. and Ozawa K. (2009) Condensation kinetics of forsterite and metal and chemical fractionation in the proto solar nebula. *Lunar Planet. Sci. XL.* Lunar Planet. Inst., Houston, #2158(abstr.).
Nagahara H., Ogawa R., Ozawa K., Tamada S., Tachibana S. and Chiba H. (2009) Laboratory condensation and reaction of silicate dust. *Astr. Soc. P.* **414**, 403–410.
Nguyen A. N. and Zinner E. (2004) Discovery of ancient silicate stardust in a meteorite. *Science* **303**, 1496–1499.
Nguyen A. N., Stadermann F. J., Zinner E., Stroud R. M., Alexander C. M. O'D., and Nittler L. R. (2007) Characterization of presolar silicate and oxide grains in primitive carbonaceous chondrites. *Astrophys. Jour.* **656**, 1223–1240.
Nguyen A. N., Nittler L. R., Stadermann F. J., Stroud R. M., and Alexander C. M. O'D. (2010) Coordinated analyses of presolar grains in the Allan Hills 77307 and Queen Elizabeth Range 99177 meteorites. *Astrophys. Jour.* **719**, 166–189.
Nguyen A. N., Nakamura-Messenger K., Messenger S., Keller L. P., and Klock W. (2014) Identification of a compound spinel and silicate presolar grain in a chondritic interplanetary dust particle. *Lunar Planet. Sci. XLV.* Lunar Planet. Inst., Houston, #2351(abstr.).

Nguyen A. N., Keller L. P., and Messenger S. (2016) Mineralogy of presolar silicate and oxide grains of diverse stellar origins. *Astrophys. Jour.* **818**.
Nittler L. R., Alexander C. M. O'D., Gao X., Walker R. M., and Zinner E. (1997) Stellar sapphires: The properties and origins of presolar Al2O3 in meteorites. *Astrophys. Jour.* **483**, 475–495.
Nittler L. R., Alexander C. M. O'D., Gallino R., Hoppe P., Nguyen A. N., Stadermann F. J., and Zinner E.K. (2008) Aluminum-, calcium-and titanium-rich oxide stardust in ordinary chondrite meteorites. *The Astrophys. Jour.* **682**, 1450.
Nittler L. R. and Ciesla F. (2016) Astrophysics with extraterrestrial materials. *Ann. Rev. Astron. Astrophys.* **54**, 53–93.
Nittler L. R., Alexander C. M. O'D., Davidson J., Riebe M. E. I., Stroud R. M., and Wang J. H. (2018) High abundances of presolar grains and N-15-rich organic matter in CO3.0 chondrite Dominion Range 08006. *Geochim. Cosmochim. Acta* **226**, 107–131.
Nittler L. R., Stroud R. M., Alexander C. M. O'D., and Howell K. (2020) Presolar grains in primitive ungrouped carbonaceous chondrite Northwest Africa 5958. *Meteor. Planet. Sci.* **55**, 1160–1175.
Nittler L. R., Alexander C. M. O'D., Patzer A., and Verdier-Paoletti M. J. (2021) Presolar stardust in highly pristine CM chondrites Asuka 12169 and Asuka 12236. *Meteor. Planet. Sci.* **56**, 260–276.
Nittler L. R., Barosch J., De Gregorio B. T., Stroud R. M., Yabuta H., Yurimoto H., Nakamura T., Noguchi T., Okazaki R., Naraoka H., Sakamoto K., Tachibana S., Watanabe S., Tsuda Y., and the Hayabusa2 Organic Macromolecular Initial Analysis Team. (2022) Carbonaceous presolar grains in asteroid Ryugu. *Lunar Planet. Sci. LIII.* Lunar Planet. Inst., Houston, #1423(abstr.).
Sanghani M. N., Lajaunie L., Marhas K. K., Rickard W. D. A., Hsiao S. S.-Y., Peeters Z., Shang H., Lee D.-C., Calvino J. J., and Bizzarro M. (2022) Microstructural and chemical investigations of presolar silicates from diverse stellar environments. *Astrophys. Jour.* **925**, 110.
Scott E. R. D. and Krot A. N. (2005) Chondritic meteorites and the high-temperature nebular origins of their components. In *Chondrites and the Protoplanetary Disk*. (eds. A. N. Krot, E. R. D. Scott, and B. Reipurth). ASP Conference Series. p. 15–53.
Sharp C. M. and Wasserburg G. J. (1995) Molecular Equilibria and Condensation Temperatures in Carbon-Rich Gases. *Geochim. Cosmochim. Acta* **59**, 1633–1652.
Singerling S. A., Liu N., Nittler L. R., Alexander C. M. O'D., and Stroud R. M. (2021) TEM analyses of unusual presolar silicon carbide: Insights into the range of circumstellar dust condensation conditions. *Astrophys. Jour.* **913**.
Speck A. K., Barlow M. J., Sylvester R. J., and Hofmeister A. M. (2000) Dust features in the 10-µm infrared spectra of oxygen-rich evolved stars. *Astron. Astrophys. Sup* **146**, 437–464.
Stroud R. M. (2003) Focused ion beam microscopy of extraterrestrial materials: Advances and limitations. In *Workshop on Cometary Dust in Astrophysics* (eds. D. E. Brownlee, L. P. Keller, andS. R. Messenger). Lunar Planet. Inst., p. A6011.
Stroud R. M., Nittler L. R., Alexander C. M. O'D., Bernatowicz T. J., and Messenger S. R. (2003) Transmission electron microscopy of non-etched presolar silicon carbide. *Lunar Planet. Sci. XXXIV.* Lunar Planet. Inst., Houston, #1852(abstr.).
Stroud R. M., Nittler L. R., and Alexander C. M. O'D. (2004) Polymorphism in presolar Al2O3 grains from asymptotic giant branch stars. *Science* **305**, 1455–1457.

Stroud R. M. (2005) Microstructural investigations of the cosmochemical histories of presolar grains. In *Chondrites and the Protoplanetary Disk* (eds. A. N. Krot, E. R. D. Scott, and B. Reipurth). ASP Conference Series. pp. 645–656.

Stroud R. M., Nittler L. R., Alexander C. M. O'D., and Zinner E. (2007) Transmission electron microscopy and secondary ion mass spectrometry of an unusual Mg-rich presolar Al2O3 grain. *Lunar Planet. Sci. XXXVIII.* Lunar Planet. Inst., Houston, #2203(abstr.).

Stroud R. M., Floss C., and Stadermann F. J. (2009) Structure, elemental composition and isotopic composition of presolar silicates in MET 00426. *Lunar Planet. Sci. XL.* Lunar Planet. Inst., Houston, #1063(abstr.).

Sun K., Wang L. M., Ewing R. C., and Weber W. J. (2004) Electron irradiation induced phase separation in a sodium borosilicate glass. *NIMPA* **218**, 368–374.

Takigawa A., Tachibana S., Huss G. R., Nagashima K., Makide K., Krot A. N., and Nagahara H. (2014) Morphology and crystal structures of solar and presolar Al2O3 in unequilibrated ordinary chondrites. *Geochim. Cosmochim. Acta* **124**, 309–327.

Takigawa A., Stroud R. M., Nittler L. R., Alexander C. M. O'D., and Miyake A. (2018) High-temperature dust condensation around an AGB star: Evidence from a highly pristine presolar corundum. *Astrophys. Jour. Lett.* **862**.

Tielens A. G. G. M., Mckee C. F., Seab C. G., and Hollenbach D. J. (1994) The physics of grain-grain collisions and gas-grain sputtering in interstellar shocks. *Astrophys. Jour.* **431**, 321–340.

van Dishoeck E. F. (2004) ISO spectroscopy of gas and dust: From molecular clouds to protoplanetary disks. *Ann. Rev. Astron. Astrophys.* **42**, 119–167.

Verhoelst T., Van der Zypen N., Hony S., Decin L., Cami J., and Eriksson K. (2009) The dust condensation sequence in red supergiant stars. *Astron. Astrophys.* **498**, 127–138.

Vollmer C., Hoppe P., Brenker F. E., and Holzapfel C. (2007) Stellar MgSiO3 perovskite: A shock-transformed stardust silicate found in a meteorite. *Astrophys. Jour.* **666**, L49–L52.

Vollmer C., Hoppe P., and Brenker F. E. (2008) Si isotopic compositions of presolar silicate grains from red giant stars and supernovae. *Astrophys. Jour.* **684**, 611.

Vollmer C., Brenker F. E., Hoppe P., and Stroud R. M. (2009a) Direct laboratory analysis of silicate stardust from red giant stars. *Astrophys. Jour.* **700**, 774–782.

Vollmer C., Hoppe P., Stadermann F. J., Floss C., and Brenker F. E. (2009b) NanoSIMS analysis and Auger electron spectroscopy of silicate and oxide stardust from the carbonaceous chondrite Acfer 094. *Geochim. Cosmochim. Acta* **73**, 7127–7149.

Vollmer C., Hoppe P., and Brenker F. E. (2013) Transmission electron microscopy of Al-rich silicate stardust from aymptotic giant branch stars. *Astrophys. Jour.* **769**.

Wang L. M., Wang S. X., Ewing R. C., Meldrum A., Birtcher R. C., Provencio P. N., Weber W. J., and Matzke H. (2000) Irradiation-induced nanostructures. *Mat. Sci. Eng. a-Struct.* **286**, 72–80.

Warren P. H. (2011) Stable-isotopic anomalies and the accretionary assemblage of the Earth and Mars: A subordinate role for carbonaceous chondrites. *Earth Planet. Sci. Lett.* **311**, 93–100.

Watson D. M., Leisenring J. M., Furlan E., Bohac C. J., Sargent B., Forrest W. J., Calvet N., Hartmann L., Nordhaus J. T., and Green J. D. (2008) Crystalline silicates and dust processing in the protoplanetary disks of the Taurus young cluster. *Astrophy. Jour. Supp. Series* **180**, 84.

Weber I., Stojic A. N., Morlok A., Reitze M. P., Markus K., Hiesinger H., Pavlov S. G., Wirth R., Schreiber A., Sohn M., Hubers H. W., and Helbert J. (2020) Space weathering by simulated micrometeorite bombardment on natural olivine and pyroxene: A coordinated IR and TEM study. *Earth Planet. Sci. Lett.* **530**.

Yoneda S. and Grossman L. (1995) Condensation of CaO-MgO-Al2O3-SiO2 liquids from cosmic gases. *Geochim. Cosmochim. Acta* **59**, 3413–3444.

Zega T. J., Alexander C. M. O'D., Nittler L. R., and Stroud R. M. (2011) A transmission electron microscopy study of presolar hibonite. *Astrophys. Jour.* **730**.

Zega T. J. and Floss C. (2013) Extraction and analysis of a presolar oxide grain from the Adelaide ungrouped C2 chondrite. *Lunar Planet. Sci. XLIV.* Lunar Planet. Inst., Houston, #1287(abstr.).

Zega T. J., Nittler L. R., Gyngard F., Alexander C. M. O'D., Stroud R. M., and Zinner E. K. (2014) A transmission electron microscopy study of presolar spinel. *Geochim. Cosmochim. Acta* **124**, 152–169.

Zega T. J., Haenecour P., Floss C., and Stroud R. M. (2015) Circumstellar magnetite from the LAP 031117 CO3.0 Chondrite. *Astrophys. Jour.* **808**.

Ziegler J. F., Ziegler M. D., and Biersack J. P. (2010) SRIM - The stopping and range of ions in matter (2010). *Nucl. Instrum. Meth. B.* **268**, 1818–1823.

Zinner E. (2014) Presolar grains. In *Treatise on Geochemistry 2nd Edition* (eds. H. D. Holland and K. K. Turekian). pp. 181–213.

# Supplementary Figures to TEM analyses of in situ presolar grains from unequilibrated ordinary chondrite LL3.0 Semarkona

S. A. Singerling[1], L. R. Nittler[2], J. Barosch[2], E. Dobrică[3], A. J. Brearley[4], and R. M. Stroud[1]
[1]U.S. Naval Research Laboratory, Code 6366, Washington, DC 20375, USA
[2]Carnegie Institution of Washington, Washington, DC 20015, USA
[3]University of Hawai'i at Mānoa, Honolulu, HI, 96822, USA
[4]University of New Mexico, Albuquerque, NM, 87131, USA

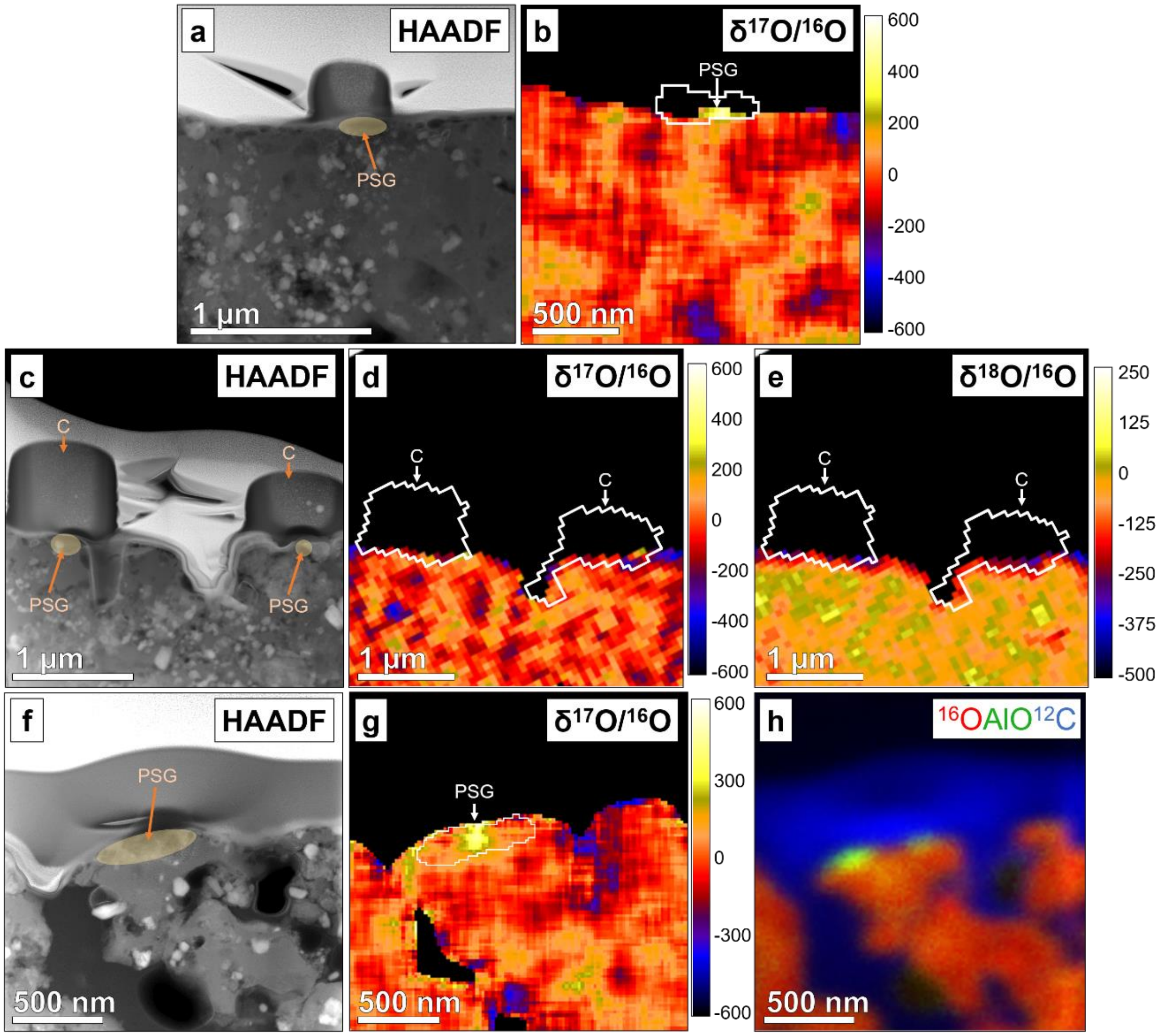

**Figure S1.** STEM HAADF images and isotopic maps showing the results from the NanoSIMS analyses of the FIB sections following TEM work. The suspected locations of the presolar grains are indicated by the yellow ellipses and labeled PSG in the STEM HAADF images. (a–b) present data on F2-9, with the white arrow in (b) indicating a region with anomalous O isotopic ratios. (c–e) present data on F2-30a and F2-30b, with the white outlines in (d) and (e) corresponding to the C fiducial markers. We did not observe isotopic anomalies for either grain; however, overlapping C from the fiducial marker obscured possible anomalous C in SiC F2-30a. Silicate F2-30b may be too small to be resolved by the NanoSIMS beam. (f–h) present data on silicate F1-1, with the white arrow in (g) a region with an anomalous O isotopic ratio.

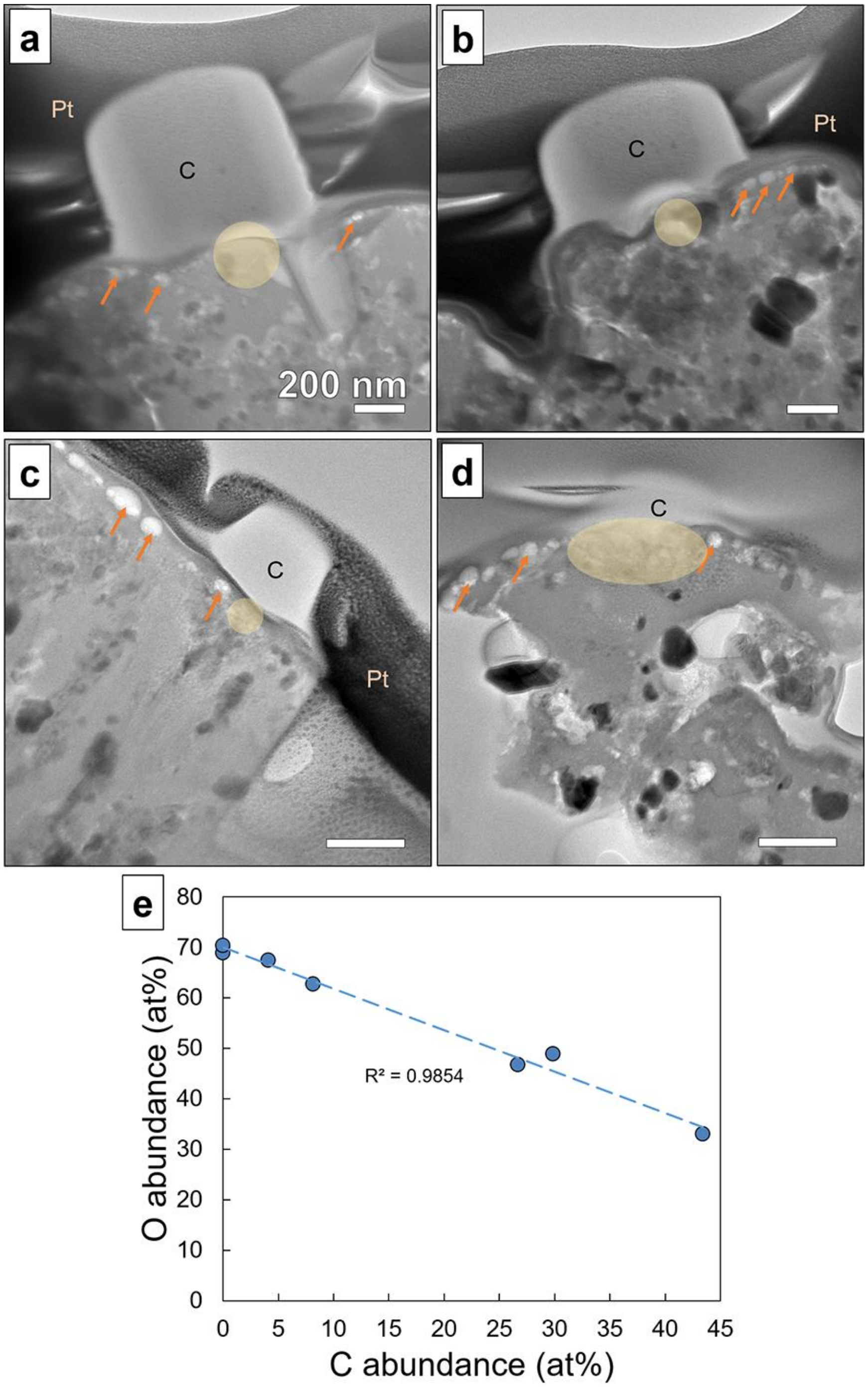

**Figure S2.** Bright field TEM images (a–d) and a plot (c) of vesicles observed in FIB sections with presolar grains: (a) F2-30a, (b) F2-30b, (e) F2-23, and (f) F1-1. Scale bars are 200 nm. The locations of the presolar grains are shown by the yellow ellipses. The plot (e) shows that there is a strong correlation between increasing C abundance and decreasing O abundance. The latter implies that C contamination can cause the use of total O content of the vesicle to be misleading, hence the use of the O+C in the ratio in Figure 2.

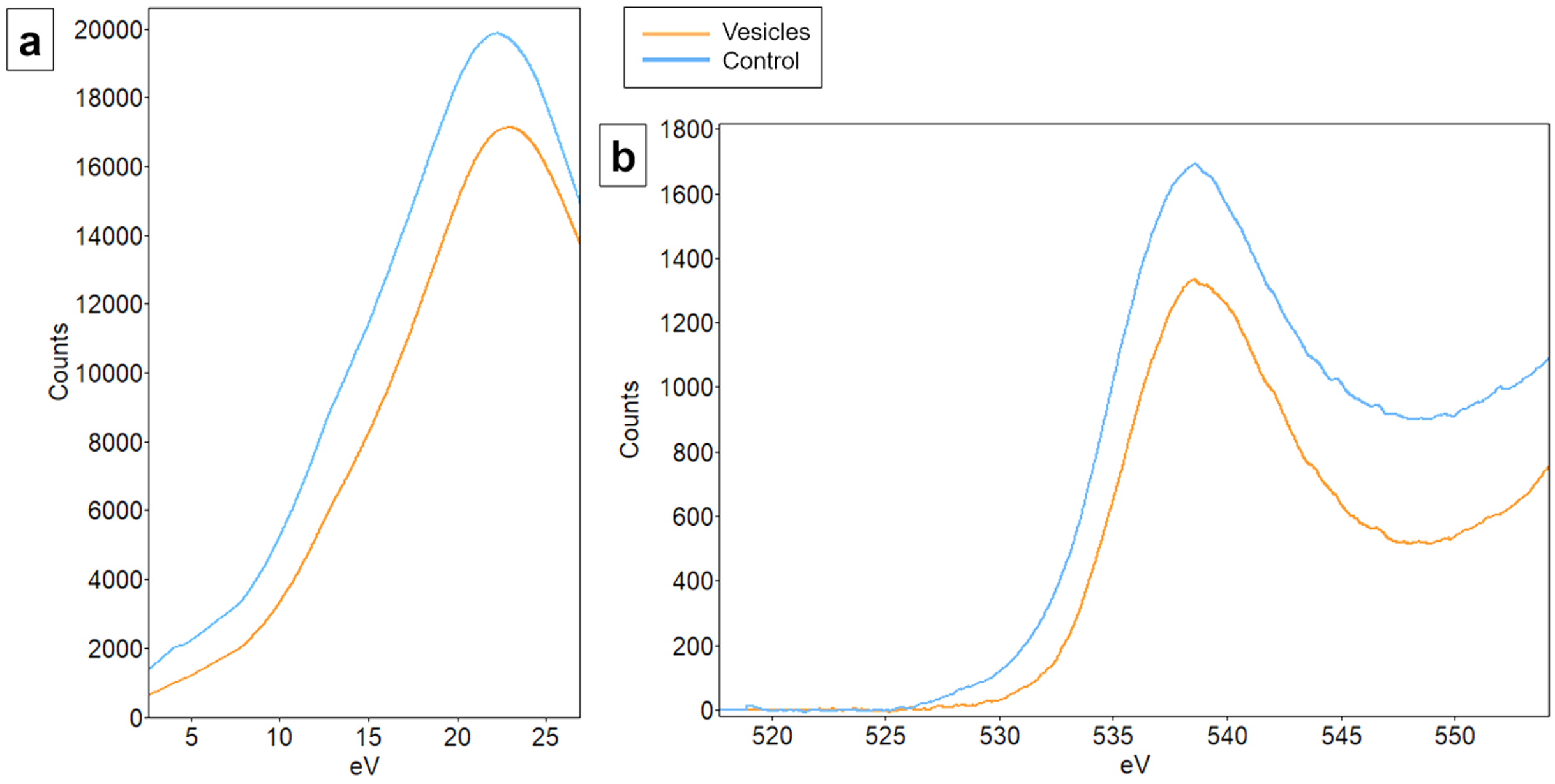

**Figure S3.** EELS spectra of vesicles and adjacent material (control) for the (a) low loss region and (b) O K edge. No significant differences are apparent between the spectra of the vesicles and adjacent material, which does not necessarily preclude implantation of $O_2$ or $H_2O$. Instead, the spectral resolution could be too low to distinguish features and/or O contamination from FIB work could obscure features resulting from NanoSIMS $O^-$ ion beam damage.

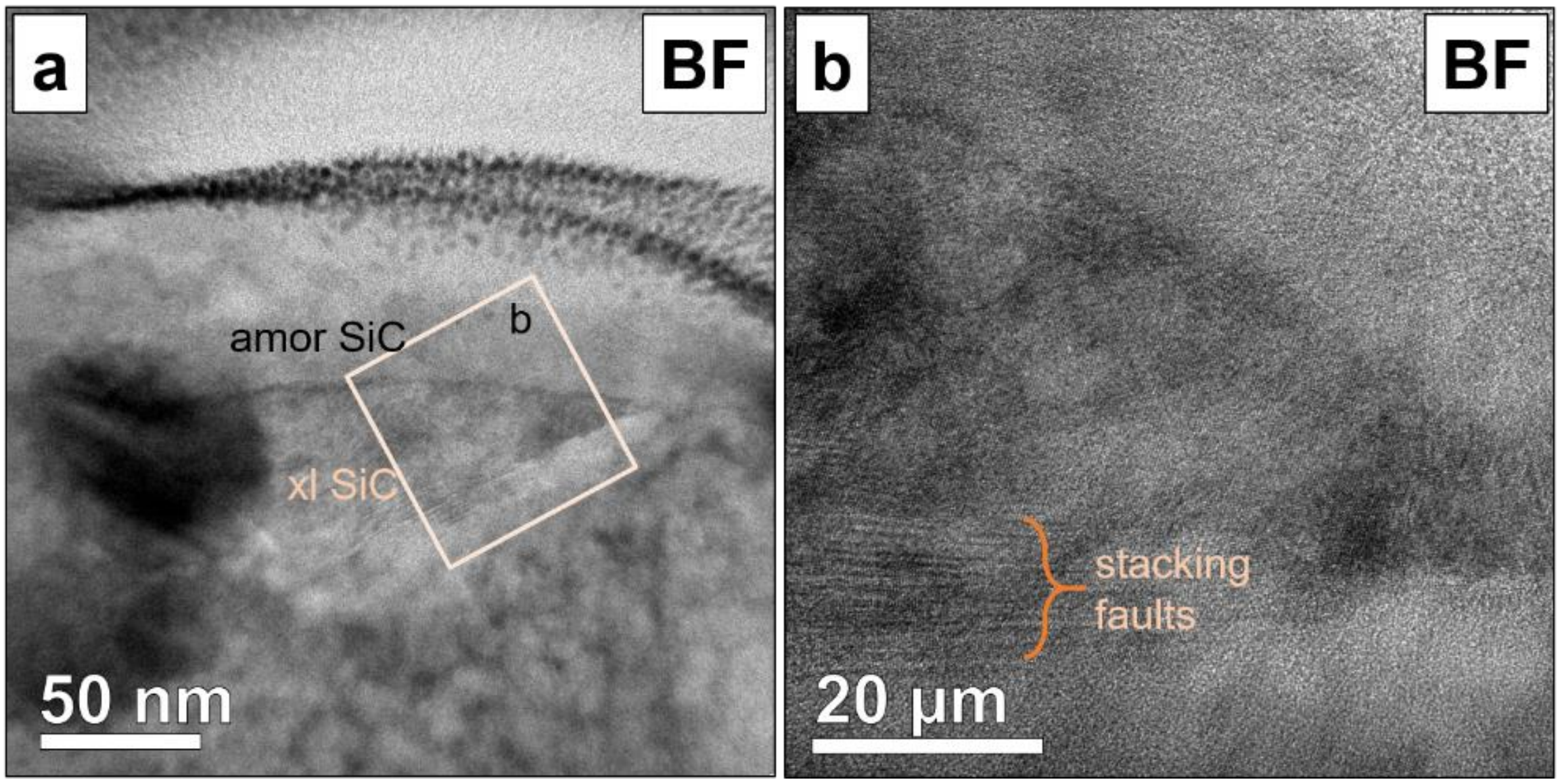

**Figure S4.** Bright field (BF) TEM images of SiC grain F2-30a showing (a) the overall grain and (b) a high-resolution image containing stacking faults.

**Table S1.** HRTEM FFT diffraction indexing

| Sample | Description | Measured d spacing (Å) | Phase | Best Fit Theoretical hkl | d spacing (Å) |
|---|---|---|---|---|---|
| F2-30a | SiC | 2.451 | 3C SiC | -1 -1 -1 | 2.510 |
| | | 2.246 | | -2 0 0 | 2.174 |
| | | 1.444 | | 0 2 2 | 1.537 |
| | | 2.451 | 2H SiC | 0 0 2 | 2.524 |
| | | 2.246 | | 1 1 1 | 2.356 |
| | | 1.444 | | 0 1 3 | 1.423 |
| F2-37 | Mg-Al spinel | 2.767 | Mg-Al spinel | 2 2 0 | 2.858 |
| | | 2.333 | | 2 2 2 | 2.334 |
| | | 1.974 | | 4 0 0 | 2.021 |
| | | 1.462 | | 4 4 0 | 1.429 |
| | | 1.35 | | 5 3 1 | 1.367 |
| F2-9 | NS silicate | 2.312 | | N/A | |
| | | 1.967 | | | |
| | | 1.391 | | | |
| | | 1.164 | | | |
| F2-30b | NS silicate | 2.540 | | N/A | |
| | | 2.032 | | | |
| | | 1.735 | | | |
| | | 1.482 | | | |
| F1-1 | Cr-rich NS silicate | 1.9957 | | N/A | |
| | | 1.517 | | | |
| | | 1.411 | | | |
| | | 1.369 | | | |
| | Al-rich NS silicate | 2.161 | | N/A | |
| | | 2.070 | | | |
| | | 1.484 | | | |
| | | 1.449 | | | |
| | | 1.207 | | | |
| | Si-rich NS silicate | 2.082 | | N/A | |
| | | 1.485 | | | |

NS – non-stoichiometric, N/A – numerous phases match the measured d spacings to within 5% difference, but EDS data do not allow us to determine the most likely candidate phases

**Table S2.** TEM EDS raw data for vesicles, amorphous silicate matrix material, and presolar grains

| | | | C | | O[1] | | Na | | Mg | | Al | | Si | | S | | K | | Ca | | Ti | | Cr | | Fe | | Ni | |
|---|---|---|---|---|---|---|---|---|---|---|---|---|---|---|---|---|---|---|---|---|---|---|---|---|---|---|---|---|
| **Type** | **Sample** | **Description** | **at%** | **err** | **at%** | **err** | **at%** | **err** | **at%** | **err** | **at%** | **err** | **at%** | **err** | **at%** | **err** | **at%** | **err** | **at%** | **err** | **at%** | **err** | **at%** | **err** | **at%** | **err** | **at%** | **err** |
| Vesicles | F2-37 | Vesicle 1 | 29.88 | 0.99 | 48.86 | 1.49 | nd | nd | 3.29 | 0.31 | 0.36 | 0.12 | 7.23 | 0.46 | nd | nd | nd | nd | nd | nd | nd | nd | nd | nd | 10.38 | 1.27 | nd | nd |
| | | Vesicle 2 | 4.07 | 0.27 | 67.45 | 1.65 | nd | nd | 4.29 | 0.30 | 0.39 | 0.10 | 9.04 | 0.41 | nd | nd | nd | nd | nd | nd | nd | nd | nd | nd | 14.76 | 1.38 | nd | nd |
| | | Vesicle 3 | 26.69 | 0.69 | 46.71 | 1.20 | 0.77 | 0.11 | 2.74 | 0.21 | 0.54 | 0.10 | 8.90 | 0.34 | 0.36 | 0.10 | nd | nd | nd | nd | nd | nd | nd | nd | 13.30 | 1.25 | nd | nd |
| | | Controls vesicles 1–3 | nd | nd | 67.91 | 1.47 | 0.65 | 0.08 | 4.41 | 0.23 | 0.48 | 0.08 | 9.65 | 0.28 | 0.29 | 0.07 | nd | nd | nd | nd | 0.21 | 0.08 | nd | nd | 16.40 | 1.29 | nd | nd |
| | | Vesicle 4 | 43.39 | 1.37 | 32.98 | 1.20 | nd | nd | 2.77 | 0.30 | 11.43 | 0.79 | 5.91 | 0.44 | nd | nd | nd | nd | nd | nd | nd | nd | nd | nd | 3.53 | 0.67 | nd | nd |
| | | Control vesicle 4 | nd | nd | 64.23 | 1.55 | nd | nd | 7.30 | 0.33 | 18.39 | 0.79 | 5.68 | 0.19 | 0.21 | 0.06 | 0.09 | 0.05 | 0.11 | 0.05 | nd | nd | nd | nd | 4.00 | 0.42 | nd | nd |
| | F2-23 | Vesicle 1 | nd | nd | 68.91 | 3.62 | 3.30 | 0.80 | 4.81 | 1.02 | nd | nd | 11.16 | 1.71 | nd | nd | nd | nd | nd | nd | nd | nd | nd | nd | 11.82 | 3.12 | nd | nd |
| | | Control vesicle 1 | nd | nd | 71.69 | 3.60 | 1.48 | 0.52 | 3.97 | 0.89 | nd | nd | 14.32 | 1.85 | nd | nd | nd | nd | nd | nd | nd | nd | nd | nd | 8.54 | 2.53 | nd | nd |
| | | Vesicle 2 | nd | nd | 70.33 | 2.03 | 1.88 | 0.27 | 4.52 | 0.44 | 0.90 | 0.22 | 7.61 | 0.60 | 1.26 | 0.28 | nd | nd | nd | nd | nd | nd | nd | nd | 13.50 | 1.66 | nd | nd |
| | | Control vesicle 2 | nd | nd | 66.97 | 1.70 | 2.12 | 0.21 | 5.16 | 0.36 | 0.92 | 0.16 | 11.54 | 0.51 | 1.14 | 0.20 | nd | nd | nd | nd | nd | nd | nd | nd | 12.14 | 1.26 | nd | nd |
| | F1-1 | Vesicle 1 | 8.16 | 0.68 | 62.68 | 2.13 | nd | nd | 4.66 | 0.53 | 2.02 | 0.38 | 12.36 | 0.90 | nd | nd | nd | nd | 4.80 | 0.79 | nd | nd | nd | nd | 5.32 | 1.10 | nd | nd |
| | | Control vesicle 1 | nd | nd | 67.68 | 1.91 | nd | nd | 4.50 | 0.40 | 2.17 | 0.29 | 16.59 | 0.76 | nd | nd | nd | nd | 2.42 | 0.41 | nd | nd | nd | nd | 6.65 | 0.97 | nd | nd |
| Matrix | F2-30a | Amorphous silicate | nd | nd | 56.09 | 1.13 | 1.08 | 0.08 | 3.44 | 0.16 | 0.77 | 0.08 | 20.69 | 0.29 | 0.96 | 0.10 | 0.28 | 0.07 | 0.35 | 0.07 | nd | nd | 0.11 | 0.06 | 16.15 | 1.15 | 0.09 | 0.06 |
| | F2-37 | Amorphous silicate | nd | nd | 69.38 | 1.60 | 1.06 | 0.09 | 3.98 | 0.20 | 0.68 | 0.08 | 15.13 | 0.27 | 0.56 | 0.08 | 0.36 | 0.08 | 0.22 | 0.07 | nd | nd | nd | nd | 8.64 | 0.76 | nd | nd |
| | F2-9 | Amorphous silicate | nd | nd | 63.23 | 1.39 | 1.30 | 0.09 | 5.72 | 0.24 | 3.03 | 0.16 | 15.56 | 0.22 | 1.26 | 0.11 | 0.45 | 0.07 | 0.24 | 0.06 | 0.11 | 0.05 | 0.13 | 0.05 | 8.72 | 0.71 | 0.16 | 0.06 |
| | F2-30b | Amorphous silicate | nd | nd | 61.05 | 1.29 | 1.05 | 0.09 | 3.37 | 0.18 | 0.74 | 0.09 | 18.01 | 0.33 | 0.56 | 0.09 | 0.19 | 0.06 | 0.36 | 0.08 | nd | nd | nd | nd | 14.67 | 1.12 | nd | nd |
| | F2-23 | Amorphous silicate | nd | nd | 66.01 | 1.49 | 2.45 | 0.13 | 4.68 | 0.21 | 0.84 | 0.08 | 16.42 | 0.24 | 0.53 | 0.07 | 0.21 | 0.06 | 0.18 | 0.06 | 0.10 | 0.05 | nd | nd | 8.57 | 0.72 | nd | nd |
| | F1-1 | Amorphous silicate | nd | nd | 66.70 | 1.48 | 0.88 | 0.08 | 3.66 | 0.18 | 0.61 | 0.07 | 16.91 | 0.25 | 0.51 | 0.07 | 0.74 | 0.10 | 0.26 | 0.06 | 0.15 | 0.06 | nd | nd | 9.58 | 0.79 | nd | nd |
| Presolar grains | F2-30a | SiC | 35.74 | 0.74 | 27.23 | 0.73 | nd | nd | 0.09 | 0.04 | 0.31 | 0.05 | 35.92 | 0.26 | 0.22 | 0.05 | nd | nd | nd | nd | nd | nd | nd | nd | 0.43 | 0.08 | 0.05 | 0.04 |
| | | Al-rich SiC | 37.48 | 0.90 | 22.13 | 0.68 | nd | nd | nd | nd | 1.53 | 0.14 | 37.66 | 0.53 | 0.21 | 0.07 | nd | nd | nd | nd | nd | nd | nd | nd | 0.53 | 0.14 | 0.09 | 0.07 |
| | | Amorphous SiC | 32.52 | 1.16 | 30.82 | 1.13 | nd | nd | nd | nd | 0.92 | 0.18 | 31.96 | 0.93 | 1.00 | 0.20 | nd | nd | nd | nd | nd | nd | nd | nd | 2.77 | 0.53 | nd | nd |
| | | Crystalline SiC | 39.18 | 1.22 | 15.77 | 0.70 | nd | nd | nd | nd | 0.87 | 0.16 | 40.39 | 0.93 | 1.10 | 0.19 | nd | nd | nd | nd | nd | nd | nd | nd | 2.70 | 0.48 | nd | nd |
| | F2-37 | Mg-Al spinel | nd | nd | 64.19 | 1.51 | 0.20 | 0.04 | 8.38 | 0.33 | 15.62 | 0.64 | 6.54 | 0.13 | 0.22 | 0.05 | nd | nd | 0.13 | 0.05 | 0.07 | 0.04 | nd | nd | 4.65 | 0.43 | nd | nd |
| | | Al-rich spinel | nd | nd | 64.91 | 1.65 | nd | nd | 6.60 | 0.35 | 20.15 | 0.92 | 4.91 | 0.24 | 0.23 | 0.08 | nd | nd | nd | nd | nd | nd | nd | nd | 3.21 | 0.42 | nd | nd |
| | | Mg-rich spinel | nd | nd | 62.09 | 1.68 | nd | nd | 13.44 | 0.70 | 10.49 | 0.64 | 6.60 | 0.42 | 0.22 | 0.10 | nd | nd | nd | nd | nd | nd | nd | nd | 7.17 | 0.92 | nd | nd |
| | F2-9 | Si-rich silicate | nd | nd | 61.57 | 1.71 | 0.35 | 0.10 | 9.68 | 0.55 | 0.38 | 0.11 | 27.25 | 0.76 | 0.18 | 0.09 | nd | nd | nd | nd | 0.17 | 0.05 | nd | nd | 0.58 | 0.22 | nd | nd |
| | | Mg-rich silicate | nd | nd | 59.71 | 1.87 | 0.77 | 0.17 | 23.93 | 1.23 | 0.40 | 0.14 | 14.41 | 0.75 | 0.38 | 0.15 | nd | nd | nd | nd | nd | nd | nd | nd | 0.39 | 0.23 | nd | nd |
| | | Mg-rich rim | nd | nd | 56.41 | 1.21 | 0.80 | 0.07 | 16.85 | 0.58 | 2.81 | 0.15 | 11.36 | 0.19 | 0.65 | 0.07 | 0.28 | 0.06 | 0.39 | 0.07 | nd | nd | 0.34 | 0.08 | 9.80 | 0.77 | 0.30 | 0.08 |
| | F2-30b | Silicate | 2.86 | 0.32 | 54.10 | 1.50 | 1.71 | 0.24 | 3.57 | 0.36 | nd | nd | 17.07 | 0.81 | 0.77 | 0.21 | nd | nd | nd | nd | nd | nd | nd | nd | 19.92 | 1.95 | nd | nd |
| | F2-23 | Silicate | nd | nd | 65.29 | 1.41 | 1.82 | 0.12 | 4.86 | 0.22 | 0.95 | 0.10 | 12.63 | 0.26 | 0.76 | 0.09 | 0.16 | 0.06 | 0.18 | 0.06 | nd | nd | 0.13 | 0.06 | 13.12 | 1.04 | 0.11 | 0.07 |
| | F1-1 | Al-rich silicate | 1.31 | 0.18 | 62.03 | 1.72 | nd | nd | 5.97 | 0.40 | 4.79 | 0.39 | 22.26 | 0.70 | nd | nd | nd | nd | 2.53 | 0.36 | nd | nd | nd | nd | 1.10 | 0.30 | nd | nd |
| | | Cr-rich silicate | 14.19 | 0.60 | 53.70 | 1.52 | 0.37 | 0.10 | 5.58 | 0.40 | 2.44 | 0.28 | 12.07 | 0.56 | 0.43 | 0.14 | nd | nd | 5.25 | 0.57 | nd | nd | 1.74 | 0.37 | 4.24 | 0.66 | nd | nd |
| | | Mg-rich silicate | nd | nd | 61.64 | 0.08 | 0.16 | 1.47 | 13.44 | 0.05 | 0.53 | 0.09 | 15.61 | 0.34 | 0.31 | 0.34 | nd | nd | 3.86 | 0.47 | nd | nd | nd | nd | 4.45 | 0.55 | nd | nd |
| | | Si-rich silicate | 0.02 | 0.02 | 61.49 | 1.83 | nd | nd | 3.99 | 0.38 | 1.30 | 0.23 | 30.25 | 1.03 | nd | nd | nd | nd | 2.45 | 0.43 | nd | nd | nd | nd | 0.50 | 0.25 | nd | nd |

at% – atomic percent element, err – error, control – adjacent material with which to compare the vesicle to, nd – not detected

Excludes elements for which all analyses were less than 0.1 at.%

[1]Oxygen excesses were from overlapping meteoritic materials, O implantation by the $O^-$ beam used for NanoSIMS analsyes, and/or oxidation of surfaces irradiated by the $Ga^+$ beam used for FIB section preparations.

**Table S3.** Literature data (converted to atomic %) for amorphous silicates in various meteorite groups

| Source[1] | Meteorite Group | O | Na | Mg | Al | Si | P | S | K | Ca | Ti | Cr | Mn | Fe | Ni |
|---|---|---|---|---|---|---|---|---|---|---|---|---|---|---|---|
| Brearley 1993 | CO | 60.04 | 0.00 | 5.89 | 2.72 | 16.30 | 0.30 | 1.91 | 0.00 | 0.20 | 0.00 | 0.13 | 0.16 | 9.51 | 2.86 |
| Brearley 1993 | CO | 61.97 | 0.00 | 3.57 | 3.18 | 17.35 | 2.11 | 1.78 | 0.00 | 0.50 | 0.00 | 0.11 | 0.02 | 6.96 | 2.44 |
| Brearley 1993 | CO | 60.90 | 0.00 | 5.48 | 2.34 | 19.14 | 0.00 | 1.35 | 0.00 | 0.06 | 0.00 | 0.27 | 0.20 | 8.30 | 1.96 |
| Brearley 1993 | CO | 62.38 | 0.00 | 4.03 | 2.57 | 19.88 | 1.29 | 1.60 | 0.00 | 0.21 | 0.00 | 0.13 | 0.09 | 6.54 | 1.27 |
| Brearley 1993 | CO | 62.04 | 0.00 | 5.87 | 1.67 | 19.46 | 0.00 | 3.73 | 0.00 | 0.00 | 0.00 | 0.13 | 0.05 | 6.71 | 0.34 |
| Brearley 1993 | CO | 63.14 | 0.00 | 3.49 | 3.25 | 21.42 | 1.42 | 1.07 | 0.00 | 0.08 | 0.00 | 0.10 | 0.02 | 4.94 | 1.07 |
| Brearley 1993 | CO | 62.35 | 0.00 | 4.73 | 2.38 | 22.49 | 0.00 | 0.99 | 0.00 | 0.09 | 0.00 | 0.07 | 0.04 | 5.69 | 1.16 |
| Brearley 1993 | CO | 63.73 | 0.00 | 5.10 | 0.97 | 25.24 | 0.80 | 0.48 | 0.00 | 0.06 | 0.00 | 0.12 | 0.04 | 2.73 | 0.73 |
| Chizmadia 2004 | CM | 58.62 | 0.00 | 9.64 | 1.90 | 14.87 | 0.27 | 0.93 | 0.06 | 0.13 | 0.03 | 0.17 | 0.00 | 13.35 | 0.04 |
| Chizmadia 2004 | CM | 59.62 | 0.00 | 9.33 | 0.87 | 14.55 | 0.21 | 3.79 | 0.01 | 0.16 | 0.01 | 0.32 | 0.09 | 10.41 | 0.63 |
| Chizmadia 2004 | CM | 58.94 | 0.00 | 9.93 | 1.54 | 14.82 | 0.14 | 2.03 | 0.09 | 0.19 | 0.00 | 0.17 | 0.06 | 11.43 | 0.66 |
| Chizmadia 2004 | CM | 60.26 | 0.00 | 9.89 | 0.61 | 16.91 | 0.21 | 2.73 | 0.06 | 0.09 | 0.12 | 0.36 | 0.11 | 8.21 | 0.45 |
| Chizmadia 2004 | CM | 59.66 | 0.00 | 7.24 | 0.40 | 9.57 | 1.50 | 7.11 | 0.02 | 0.06 | 0.05 | 0.34 | 0.00 | 13.00 | 1.05 |
| Chizmadia 2004 | CM | 58.75 | 0.00 | 9.76 | 1.44 | 14.57 | 0.30 | 1.68 | 0.02 | 0.06 | 0.03 | 0.13 | 0.00 | 12.52 | 0.73 |
| Chizmadia & Brearley 2008 | CM | 58.91 | 0.00 | 10.02 | 2.97 | 14.16 | 0.17 | 1.82 | 0.08 | 0.13 | 0.05 | 0.18 | 0.05 | 11.44 | 0.00 |
| Le Guillou & Brearely 2014 | CR | 59.15 | 0.00 | 9.81 | 1.16 | 15.82 | 0.00 | 1.84 | 0.00 | 0.21 | 0.00 | 0.11 | 0.00 | 11.91 | 0.00 |
| Le Guillou & Brearely 2014 | CR | 59.21 | 0.00 | 9.86 | 1.41 | 15.71 | 0.00 | 1.89 | 0.00 | 0.00 | 0.00 | 0.22 | 0.11 | 11.59 | 0.00 |
| Le Guillou & Brearely 2014 | CR | 59.13 | 0.00 | 10.04 | 1.30 | 17.60 | 0.00 | 0.00 | 0.00 | 0.00 | 0.00 | 0.00 | 0.00 | 11.93 | 0.00 |
| Le Guillou & Brearely 2014 | CR | 59.66 | 0.00 | 7.42 | 1.20 | 16.81 | 0.00 | 1.91 | 0.00 | 0.22 | 0.00 | 0.00 | 0.00 | 12.77 | 0.00 |
| Le Guillou & Brearely 2014 | CR | 58.63 | 0.00 | 12.79 | 1.42 | 14.54 | 0.00 | 1.91 | 0.00 | 0.11 | 0.00 | 0.22 | 0.00 | 10.38 | 0.00 |
| Le Guillou & Brearely 2014 | CR | 59.99 | 0.00 | 6.35 | 1.27 | 17.10 | 0.00 | 2.20 | 0.00 | 0.20 | 0.00 | 0.10 | 0.10 | 12.70 | 0.00 |
| Le Guillou & Brearely 2014 | CR | 60.08 | 0.00 | 7.85 | 1.33 | 18.00 | 0.00 | 1.51 | 0.00 | 0.00 | 0.00 | 0.00 | 0.00 | 11.23 | 0.00 |
| Le Guillou & Brearely 2014 | CR | 60.81 | 0.00 | 5.52 | 1.59 | 19.65 | 0.00 | 1.06 | 0.00 | 0.00 | 0.00 | 0.21 | 0.00 | 11.15 | 0.00 |
| Le Guillou & Brearely 2014 | CR | 59.10 | 0.00 | 15.56 | 0.74 | 17.25 | 0.00 | 0.53 | 0.00 | 0.11 | 0.00 | 0.11 | 0.00 | 6.62 | 0.00 |
| Le Guillou & Brearely 2014 | CR | 59.34 | 0.00 | 7.53 | 2.58 | 16.82 | 0.00 | 0.52 | 0.00 | 0.21 | 0.00 | 0.10 | 0.10 | 12.80 | 0.00 |
| Le Guillou & Brearely 2014 | CR | 60.09 | 0.00 | 11.56 | 1.45 | 18.90 | 0.00 | 0.56 | 0.00 | 0.00 | 0.00 | 0.00 | 0.00 | 7.45 | 0.00 |
| Le Guillou & Brearely 2014 | CR | 60.24 | 0.00 | 11.17 | 1.64 | 19.61 | 0.00 | 0.00 | 0.00 | 0.00 | 0.00 | 0.11 | 0.00 | 7.23 | 0.00 |
| Le Guillou & Brearely 2014 | CR | 59.58 | 0.00 | 7.37 | 2.32 | 17.37 | 0.00 | 0.53 | 0.00 | 0.21 | 0.00 | 0.21 | 0.00 | 12.42 | 0.00 |
| Le Guillou & Brearely 2014 | CR | 59.46 | 0.00 | 13.29 | 1.58 | 18.13 | 0.00 | 0.00 | 0.00 | 0.00 | 0.00 | 0.00 | 0.00 | 7.55 | 0.00 |
| Le Guillou & Brearely 2014 | CR | 59.37 | 0.00 | 12.76 | 1.68 | 17.91 | 0.00 | 0.00 | 0.00 | 0.00 | 0.00 | 0.00 | 0.00 | 8.28 | 0.00 |
| Le Guillou & Brearely 2014 | CR | 58.40 | 0.00 | 11.09 | 2.74 | 15.37 | 0.00 | 0.00 | 0.00 | 0.00 | 0.00 | 0.11 | 0.00 | 12.29 | 0.00 |
| Le Guillou & Brearely 2014 | CR | 59.80 | 0.00 | 10.41 | 0.95 | 18.11 | 0.00 | 1.01 | 0.00 | 0.00 | 0.00 | 0.00 | 0.00 | 9.73 | 0.00 |
| Le Guillou & Brearely 2014 | CR | 58.94 | 0.00 | 13.06 | 0.73 | 16.60 | 0.00 | 0.91 | 0.00 | 0.24 | 0.00 | 0.00 | 0.12 | 9.40 | 0.00 |
| Le Guillou & Brearely 2014 | CR | 58.72 | 0.00 | 12.44 | 0.85 | 16.10 | 0.00 | 0.91 | 0.00 | 0.37 | 0.00 | 0.00 | 0.00 | 10.61 | 0.00 |
| Gresake 1997 | Ungrouped C3 | 59.03 | 0.00 | 9.57 | 2.00 | 16.78 | 0.00 | 0.37 | 0.40 | 0.59 | 0.00 | 0.22 | 0.00 | 10.41 | 0.63 |
| Gresake 1997 | Ungrouped C3 | 60.55 | 0.00 | 5.36 | 2.37 | 17.88 | 0.00 | 1.90 | 0.00 | 0.33 | 0.00 | 0.27 | 0.29 | 8.40 | 2.64 |
| Gresake 1997 | Ungrouped C3 | 61.02 | 0.00 | 4.97 | 1.74 | 18.79 | 0.00 | 2.31 | 0.00 | 0.61 | 0.00 | 0.15 | 0.00 | 8.40 | 2.01 |
| Gresake 1997 | Ungrouped C3 | 61.20 | 0.00 | 4.47 | 1.40 | 20.80 | 0.00 | 0.99 | 0.44 | 0.54 | 0.00 | 0.27 | 0.26 | 8.33 | 1.30 |
| Gresake 1997 | Ungrouped C3 | 60.05 | 0.00 | 5.34 | 1.90 | 17.84 | 0.00 | 1.48 | 0.35 | 0.76 | 0.00 | 0.00 | 0.23 | 10.21 | 1.84 |
| Gresake 1997 | Ungrouped C3 | 60.96 | 0.00 | 6.35 | 1.24 | 19.44 | 0.00 | 1.76 | 0.00 | 0.77 | 0.00 | 0.21 | 0.00 | 7.18 | 2.09 |
| Dobrica & Brearley 2020 | Semarkona | 60.56 | 1.94 | 6.49 | 3.68 | 19.57 | 0.00 | 0.79 | 0.28 | 0.11 | 0.00 | 0.08 | 0.03 | 6.46 | 0.01 |
| Dobrica & Brearley 2020 | Semarkona | 60.26 | 2.44 | 6.48 | 3.73 | 19.44 | 0.00 | 0.65 | 0.44 | 0.04 | 0.00 | 0.00 | 0.02 | 6.51 | 0.00 |
| Dobrica & Brearley 2020 | Semarkona | 60.89 | 0.60 | 7.37 | 3.51 | 19.29 | 0.00 | 1.17 | 0.29 | 0.08 | 0.00 | 0.00 | 0.04 | 6.77 | 0.00 |
| Dobrica & Brearley 2020 | Semarkona | 61.12 | 0.09 | 7.17 | 3.84 | 19.60 | 0.00 | 0.88 | 0.25 | 0.13 | 0.00 | 0.05 | 0.04 | 6.79 | 0.06 |
| Dobrica & Brearley 2020 | Semarkona | 60.85 | 0.21 | 7.49 | 0.27 | 21.19 | 0.00 | 0.60 | 0.25 | 0.20 | 0.00 | 0.03 | 0.07 | 8.76 | 0.07 |
| Dobrica & Brearley 2020 | Semarkona | 60.85 | 0.21 | 7.41 | 0.27 | 21.18 | 0.00 | 0.60 | 0.25 | 0.20 | 0.00 | 0.03 | 0.07 | 8.85 | 0.07 |
| Dobrica & Brearley 2020 | Semarkona | 59.85 | 2.41 | 3.30 | 0.29 | 20.45 | 0.00 | 0.51 | 0.39 | 0.20 | 0.00 | 0.00 | 0.05 | 12.53 | 0.00 |
| Dobrica & Brearley 2020 | Semarkona | 60.56 | 0.52 | 7.44 | 0.56 | 20.76 | 0.00 | 0.45 | 0.20 | 0.27 | 0.00 | 0.00 | 0.03 | 9.18 | 0.04 |
| Dobrica & Brearley 2020 | Semarkona | 60.56 | 0.52 | 7.35 | 0.55 | 20.75 | 0.00 | 0.45 | 0.20 | 0.28 | 0.00 | 0.00 | 0.03 | 9.27 | 0.04 |
| Dobrica & Brearley 2020 | Semarkona | 60.45 | 0.00 | 13.06 | 0.17 | 20.51 | 0.00 | 0.29 | 0.03 | 0.13 | 0.00 | 0.06 | 0.05 | 5.24 | 0.00 |
| Dobrica & Brearley 2020 | Semarkona | 61.07 | 0.31 | 4.49 | 0.61 | 21.19 | 0.00 | 0.84 | 0.15 | 0.23 | 0.00 | 0.07 | 0.03 | 10.98 | 0.03 |
| Dobrica & Brearley 2020 | Semarkona | 60.88 | 1.61 | 2.84 | 0.34 | 21.89 | 0.00 | 0.63 | 0.29 | 0.18 | 0.00 | 0.02 | 0.08 | 11.23 | 0.02 |
| Dobrica & Brearley 2020 | Semarkona | 61.09 | 0.27 | 4.98 | 0.67 | 21.17 | 0.00 | 0.84 | 0.09 | 0.28 | 0.00 | 0.03 | 0.02 | 10.56 | 0.01 |
| Dobrica & Brearley 2020 | Semarkona | 60.92 | 0.27 | 5.30 | 0.68 | 20.85 | 0.00 | 0.86 | 0.14 | 0.30 | 0.00 | 0.01 | 0.02 | 10.62 | 0.01 |
| Dobrica & Brearley 2020 | Semarkona | 60.60 | 1.34 | 3.99 | 0.48 | 21.03 | 0.00 | 0.67 | 0.15 | 0.30 | 0.00 | 0.01 | 0.05 | 11.38 | 0.00 |
| Dobrica & Brearley 2020 | Semarkona | 60.51 | 1.79 | 4.09 | 0.28 | 21.31 | 0.00 | 0.57 | 0.22 | 0.30 | 0.00 | 0.00 | 0.08 | 10.86 | 0.00 |
| Dobrica & Brearley 2020 | Semarkona | 60.49 | 1.22 | 4.27 | 0.54 | 20.43 | 0.00 | 1.05 | 0.31 | 0.21 | 0.00 | 0.02 | 0.03 | 11.36 | 0.06 |
| Dobrica & Brearley 2020 | Semarkona | 60.71 | 1.06 | 3.75 | 0.45 | 20.84 | 0.00 | 0.93 | 0.19 | 0.30 | 0.00 | 0.09 | 0.00 | 11.62 | 0.06 |

**Table S3 cont.**

| Source[1] | Meteorite Group | O | Na | Mg | Al | Si | P | S | K | Ca | Ti | Cr | Mn | Fe | Ni |
|---|---|---|---|---|---|---|---|---|---|---|---|---|---|---|---|
| Dobrica & Brearley 2020 | Semarkona | 60.72 | 0.94 | 3.81 | 0.45 | 20.89 | 0.00 | 0.90 | 0.24 | 0.27 | 0.00 | 0.03 | 0.08 | 11.68 | 0.00 |
| Dobrica & Brearley 2020 | Semarkona | 59.86 | 2.34 | 3.25 | 0.28 | 20.43 | 0.00 | 0.52 | 0.39 | 0.20 | 0.00 | 0.00 | 0.05 | 12.67 | 0.00 |
| Dobrica & Brearley 2020 | Semarkona | 60.46 | 0.00 | 12.96 | 0.17 | 20.53 | 0.00 | 0.30 | 0.03 | 0.13 | 0.00 | 0.06 | 0.06 | 5.30 | 0.00 |
| Dobrica & Brearley 2020 | Semarkona | 61.06 | 0.31 | 4.43 | 0.62 | 21.16 | 0.00 | 0.85 | 0.14 | 0.23 | 0.00 | 0.07 | 0.03 | 11.08 | 0.03 |
| Dobrica & Brearley 2020 | Semarkona | 60.87 | 1.57 | 2.80 | 0.33 | 21.87 | 0.00 | 0.63 | 0.29 | 0.18 | 0.00 | 0.02 | 0.08 | 11.34 | 0.02 |
| Dobrica & Brearley 2020 | Semarkona | 59.39 | 4.59 | 5.40 | 0.85 | 19.90 | 0.00 | 0.87 | 0.24 | 0.15 | 0.00 | 0.00 | 0.07 | 8.54 | 0.00 |
| Dobrica & Brearley 2020 | Semarkona | 58.67 | 6.65 | 5.21 | 0.85 | 19.44 | 0.00 | 0.92 | 0.24 | 0.21 | 0.00 | 0.00 | 0.00 | 7.81 | 0.00 |
| Dobrica & Brearley 2020 | Semarkona | 59.92 | 3.39 | 5.75 | 0.84 | 20.25 | 0.00 | 0.98 | 0.23 | 0.21 | 0.00 | 0.00 | 0.08 | 8.35 | 0.00 |
| Dobrica & Brearley 2020 | Semarkona | 59.80 | 3.29 | 6.29 | 0.79 | 20.26 | 0.00 | 0.74 | 0.31 | 0.24 | 0.00 | 0.00 | 0.00 | 8.28 | 0.00 |
| Dobrica & Brearley 2020 | Semarkona | 59.60 | 3.56 | 6.05 | 0.99 | 19.72 | 0.00 | 0.87 | 0.22 | 0.20 | 0.00 | 0.00 | 0.00 | 8.79 | 0.00 |
| Dobrica & Brearley 2020 | Semarkona | 59.33 | 4.70 | 5.48 | 0.76 | 19.89 | 0.00 | 0.82 | 0.17 | 0.22 | 0.00 | 0.00 | 0.00 | 8.62 | 0.00 |
| Dobrica & Brearley 2020 | Semarkona | 60.66 | 0.59 | 7.68 | 0.89 | 20.60 | 0.00 | 0.66 | 0.19 | 0.14 | 0.00 | 0.00 | 0.00 | 8.59 | 0.00 |

[1]Full citations for the sources are in the references section

**Table S4.** Oxide and silicate literature TEM data

| Source[1] | Analysis Technique | In/Ex Situ | Host | Host Details | Grain | Type | Phase(s) | Isotopic Group | Progenitor Star | Maximum Size (nm) | Crystallinity | Compositions | Stoichiometry |
|---|---|---|---|---|---|---|---|---|---|---|---|---|---|
| Bose et al 2012 | FIB TEM | In situ | Meteorite | CO | 21420 | Silicate | Silica | 1 | R/AGB | 140 | Amorphous | N/A | O/Si = 2.2 (from Auger) |
| Busemann et al 2009 | FIB TEM | In situ | IDP | Anhy | E1-B #9 | Silicate | Olivine (Fo) | 1 | R/AGB | 50 | Single xl | N/A | N/A |
| Leitner et al 2012 | FIB TEM | In situ | Meteorite | CR | 6_8 | Oxide | Hibonite | 1 | R/AGB | 1750 | Single xl | Subgrain (perovskite) | Ca/(Mg,Ti,Al) from 0.07 to 0.13 |
| Leitner et al 2020 | FIB TEM | In situ | Meteorite | CM | JW01_SA4_54 | Silicate | Akermanite/diopside | 1 | R/AGB | 735 | Single xl? | N/A | N/A |
| Messenger et al 2003 | Ultramicrotome TEM | In situ | IDP | Anhy | L2005 C13 | Silicate | GEMS | 3 | Low mass, low Z AGB | 400 | N/A | N/A | N/A |
| | Ultramicrotome TEM | In situ | IDP | Anhy | L2036 C4 | Silicate | GEMS | 1 | R/AGB | 630 | N/A | N/A | N/A |
| | Ultramicrotome TEM | In situ | IDP | Anhy | L2005 C13 | Silicate | N/A | 1 | R/AGB | 300 | N/A | N/A | N/A |
| | Ultramicrotome TEM | In situ | IDP | Anhy | L2036 C4 | Silicate | N/A | 1 | R/AGB | 940 | N/A | N/A | N/A |
| | Ultramicrotome TEM | In situ | IDP | Anhy | L2036 C4 | Silicate | N/A | 1 | R/AGB | 300 | N/A | N/A | N/A |
| | Ultramicrotome TEM | In situ | IDP | Anhy | L2005 C13 | Silicate | Olivine (Fo) | 1 | R/AGB | 330 | N/A | N/A | N/A |
| Messenger et al 2005 | Ultramicrotome TEM | In situ | IDP | Anhy | B10A | Silicate | Olivine (Fo) | ung | SNe | 500 | Poly xl | Subgrains | Mg/(Mg+Fe) = 0.83 |
| Nguyen et al 2007 | FIB TEM | In situ | Meteorite | C2 ung | Acfer 094 grain 1 | Silicate | NS | 2 | Low mass or CBP AGB | 500 | Amorphous | No rims or subgrains | Mg/Fe = 1.25 |
| Nguyen et al 2010 | FIB TEM | In situ | Meteorite | CO | AH-65a | Silicate | NS | 1 | R/AGB | 615 | Amorphous (from NanoSIMS?) | N/A | (Ca+Mg+Fe)/Si from 1.23 to 1.70 |
| | FIB TEM | In situ | Meteorite | CO | AH-166a | Silicate | NS | 1 | R/AGB | 305 | Nanocrystalline | Ca- and Al-rich interior, Si- and Mg-rich exterior | (Ca+Mg+Fe)/Si = 0.27 |
| | FIB TEM | In situ | Meteorite | CO | AH-139a | Silicate | NS | 1 | R/AGB | 520 | N/A | Left vs right heterogeneities in Al, Si, Fe, Ni, and O | (Ca+Mg+Fe)/Si from 0.10 to 0.23 |
| Nguyen et al 2014 | Ultramicrotome TEM | In situ | IDP | Anhy | W7027E6 | Composite | Mg-Al spinel-NS | 1? | R/AGB? | 345 | N/A | Sp core with minor Ti and Fe, NS Mg,Si-rich rim | |
| Nguyen et al 2016 | FIB TEM | Ex situ | Meteorite | C ung | 3_13b | Oxide | Fe oxide | 1 | R/AGB | 200 | Amorphous | Fe2+ | Fe 24 at%, O 40 at% |
| | FIB TEM | Ex situ | Meteorite | C ung | 1_10d | Silicate | Al-rich silicate | 1 | R/AGB | 280 | Amorphous | Minor Fe, no Ca or Na | Al/Si = 0.7 |
| | FIB TEM | Ex situ | Meteorite | C ung | 4_2 | Silicate | Al-rich silicate | Extreme 1 | Nova | 200 | Amorphous | Trace Mg; no Ca, Na, K | Al/Si = 0.33 |
| | FIB TEM | Ex situ | Meteorite | C ung | 2_33b | Silicate | GEMS-like | 4 | SNe | 240 | Amorphous | Minor Mg, Ca | Fe/Si = 1.4, Mg/Si = 0.15 |
| | FIB TEM | Ex situ | Meteorite | C ung | 6_16 | Silicate | NS | 4 | SNe | 290 | Amorphous | More Fe at the base | (Mg+Fe)/Si = 1.33 |
| | FIB TEM | Ex situ | Meteorite | C ung | 3_13a | Silicate | NS | 3 | SNe | 155 | Amorphous | Homogeneous | (Ca+Mg+Fe)/Si = 0.85 |
| | FIB TEM | Ex situ | Meteorite | CR | 3_59_9 | Silicate | Olivine (Fo) | 1 | R/AGB | 130 | Single xl (euhedral) | N/A | Fo80 |
| | FIB TEM | Ex situ | Meteorite | C ung | 1_10c | Silicate | Pyroxene (En) | 1 | R/AGB | 145 | Twinned | Trace Cr, Mg, Fe | En99 |
| | FIB TEM | Ex situ | Meteorite | C ung | 2_4 | Silicate | Pyroxene (En) | 4 | SNe | 500 | Amorphous | Minor Fe | En100 |
| | FIB TEM | Ex situ | Meteorite | CR | 3_59_4 | Silicate | Pyroxene (En) | 1 | R/AGB | 360 | Nanocrystalline core, amorphous shell | Both core and shell same | N/A |
| Nittler et al 2018 | FIB TEM | In situ | Meteorite | CO | DOM-77 | Silicate | AOA-like | 1 | Low mass AGB | 349 | Poly xl | Ca- and Al-rich core, olivine rim | Rim: Mg/(Mg+Fe) = 0.93, core: Ca/Al = 0.5 |
| | FIB TEM | In situ | Meteorite | CO | DOM-3 | Silicate | NS | 1 | R/AGB | 532 | Amorphous | N/A | (Mg+Fe+Ca)/Si = 1.5; Fe-rich (Mg# = 0.38) |
| | FIB TEM | In situ | Meteorite | CO | DOM-17 | Silicate | NS | 1 | R/AGB | 343 | Amorphous | N/A | (Mg+Fe)/Si = 1.42; Fe-rich |
| | FIB TEM | In situ | Meteorite | CO | DOM-8 | Silicate | Olivine | 1 | R/AGB | 508 | N/A | N/A | (Mg+Fe)/Si = 1.94; Mg# = 0.83 |
| | FIB TEM | In situ | Meteorite | CO | DOM-18 | Silicate | Olivine | 1 or 2 | R/AGB | 301 | Single xl | N/A | (Mg+Fe)/Si = 1.99; Mg-rich |
| Nittler et al 2020 | FIB TEM | In situ | Meteorite | C ung | A2-2-15 | Oxide | Mg-Al spinel | 1 | R/AGB | 325 | Poly xl | N/A | Mg/Al = 0.51; minor Cr (Mg0.97Cr0.11Al1.91O4) |
| | FIB TEM | In situ | Meteorite | C ung | A2-1-5 | Silicate | Pyroxene (En) | 1 | R/AGB | 275 | Single xl? | Clino-orthopyroxene intergrowths | Stoichiometric (Mg0.961Al0.013Ca0.003Cr0.005 Mn0.002Fe0.004Si1.001O3) |
| Sanghani et al 2022 | FIB TEM | In situ | Meteorite | CR | NWA 801_15 | Silicate | NS | 1 | R/AGB | 140 | Poly xl | Mg-rich core, Fe-rich rim | Rim: Mg/Si = 0.4, Fe/Si = 3.6; core: Mg/Si = 1.1, Fe/Si = 0.8 |
| | FIB TEM | In situ | Meteorite | | NWA 801_18 | Silicate | GEMS-like | 1 | R/AGB | 280 | Amorphous | Metal and sulfide grains 30-50 nm within grain and matrix | Mg/Si = 1.2, Fe/Si = 0.9 |
| | FIB TEM | In situ | Meteorite | | NWA 801_21 | Silicate | NS | 1 | R/AGB | 160 | N/A | Heterogeneities in Si, Mg, Fe | Mg/Si = 0.5, Fe/Si = 0.9 |
| | FIB TEM | In situ | Meteorite | | NWA 801_23 | Silicate | Olivine | 1 | R/AGB | 300 | Amorphous | Left vs right heterogeneities in Mg and Fe | R: Mg/Si = 1.5, Fe/Si = 0.7; L: Mg/Si - 1.1, Fe/Si = 1.0 |
| | FIB TEM | In situ | Meteorite | | NWA 801_14 | Silicate | Nontronite | 4 | SNe | 420 | Single xl? | Mg,Si-rich rim | Mg/Si = 0.3, Fe/Si = 2.4 |
| | FIB TEM | In situ | Meteorite | | NWA 801_17 | Silicate | GEMS-like | 4 | SNe | 180 | Poly xl | Heterogeneities in F and Fe | Mg/Si = 0.7, Fe/Si = 0.3 |
| | FIB TEM | In situ | Meteorite | | NWA 801_20 | Silicate | Pyroxene (En, Fs) | 4 | SNe | 70 | N/A | | Mg/Si = 0.7, Fe/Si = 0.5, Ca/Si = 0.1 |
| | FIB TEM | In situ | Meteorite | CH3/CBb3 | Isheyevo_9 | Silicate | Pyroxene (En) | 1 | R/AGB | 370 | Amorphous | Subgrain (oldhamite) | Mg/Si = 0.9, Fe/Si = 0.01 |
| | FIB TEM | In situ | Meteorite | | Isheyevo_4 | Silicate | Antigorite | 4 | SNe | 155 | Single xl? | | Mg/Si = 1.7, Fe/Si = 0.1 |
| Stroud et al 2004 | FIB TEM | Ex situ | Meteorite | OC | T103 | Oxide | Corundum | 1 | R/AGB | 1200 | Single xl | Trace Ti, no subgrains | N/A |
| | FIB TEM | Ex situ | Meteorite | OC | T96 | Oxide | Corundum | 1 | R/AGB | 1000 | Amorphous/nano xl | No subgrains | N/A |
| Stroud et al 2007 | FIB TEM | Ex situ | Meteorite | CI | 114-12 | Oxide | Al2O3 (not cor structure) | 2 | Low mass or CBP AGB | 500 | Single xl | No heterogeneities | 95 wt% Al2O3 |
| Stroud et al 2009 | FIB TEM | In situ | Meteorite | CR | 4c_3 | Silicate | NS | 1 | R/AGB | 350 | Amorphous/nano xl | Heterogeneities in Mg and Si. FeS adjacent to the PSG | (Mg+Fe)/Si from 0.7 to 1.5 |
| | FIB TEM | In situ | Meteorite | CR | 2b_8 | Silicate | NS | 1 | R/AGB | 350 | Amorphous/nano xl | N/A | N/A |
| Takigawa et al 2014 | FIB TEM | Ex situ | Meteorite | UOC (LL) | QUE053 | Oxide | Corundum | 1 | R/AGB | 500 | N/A | N/A | N/A |
| | FIB TEM | Ex situ | Meteorite | UOC (LL) | QUE067 | Oxide | Corundum | 4 | SNe | | N/A | N/A | N/A |
| Takigawa et al 2018 | FIB TEM | Ex situ | Meteorite | UOC (LL) | QUE060 | Oxide | Corundum | 2 | R/AGB | 1400 | Single xl | Mg heterogeneities | Mg/Al = 0.01 |
| Vollmer et al 2007 | FIB TEM | In situ | Meteorite | C2 ung | 1_07 | Silicate | Perovskite | Extreme 1 | ? | 500 | Single xl | Mg-rich core | Mg:Si = 1 on edges, Mg enrichment of 24% in core |
| Vollmer et al 2009 | FIB TEM | In situ | Meteorite | C2 ung | 14_2_3a | Silicate | GEMS-like | 1 or 2 | R/AGB | 470 | Amorphous, Fe xl | Homogeneous glass | Mg/Si = 0.54, Fe/Si =0.88–0.97, S = 2.6–3.8, Al, Ca ~1 |
| | FIB TEM | In situ | Meteorite | C2 ung | 4_11 | Silicate | GEMS-like sans S | 1 or 2 | R/AGB | 370 | Amorphous | Fe minerals dispersed along lines in homogeneous glass | Mg/Si = 0.22–0.47, Fe/Si = 0.02–0.1, Al, Ca, S <1 |
| | FIB TEM | In situ | Meteorite | C2 ung | 8_10 | Silicate | GEMS-like sans S | 1 or 2 | R/AGB | 810 | Amorphous | Fe minerals at grain boundaries | Mg/Si = 0.52–1.74, Fe/Si = 0.12–1.49, Al = 0.1–0.5, Ca - 0.2–1.9, S <1 |
| | FIB TEM | In situ | Meteorite | C2 ung | 32_08 | Silicate | LIME (low Fe, Mn enriched) | 1 or 2 | R/AGB | 200 | Amorphous, xl subarea | Homogeneous | Mg/Si = 0.86–1.37, Fe/Si = 0.04–0.05, Mn up to 0.7 |
| | FIB TEM | In situ | Meteorite | C2 ung | 21_06 | Silicate | NS | 1 or 2 | R/AGB | 570 | Amorphous cluster of particles | N/A | Mg/Si = 0.72–1.52, Fe/Si = 0.03–0.17, Al, Ca, S <1 |
| | FIB TEM | In situ | Meteorite | C2 ung | 18_08 | Silicate | Olivine | 1 or 2 | R/AGB | 480 | Single xl, amorphous rim | Homogeneous grain with relict xl in center | Center: Mg/Si = 1.13–1.34, Fe/Si = 0.31–0.43; Rim: Mg/Si = 0.36–0.46, Fe/Si = 0.82–0.91, S = 0.1–0.9, Al = 0.1–0.7, Ca = 0.1–1.1 |
| | FIB TEM | In situ | Meteorite | C2 ung | 31_13 | Silicate | Olivine | 1 or 2 | R/AGB | 670 | Xl with amorphous region in center | Differences between xl and amorphous regions | Xl: Mg/Si = 1.75–1.78, Fe/Si = 0.15–0.26 (Mg# 0.9), amor: Mg/Si = 1.07–1.33, Fe/Si = 0.28–0.34 |
| | FIB TEM | In situ | Meteorite | C2 ung | 25_20 | Silicate | Pyroxene (En) | 1 or 2 | R/AGB | 350 | Amorphous or weakly xl | Homogeneous | Mg/Si = 0.69–1.19, Fe/Si = 0.02–0.09, Al, Ca, S <0.5 |
| Vollmer et al 2013 | FIB TEM | In situ | Meteorite | C2 ung | 32_03 | Composite | Hibonite-NS | 1 | R/AGB | 200 | Nanocrystalline | Al-rich subregion embedded in Ca-Si-rich | Al/Si = 0.03–2.59, Mg/Si = 0.29–0.94, Ca/Si = 0.17–0.61, Fe/Si = 0–0.82 |
| | FIB TEM | In situ | Meteorite | C2 ung | 32_13 | Silicate | Pyroxene (En) | 1 | R/AGB | 400 | Single xl | Homogeneous | Mg/Si = 0.9 with Al |
| Zega & Floss 2013 | FIB TEM | In situ | Meteorite | C2 | 7a-1-o1 | Oxide | Mg-Al spinel | 1 | R/AGB | 304 | Single xl | Fe-rich rim (≤50 nm) | N/A |
| Zega et al 2011 | FIB TEM | Ex situ | Meteorite | UOC (LL) | KH1 | Oxide | Hibonite | 2 | R/AGB | 1340 | Single xl | Homogeneous | Ca1.04Al11.58Si0.09Mg0.14Ti0.11Fe0.06O19 |
| | FIB TEM | Ex situ | Meteorite | UOC (LL) | KH2 | Oxide | Hibonite | 4 | SNe | 600 | Single xl | Homogeneous | Ca0.98Al11.77Si0.02Mg0.14Ti0.09Fe0.01O19 |
| | FIB TEM | Ex situ | Meteorite | UOC (LL) | KH6 | Oxide | Hibonite | 1 | R/AGB | 322 | Single xl | Homogeneous | Stoichiometric CaAl12O19 |
| | FIB TEM | Ex situ | Meteorite | UOC (LL) | KH15 | Oxide | Hibonite | 2 | R/AGB | 2400 | Single xl with segment branching off | Homogeneous | Ca1.06Al11.69Mg0.06Ti0.17Fe0.01O19 |
| | FIB TEM | Ex situ | Meteorite | UOC (LL) | KH21 | Oxide | Hibonite | 1 | R/AGB | 3700 | Single xl with defects | Stacking faults Ca- and Mg-poor, Al-rich | Ca1.01Al11.73Mg0.21Ti0.07Si0.01Fe0.01O19 |
| Zega et al 2014 | FIB TEM | Ex situ | Meteorite | CI | ORG-36-21 | Oxide | Fe-Cr spinel | 1 | R/AGB | 1300 | Poly xl (3 xl) | Host = minor Ni, Ti; spinel subgrains present | (Fe+Mg)/(Cr+Al) = 0.55 |
| | FIB TEM | Ex situ | Meteorite | CM | Murray 2-19-13 | Oxide | Mg-Al spinel | 2 | Low mass or CBP AGB | 518 | Single xl | Homogeneous, minor Fe, Cr, and Si | Mg/Al = 0.56 |
| | FIB TEM | Ex situ | Meteorite | UOC | UOC-S1 | Oxide | Mg-Al spinel | 1 | R/AGB | 395 | Single xl | Homogeneous, minor Fe, Ca, Cr | Mg/Al = 0.31 |
| | FIB TEM | Ex situ | Meteorite | UOC | UOC-S2 | Oxide | Mg-Al spinel | 2 | Low mass or CBP AGB | 420 | Single xl | Homogeneous, minor Fe, Cr | Mg/Al = 0.51 |
| Zega et al 2015 | FIB TEM | In situ | Meteorite | CO | LAP-103 | Oxide | Magnetite | 1 | R/AGB | 750 | Single xl | Homogeneous | Stoichiometric Fe3O4 |

Anhy – anhydrous, ung – ungrouped, N/A – not available, NS – non-stoichiometric, Fo – forsterite, En – enstatite, xl – crystal/crystalline

[1]Full citations for the sources are in the references section

**Table S5.** Data for plots

| Plot | Description | Axes | | |
|---|---|---|---|---|
| Figure 1g | **Series** | **Label** | **O(+C)/Cations** | |
| | Vesicles | F2-37 v1 | 3.70 | |
| | | F2-37 v2 | 2.51 | |
| | | F2-37 v3 | 2.76 | |
| | | F2-37 v4 | 3.23 | |
| | | F2-23 v1 | 2.22 | |
| | | F2-23 v2 | 2.37 | |
| | | F1-1 v1 | 2.43 | |
| | Control | F2-37 v1 | 2.12 | |
| | | F2-37 v2 | 2.12 | |
| | | F2-37 v3 | 2.12 | |
| | | F2-37 v4 | 1.80 | |
| | | F2-23 v1 | 2.53 | |
| | | F2-23 v2 | 2.03 | |
| | | F1-1 v1 | 2.09 | |
| Figure 2f | **Series** | **Fe'** | **Mg'** | **Si'** |
| | Amorphous silicates | 40.10 | 8.53 | 51.36 |
| | | 31.12 | 14.34 | 54.53 |
| | | 29.07 | 19.08 | 51.86 |
| | | 40.69 | 9.35 | 49.96 |
| | | 28.89 | 15.78 | 55.33 |
| | | 31.76 | 12.14 | 56.09 |
| Figure 11 | **Label** | **(Ca+Mg+Fe)/Si** | **Mg#** | |
| | F2-9 Si-rich | 0.38 | 94.31 | |
| | F2-9 Mg-rich | 1.69 | 98.40 | |
| | F2-9 rim | 2.38 | 63.25 | |
| | F2-30b | 1.38 | 15.20 | |
| | F2-23 | 1.44 | 27.02 | |
| | F1-1 Si-rich | 0.23 | 88.92 | |
| | F1-1 Al-rich | 0.43 | 84.42 | |
| | F1-1 Cr-rich | 1.25 | 56.81 | |
| | F1-1 Mg-rich | 1.39 | 75.12 | |
| | F2-30a amorphous silicate | 0.95 | 17.55 | |
| | F2-37 amorphous silicate | 0.85 | 31.54 | |
| | F2-9 amorphous silicate | 0.94 | 39.63 | |
| | F2-30b amorphous silicate | 1.02 | 18.69 | |
| | F2-23 amorphous silicate | 0.82 | 35.32 | |
| | F1-1 amorphous silicate | 0.80 | 27.66 | |
| Figure S2e | **Series** | **C (at%)** | **O (at%)** | |
| | Vesicles | 29.88 | 48.86 | |
| | | 4.07 | 67.45 | |
| | | 26.69 | 46.71 | |
| | | 43.39 | 32.98 | |
| | | 0.00 | 68.91 | |
| | | 0.00 | 70.33 | |
| | | 8.16 | 62.68 | |

| Plot | Description | Axes | |
|---|---|---|---|
| Figure 9 | **Label** | **Crystallinity** | **Stoichiometry** |
| | Oxides CCs | Single xl | Stoichiometric |
| | | Single xl | Stoichiometric |
| | | Single xl | Stoichiometric |
| | | Single xl | Stoichiometric |
| | | Single xl | Stoichiometric |
| | | Single xl | Stoichiometric |
| | | Polyxl | Stoichiometric |
| | | Polyxl | Stoichiometric |
| | | Weakly nanoxl | Stoichiometric |
| | | Weakly nanoxl | Stoichiometric |
| | Oxides OCs | Single xl | Non-stoichiometric |
| | | Single xl | Stoichiometric |
| | | Single xl | Stoichiometric |
| | | Single xl | Stoichiometric |
| | | Single xl | Stoichiometric |
| | | Single xl | Stoichiometric |
| | | Single xl | Stoichiometric |
| | | Single xl | Stoichiometric |
| | | Single xl | Stoichiometric |
| | Silicates IDPs | Weakly nanoxl | Non-stoichiometric |
| | | Weakly nanoxl | Non-stoichiometric |
| | | Polyxl | Stoichiometric |
| | | Single xl | Stoichiometric |
| | Silicates CCs | Weakly nanoxl | Non-stoichiometric |
| | | Weakly nanoxl | Non-stoichiometric |
| | | Weakly nanoxl | Non-stoichiometric |
| | | Weakly nanoxl | Non-stoichiometric |
| | | Weakly nanoxl | Non-stoichiometric |
| | | Weakly nanoxl | Non-stoichiometric |
| | | Weakly nanoxl | Non-stoichiometric |
| | | Weakly nanoxl | Non-stoichiometric |
| | | Weakly nanoxl | Non-stoichiometric |
| | | Weakly nanoxl | Non-stoichiometric |
| | | Weakly nanoxl | Non-stoichiometric |
| | | Weakly nanoxl | Non-stoichiometric |
| | | Weakly nanoxl | Non-stoichiometric |
| | | Weakly nanoxl | Non-stoichiometric |
| | | Weakly nanoxl | Non-stoichiometric |
| | | Weakly nanoxl | Non-stoichiometric |
| | | Weakly nanoxl | Non-stoichiometric |
| | | Weakly nanoxl | Stoichiometric |
| | | Weakly nanoxl | Stoichiometric |
| | | Weakly nanoxl | Stoichiometric |
| | | Weakly nanoxl | Stoichiometric |
| | | Weakly nanoxl | Stoichiometric |
| | | Weakly nanoxl | Stoichiometric |
| | | Weakly nanoxl | Stoichiometric |
| | | Polyxl | Non-stoichiometric |
| | | Polyxl | Non-stoichiometric |
| | | Single xl | Stoichiometric |
| | | Single xl | Stoichiometric |
| | | Single xl | Stoichiometric |
| | | Single xl | Stoichiometric |
| | | Single xl | Stoichiometric |
| | | Single xl | Stoichiometric |
| | | Single xl | Stoichiometric |
| | | Single xl | Stoichiometric |
| | | Single xl | Stoichiometric |
| | | Single xl | Stoichiometric |
| | | Single xl | Stoichiometric |
| | Silicates OCs | Weakly nanoxl | Non-stoichiometric |
| | | Weakly nanoxl | Non-stoichiometric |
| | | Weakly nanoxl | Non-stoichiometric |
| | | Weakly nanoxl | Non-stoichiometric |

Fe' – (Fe/(Fe+Mg+Si))x100, Mg' – (Mg/(Fe+Mg+Si))x100, Si' – (Si/(Fe+Mg+Si))x100, xl – crystal/crystalline